\documentclass[aps,amsfonts,twocolumn,superscriptaddress,preprintnumbers,nofootinbib]
{revtex4-1}
\usepackage{graphicx}
\usepackage{times}
\usepackage{epsfig}
\usepackage{multirow}
\usepackage{amsmath}
\usepackage{amssymb}
\usepackage{longtable} 
\usepackage{bm}
\usepackage{subfigure} 
\usepackage{float}
\begin{document}

\title{Thermal QED theory for bound states}

\author{D. Solovyev}
\email{solovyev.d@gmail.com}
\affiliation{Department of Physics, St.Petersburg State University, St.Petersburg, 198504, Russia}

\begin{abstract}
The paper presents the Quantum Electrodynamics theory for bound states at finite temperatures. To describe the thermal effects arising in a heat bath, the Hadamard form of thermal photon propagator is employed. As the form allows a simple introduction of thermal gauges in a way similar to the 'ordinary' Feynman propagator, the gauge invariance can be proved for all the considered effects. Moreover, unlike the 'standard' form of the thermal photon propagator, the Hadamard expression offers well-defined analytical properties, yet contains a divergent contribution, which requires the introduction of a regularization procedure within the framework of the constructed theory. The method and physical interpretation of the regularization are given in the paper. Correctness of the procedure is confirmed also by the gauge invariance of final results and the coincidence of the results (as exemplified by the self-energy correction) for two different forms of photon propagator. The constructed theory is used to find the thermal Coulomb potential and its asymptotic at large distances. Finally, the thermal effects of the lowest order in the fine structure constant and temperature are discussed in detail. Such effects are represented by the thermal one-photon exchange between a bound electron and the nucleus, thermal one-loop self-energy, thermal vacuum polarization, and recoil corrections and that for the finite size of the nucleus. Introduction of the regularization allows one avoid applying the renormalization procedure. To confirm this, the thermal vertex (with one, two, and three vertices) corrections is also described within the adiabatic $S$-matrix formalism. Finally, the paper discusses the influence of thermal effects on the finding the proton radius and Rydberg constant.
\end{abstract}

\maketitle

\tableofcontents

\section{Introduction}
\label{Int}
The influence of blackbody radiation (BBR) on atomic systems is an important subject in modern atomic physics. Research in the field started at the end of 70s. First, the BBR-induced effects were detected experimentally (as the BBR-induced photo-ionization from Rydberg energy levels), and then the theoretical description was given within the framework of the quantum-mechanical (QM) approach in \cite{GC,Farley}. Measurements of the BBR-induced energy shifts for Rydberg levels of Rb atom were performed in \cite{Hall} with the use of high-precision laser spectroscopic techniques. In \cite{Hall}, it was noted that measurements were consistent with the predicted finite-temperature radiative corrections to atomic energy levels induced by BBR. Now, the issue of the blackbody radiation influence on atoms is widely discussed in literature. Particular interest to such kind of investigations arises in view of the substantial progress in theoretical and experimental research of atomic clocks and determination of frequency standards \cite{Saf,Porsev,Deg,SKC,Itano,Mid,Rob}. The typical approach that is usually applied to the analysis of BBR-induced effects in atomic systems corresponds to the quantum mechanical evaluation of the Stark shifts and depopulation rates for highly-excited (Rydberg) states \cite{Farley,GC}.

Also, the middle of 70s \cite{Weinberg} saw the development of the quantum field theory at finite temperatures, where the spontaneous symmetry breaking at high temperatures and its cosmological implications were discussed. Feynman rules for scalar and vector fields and for finite-temperature Green's functions were derived in \cite{Bernard74,Dol}. A brief historical review of the field theory at finite temperatures and effects on scattering processes and decay rates can be found in \cite{Don,DHR}. Namely, a renormalization prescription for fermions at finite temperatures and a procedure for calculation of radiative corrections of free particles were given in \cite{Don}. The details of the quantum electrodynamics (QED) theory for free particles at finite temperatures can be found in \cite{DHR}. The application of the theory to bound states was demonstrated also in \cite{DHR}, where the thermal correction to the Lamb shift in a hydrogen atom was roughly estimated.

As noted in \cite{DHR}, the existing methods and the results of many of special calculations in literature do not always agree with one another. The same situation still persists, since authors typically apply methods of the thermal QED theory for free particles to describe thermal effects in atomic systems (bound particles). The extensive applications of the non-relativistic quantum electrodynamics  theory to the theoretical analysis of the BBR influence upon atomic systems abound in literature, see, e. g., \cite{Soto,Soto2}, where the electronic and muonic hydrogen atoms were treated.

Recently, a rigorous quantum electrodynamic derivation of the Stark shift and depopulation rates of atomic energy levels in the presence of the blackbody radiation was performed in \cite{SLP-QED}, where the perfect agreement between QED (in non-relativistic limit) and QM results was also demonstrated. An attempt to evaluate the relativistic correction to the one-loop BBR shift in the ground states of hydrogen and ionized helium was given in \cite{china-BBR,Soto,Soto2}. Another application of the QED description of thermal effects in atoms was reported in \cite{SLP-QED} with the use of the 'QED regularization' of the self-energy correction. Namely, such regularization was supposed to lead to the level-mixing effect in the BBR field, which strongly exceeds the QM result for depopulation rates.

However, the calculations of $\Gamma_a^{\rm mix}$ were performed incorrectly, the factor $\alpha^3$ ($\alpha$ in the fine structure constant) was left out. The present paper will show the numerical results for $\Gamma_a^{\rm mix}$ to be much smaller, but still exceed the quantum mechanical ones \cite{Farley} by several orders of magnitude. The analysis of such behavior and accurate calculations are presented in \cite{zal2018}, where the mixing effect is suggested to take place at very low temperatures only. At high temperatures, the QM depopulation rates \cite{Farley} and corresponding QED values coincide within high accuracy. The main conclusion that can be drawn here is that the QED theory at finite temperatures for bound states still requires further examination and development.

To reveal the 'new' effects and re-examine well-known ones, which arise due to the blackbody radiation, we draw on the QED theory at finite temperatures developed in \cite{Dol,Don,DHR}. Within the framework of the theory, a free-electron gas (no external field) that interacts with a photon gas is considered as being in the thermal equilibrium and is described by a grand canonical statistical operator, which modifies both the electron and the photon propagators. Since our task is to describe the blackbody radiation influence on bound states, we will retain the electron propagator in the standard form (QED theory for bound states) and treat the influence of the BBR within QED perturbation theory involving the thermal photon propagator only. This is validated by the thermal part of the fermion propagator being suppressed by the factor $exp(-\beta m_e)$ ($m_e$ is the fermion mass) \cite{DHR}.

The main difficulty in developing such a theory is that the thermal photon propagator found in \cite{Dol,Don,DHR} possesses no well-defined analytical properties, see \cite{FW,kapusta-FTFT}. Thus, we start our research from the very beginning, i.e. we re-examine the derivation of the photon propagator in case of the heated vacuum. Although the final results in our paper are given in the non-relativistic limit, the relativistic corrections can be easily found from the theory developed below.

\section{QED derivation of photon propagator at finite temperatures}
\label{pp}
\subsection{Vacuum-expectation value of the T-product}
\label{vev}
At first, we derive the thermal photon propagator and show it to be defined by the Hadamard propagation function. According to the QED theory, \cite{Akhiezer,Greiner}, the photon propagator is a time-ordered product of two photon field operators averaged over the vacuum state:
\begin{eqnarray}
\label{1}
iD^{\mu\nu}_F(x,x')=\langle 0| T\left(\hat{A}^{\mu}(x)\hat{A}^\nu(x')\right) |0\rangle,
\end{eqnarray}
where vector-potential operator $\hat{A}_{\mu} (x)$ is defined by
\begin{eqnarray}
\label{2}
\hat{A}^{\mu} (x) = \sum\limits_{\lambda}\int\frac{e_{\lambda, k}^{\mu}d^3k}{(2\pi)^{3/2}}\sqrt{\frac{4\pi}{2\omega_k}}
\left(\hat{C}_{\lambda k}e^{ik x}+\hat{C}^{\dagger}_{\lambda k}e^{-ik x}\right).\,\,
\end{eqnarray}
Here $e_{\lambda, k}^{\mu}$ is the four-dimensional polarization vector ($\lambda, \mu =0,1,2,3$), $k$ and $x$ are the ordinary four-dimensional wave and time-space vectors, $\omega_k=|\vec{k}|$ is the photon frequency, and notations $\hat{C}_{\lambda k}$, $\hat{C}^{\dagger}_{\lambda k}$ correspond to the annihilation and creation operators, respectively.

\begin{figure}[hbtp]
	\centering
	\includegraphics[scale=0.15]{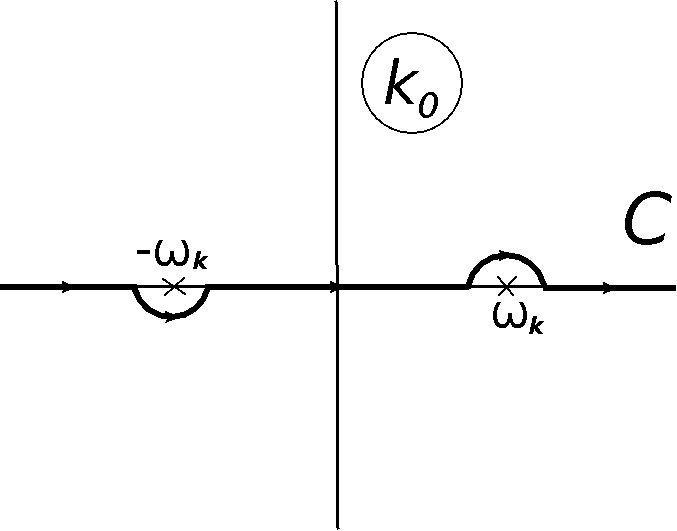}
	\caption{The integration contour $C$ (bold line) in the complex plane of $k_0$ of Eq. (\ref{5}). The $\pm\omega_k$ denote the poles of the integrand.}\label{Fig-1}
\end{figure}

Substitution of (\ref{2}) into Eq. (\ref{1}) leads to the expression, which contains four terms. Two of them are zero, since $\hat{C}_{\lambda \vec{k}}|0\rangle = 0$ and $\langle 0|\hat{C}^{\dagger}_{\lambda \vec{k}} = 0$, where $|0\rangle$ is the vacuum state. Then, for the vacuum-expectation value of the T-product, we obtain
\begin{widetext}
\begin{eqnarray}
\label{3}
\langle 0|{\rm T}(A^{\mu} (x') A^{\nu} (x'))|0\rangle = \int\frac{4\pi\, d^3k\, d^3k'}{2(2\pi)^3\sqrt{\omega_k\omega_{k'}}}
\sum\limits_{\lambda \lambda'}e_{\lambda, \vec{k}}^{\mu}e_{\lambda', \vec{k'}}^{\nu}
\langle 0|{\rm T}\left[\hat{C}_{\lambda \vec{k}}\hat{C}^{\dagger}_{\lambda' \vec{k'}}e^{i\vec{k}\vec{x}-i\omega_k t}e^{-i\vec{k'}\vec{x'}+i\omega'_k t'}
+\hat{C}^{\dagger}_{\lambda \vec{k}}\hat{C}_{\lambda' \vec{k'}}e^{-i\vec{k}\vec{x}+i\omega_k t}e^{i\vec{k'}\vec{x'}-i\omega'_k t'}\right]|0\rangle.
\end{eqnarray}
Employing relationships $\langle 0 |\hat{C}_{\lambda \vec{k}} \hat{C}^{\dagger}_{\lambda' \vec{k'}}|0\rangle = g_{\lambda \lambda'}\delta(\vec{k}-\vec{k'})$ and $g_{\lambda \lambda'}e_{\lambda, \vec{k}}^{\mu}e_{\lambda', \vec{k}}^{\nu} = g^{\mu \nu}$ ($g^{\mu \nu}$ is the pseudo-Euclidean metric tensor in Minkowski's space), we find
\begin{eqnarray}
\label{4}
\langle 0|{\rm T}(A^{\mu}(x) A^{\nu}(x'))|0\rangle = 4\pi g^{\mu \nu}\int \frac{d^3k}{(2\pi)^3 2\omega_k}\left[\theta(t-t')e^{i\vec{k}(\vec{x}-\vec{x'})-i\omega_k (t-t')}+
\theta(t'-t)e^{-i\vec{k}(\vec{x}-\vec{x'})+i\omega_k (t-t')}\right]=
\\
\nonumber
= 4\pi g^{\mu \nu}\int \frac{d^3k}{(2\pi)^3 2\omega_k}\left[e^{i\vec{k}(\vec{x}-\vec{x}')}
+
e^{-i\vec{k}(\vec{x}-\vec{x'})}\right]e^{-i\omega_k |t-t'|}
= 4\pi g^{\mu \nu}\int \frac{d^3k}{(2\pi)^3 \omega_k} e^{i\vec{k}(\vec{x}-\vec{x'})} e^{-i\omega_k |t-t'|}.
\end{eqnarray}
\end{widetext}
Here $\theta(\tau)$ is the Heaviside theta-function. With the use of
\begin{eqnarray}
\label{5}
\frac{1}{\omega_k}e^{-i\omega_k |\tau|} = \frac{1}{\pi i}\int\limits_C \frac{e^{-ik_0\tau}}{k_0^2-\omega_k^2}dk_0,
\end{eqnarray}
the vacuum-expectation value of the T-product reduces to
\begin{eqnarray}
\label{6}
\langle 0|{\rm T}(A^{\mu}(x_1) A^{\nu}(x_2))|0\rangle =
-i4\pi g^{\mu \nu}\int\limits_C \frac{d^4k}{(2\pi)^4}
\frac{e^{ik(x-x')}}{k^2},\qquad
\end{eqnarray}
where $k^2 =  k_0^2 - \vec{k}^2 =  k_0^2 - \omega_k^2$ and integration contour $C$, see  \cite{Berest}, is depicted in Fig.~\ref{Fig-1}.

Thus, the photon propagator in the Feynman gauge is
\begin{eqnarray}
\label{7}
D^{\mu\nu}_F(x,x') = -4\pi g^{\mu \nu}\int \frac{d^4k}{(2\pi)^4}\frac{e^{ik(x-x')}}{k_0^2-\omega_k^2-i0}.
\end{eqnarray}
The integration over the angles and over the poles in the complex plane $k_0$ of Eq. (\ref{7}) leads to the well-known expression:
\begin{eqnarray}
\label{8}
D^{\mu\nu}_F(x,x')= \frac{g_{\mu \nu}}{2\pi i r_{12}}\int\limits_{-\infty}^{\infty}e^{i|\omega|r_{12}-i\omega(t-t')}d\omega,
\end{eqnarray}
where $r_{12}\equiv |\vec{x}-\vec{x'}|$ and integration over frequency $\omega$ runs through all possible values including the negative half-axes.

However, the photon propagator can be presented also in another form. To this end, the chronological product Eq. (\ref{1}) can be rewritten as a sum of a commutator and an anti-commutator:
\begin{eqnarray}
\label{9}
iD^{\mu\nu}_F(x,x')\equiv i D^{\mu\nu}(x) + i D^{\mu\nu}_1(x)= 
\frac{1}{2}\langle 0|\rm{sgn}(t-t')\left[\hat{A}^{\mu}(x),\hat{A}^\nu(x')\right]
+\left\{\hat{A}^\mu(x),\hat{A}^{\nu}(x')\right\}|0\rangle,
\end{eqnarray}
where the first term represents the Pauli-Jordan function, and the second one is the Hadamard function \cite{Greiner}.

\begin{figure}[hbtp]
	\centering
	\includegraphics[scale=0.18]{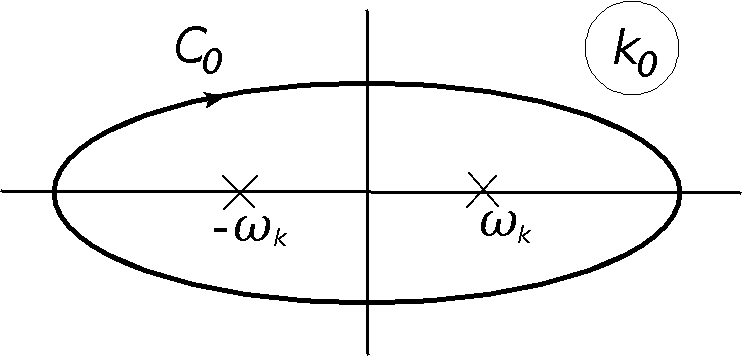}
	\caption{Integration contour $C_0$ in $k_0$ plane of Eq. (\ref{10}). Arrows on the contour define the pole-bypass rule. The poles $\pm\omega_k$ are denoted with X marks.}
	\label{Fig2}
\end{figure}

\begin{figure}[hbtp]
	\centering
	\includegraphics[scale=0.2]{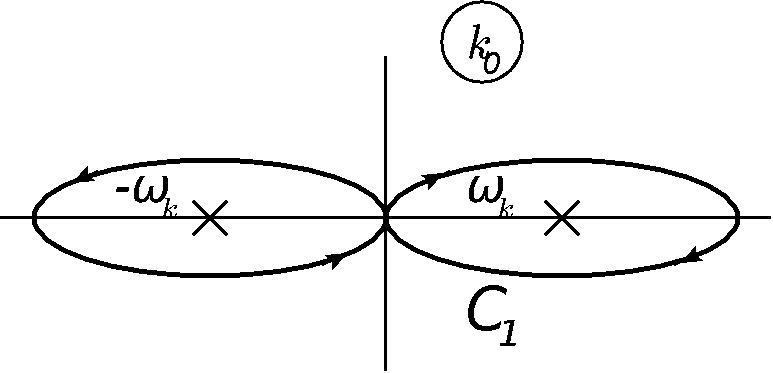}
	\caption{Integration contour $C_1$ in $k_0$ plane of Eq. (\ref{11}). Notations are those of Fig.~\ref{Fig2}.}
	\label{Fig3}
\end{figure}

In the four-dimensional integration form (see \cite{Akhiezer}), Pauli-Jordan function $D^{\mu\nu}(x)$ is
\begin{eqnarray}
\label{10}
D^{\mu\nu}(x,x') 
 = -4\pi\frac{g^{\mu\nu}}{(2\pi)^4}\int\limits_{C_0}\frac{e^{ik(x-x')}}{k^2}d^4k,\qquad
\end{eqnarray}
and the Hadamard propagation function is defined by
\begin{eqnarray}
\label{11}
D_1^{\mu\nu}(x,x') 
= - 4\pi\frac{g^{\mu\nu}}{(2\pi)^4 i}\int\limits_{C_1}\frac{e^{ik(x-x')}}{k^2}d^4k\qquad
\end{eqnarray}
and can be also given in the equivalent form:
\begin{eqnarray}
\label{12}
D_1^{\mu\nu}(x,x') = - 4\pi g^{\mu\nu}\int\frac{d^4k}{(2\pi)^3} \delta(k^2)e^{ik(x-x')}.
\end{eqnarray}
The contours of integration in $k_0$-plane for Eqs. (\ref{10}) and (\ref{11}) are given in Figs.~\ref{Fig2}, \ref{Fig3}.
We should remind that these two equivalent forms of the Hadamard function (Eqs. (\ref{11}), (\ref{12})) are related directly to the thermal averaged photon propagator, which is derived in the next section.

\subsection{The thermal part of the T-product and thermal photon propagator}
\label{thev}

To determine the heated vacuum expectation value, we are to examine the ensemble-averaged chronological product of vector potentials of Eq. (\ref{2}): 
\begin{eqnarray}
\label{13}
iD^{\mu \nu}_\beta(x, x') = \langle {\rm T}(A^{\mu}(x) A^{\nu}(x'))\rangle_\beta
\equiv{\rm Tr}\left\{\rho\,[{\rm T}(A_{\mu} (x) A_{\nu} (x'))]\right\},
\end{eqnarray}
where $\rho$ denotes (in the zeroth approximation) the statistical operator for the non-interacting photons, electrons, and positrons. We consider the bound electrons case, when the heat bath influence is much smaller in comparison to the Coulomb interaction of charges. Then, Eq. (\ref{13}) holds valid at temperatures $k_BT\ll (\alpha Z)^2m_ec^2$.

With the use of decomposition Eq. (\ref{9}), the heated vacuum expectation value of the chronological product is given by
\begin{eqnarray}
\label{14}
\langle {\rm T}(A^{\mu}(x) A^{\nu}(x'))\rangle_\beta = \frac{1}{2}\langle\left\{A^{\mu}(x), A^{\nu}(x')\right\}\rangle_\beta
\qquad
\\
\nonumber
+{\rm sgn(t-t')}\frac{1}{2}\langle[A^{\mu}(x), A^{\nu}(x')]\rangle_\beta.
\end{eqnarray}
Since the commutator $[A^{\mu}(x), A^{\nu}(x')]$ after T-ordering and averaging over the vacuum becomes a $c$-number, we can rewrite Eq. (\ref{14}) in the form:
\begin{eqnarray}
\label{15}
iD^{\mu \nu}_{F,\beta}(x, x') = \frac{1}{2}{\rm sgn}(t-t')\langle 0| \left[\hat{A}^\mu(x),\hat{A}^{\nu}(x')\right] |0\rangle
\qquad
\\
\nonumber
+\frac{1}{2}\left\langle \left\{\hat{A}^\mu(x),\hat{A}^{\nu}(x')\right\} \right\rangle_\beta.\qquad
\end{eqnarray}
Then, combining Eqs. (\ref{16}) and (\ref{9}),
\begin{eqnarray}
\label{16}
iD^{\mu \nu}_{F,\beta}(x, x') = iD^{\mu \nu}_F(x, x')
\\
\nonumber
+\frac{1}{2}\left\langle \left\{\hat{A}^\mu(x),\hat{A}^{\nu}(x')\right\} \right\rangle_\beta - i D_1^{\mu\nu}(x,x'),
\end{eqnarray}
where $D^{\mu \nu}_F(x, x')$ is defined by Eq. (\ref{9}) and represents the zeroth (unheated) vacuum Feynman propagator.

Introducing the definition
\begin{eqnarray}
\label{17}
D^{\mu \nu}_{F,\beta}(x, x') = D^{\mu \nu}_F(x, x')+D^{\mu \nu}_\beta(x, x'),
\end{eqnarray}
we find the thermal part of photon propagator, $D^{\mu \nu}_\beta(x, x')$, to be defined by Hadamard function $D_1$ as
\begin{eqnarray}
\label{18}
D^{\mu \nu}_\beta(x, x')
= \langle :A_{\mu}(x) A_{\nu}(x'):\rangle_\beta
\\
\nonumber
=\frac{1}{2}\left(D_{1,\beta}^{\mu\nu}(x,x')-D_1^{\mu\nu}(x,x')\right),
\end{eqnarray}
i.e. the thermal part of the photon propagator can be given in the form of the four-dimensional integration with the contour of Fig.~\ref{Fig3}.

Then, with the relationships
\begin{eqnarray}
\label{19}
\left\langle \left\{\hat{C}^{\dagger}_{\lambda k},\hat{C}_{\lambda' q}\right\}\right\rangle_\beta &=& 2\left\langle\hat{C}^{\dagger}_{\lambda k}\hat{C}_{\lambda' q}\right\rangle_\beta
+\left\langle \left[\hat{C}_{\lambda' q},\hat{C}^{\dagger}_{\lambda k}\right]\right\rangle_\beta,
\nonumber
\\
\left[\hat{C}_{\lambda' q},\hat{C}^{\dagger}_{\lambda k}\right] &=& -g_{\lambda\lambda'}\delta(\vec{k}-\vec{q}),
\\
\nonumber
\left\langle\hat{C}^{\dagger}_{\lambda k}\hat{C}_{\lambda q}\right\rangle_\beta &=& -g_{\lambda\lambda'}\delta(\vec{k}-\vec{q})n_\beta(\omega_k),
\end{eqnarray}
we find
\begin{eqnarray}
\label{20}
iD^{\mu \nu}_{\beta}(x, x') = -4\pi\int\frac{d^3k}{(2\pi)^3}\frac{e^{i\vec{k}(\vec{x}-\vec{x}')}}{2\omega_k}\sum\limits_{\lambda=0}^3 e_{\lambda, \vec{k}}^{\mu}e_{\lambda, \vec{k}}^{\nu}g_{\lambda\lambda}
\qquad
\nonumber
\\
\times
\left(e^{i\omega_k (t-t')}+e^{-i\omega_k (t-t')}\right)\left(2n_\beta(\omega_k)+1\right)+\qquad
\\
\nonumber
\int\frac{d^3k}{(2\pi)^3}\frac{e^{i\vec{k}(\vec{x}-\vec{x}')}}{2\omega_k}\sum\limits_{\lambda=0}^3 e_{\lambda, \vec{k}}^{\mu}e_{\lambda, \vec{k}}^{\nu}
\left(e^{i\omega_k (t-t')}+e^{-i\omega_k (t-t')}\right),
\end{eqnarray}
where $n_\beta(\omega_k) = (exp(\beta\omega_k)-1)^{-1}$, $\beta^{-1}=k_B T$, $k_B$ is the Boltzmann constant and $T$ is the temperature in Kelvins. This expression can be obviously transformed to
\begin{eqnarray}
\label{21}
iD^{\mu \nu}_{\beta}(x, x') = -4\pi\sum\limits_{\lambda=0}^3 e_{\lambda, k}^{\mu}e_{\lambda, k}^{\nu}g_{\lambda\lambda}
\\
\nonumber
\times
\int\frac{d^3k}{(2\pi)^3}e^{i\vec{k}(\vec{x}-\vec{x}')}\frac{\cos\omega_k (t-t')}{\omega_k} n_\beta(\omega_k).
\end{eqnarray}
We can also use here the relationship $2n_\beta(\omega) = \coth(\frac{\beta\omega}{2})-1$, thus, arriving at the expression
\begin{eqnarray}
\label{22}
iD^{\mu \nu}_{\beta}(x, x') = -4\pi\sum\limits_{\lambda=0}^3 e_{\lambda, k}^{\mu}e_{\lambda, k}^{\nu}g_{\lambda\lambda}
\qquad\qquad
\\
\nonumber
\times
\int\frac{d^3k}{(2\pi)^3}e^{i\vec{k}(\vec{x}-\vec{x}')}\frac{\cos\omega_k (t-t')}{2\omega_k} \left(\coth\left(\frac{\beta\omega_k}{2}\right)-1\right),
\end{eqnarray}
which coincides with the result of \cite{Don,DHR}.

Two different forms of the thermal photon propagator can be obtained from Eq. (\ref{21}). The first one corresponds to the four-dimensional integration arising via the relationship
\begin{eqnarray}
\label{23}
\frac{\cos\omega_k (t-t')}{\omega_k} = -\frac{1}{2\pi i}\int\limits_{C_1}dk_0 \frac{e^{-ik_0(t-t')}}{k^2}.
\end{eqnarray}
Employing the completeness relation \cite{GrRe} for the polarization vectors, $\sum\limits_{\lambda=0}^3 e_{\lambda, k}^{\mu}e_{\lambda, k}^{\nu}g_{\lambda\lambda} = g^{\mu\nu}$, we obtain
\begin{eqnarray}
\label{24}
i D^{\mu \nu}_{\beta}(x, x') =
- 4\pi i g^{\mu\nu}\int\limits_{C_1}\frac{d^4k}{(2\pi)^4} \frac{e^{ik(x-x')}}{k^2}n_\beta(\omega_k).\qquad
\end{eqnarray}
Thus, the thermal photon propagator can be defined by the Hadamard propagation function.

The form of Eq. (\ref{21}) allows analytical integration in $k$-space. To this end, we integrate first over angles:
\begin{eqnarray}
\label{24a}
i D^{\mu \nu}_{\beta}(x, x') = - 4\pi g^{\mu\nu}\frac{1}{2\pi^2}\int\limits_0^\infty d\kappa\frac{\sin\kappa r}{r}\frac{\cos\kappa\tau}{e^{\beta\kappa}-1},
\end{eqnarray}
where notations $r\equiv |\vec{r}-\vec{r}'|$, $\tau\equiv t-t'$ were introduced. Next, the analytical integration over $\kappa$ yields
\begin{eqnarray}
\label{24b}
i D^{\mu \nu}_{\beta}(x, x') = - 4\pi g^{\mu\nu}\frac{1}{2\pi^2}
\times
\\
\nonumber
\frac{\beta r\left(\cosh\frac{2\pi\tau}{\beta}-\cosh\frac{2\pi r}{\beta}\right)+\pi\left(r^2-\tau^2\right)\sinh\frac{2\pi r}{\beta}}{4\beta r\left(r^2-\tau^2\right)\sinh\frac{\pi(r-\tau)}{\beta}\sinh\frac{\pi(r+\tau)}{\beta}}
\end{eqnarray}
Then, the use of $\cosh x-\cosh y = 2\sinh\frac{x+y}{2}\sinh\frac{x-y}{2}$ allows transforming Eq. (\ref{24b}) to
\begin{eqnarray}
\label{25}
iD^{\mu \nu}_{\beta}(x, x') = \frac{4\pi g^{\mu\nu}}{(2\pi)^2}\left[\frac{1}{r^2-\tau^2}-\frac{\frac{\pi}{2r\beta}\sinh\frac{2\pi r}{\beta}}{\cosh{\frac{2\pi r}{\beta}}-\cosh{\frac{2\pi \tau}{\beta}}}\right],
\end{eqnarray}
which gives the space-time representation of the thermal photon propagator.

The second equivalent form for the thermal part of photon propagator can be found with the use of the following relationships:
\begin{eqnarray}
\label{26}
e^{\pm i\omega_k\tau} = \int\limits_{-\infty}^\infty dk_0 e^{ik_0\tau}\delta(k_0\mp \omega_k),
\\
\nonumber
\delta(x^2-a^2)=\frac{1}{2|a|}\left(\delta(x-a)+\delta(x+a))\right).
\end{eqnarray}
Substitution of Eq. (\ref{26}) into Eq. (\ref{21}) gives
\begin{eqnarray}
\label{27}
iD^{\mu \nu}_{\beta}(x, x') = -4\pi g^{\mu\nu}\int\frac{d^3k}{(2\pi)^3}n_\beta(\omega)e^{i\vec{k}(\vec{r}-\vec{r}')}
\times
\qquad
\\
\nonumber
\frac{1}{2\omega}\int\limits_{-\infty}^\infty dk_0\left[e^{ik_0(t-t')}\delta(k_0-\omega)+e^{ik_0(t-t')}\delta(k_0+\omega)\right].
\end{eqnarray}
The final expression
\begin{eqnarray}
\label{28}
iD^{\mu \nu}_{\beta}(x, x') =
-4\pi g^{\mu\nu}\int\frac{d^4k}{(2\pi)^3}n_\beta(\omega_k)\delta(k^2)e^{ik(x-x')}.\qquad
\end{eqnarray}
coincides with the result given in \cite{Dol,Don,DHR}. After the angular integration and some algebraic transformations (see \cite{SLP-QED}), we can also find
\begin{eqnarray}
\label{29}
D^{\mu\, \nu}_{F,\beta}(x_1,\, x_2) = \frac{g_{\mu\, \nu}}{2\pi i r_{12}}\int\limits_{-\infty}^{+\infty}d\omega e^{i|\omega|r_{12} - i\omega(t_1-t_2)}
\\
\nonumber
- \frac{g_{\mu\, \nu}}{\pi r_{12}}\int\limits_{-\infty}^{+\infty}d\omega n_{\beta}(|\omega|) \sin{|\omega|r_{12}} e^{-i\omega(t_1-t_2)}.
\end{eqnarray}

Thus, we have found two equivalent forms of the thermal photon propagator, Eqs. (\ref{24}) and (\ref{28}). The latter coincides with the result of the theory presented in \cite{Dol,Don,DHR}. However, the form Eq. (\ref{28}) has no well-defined analytical properties \cite{Land}. It was underscored in \cite{Land} that the delta function in Eq. (\ref{28}) should be regarded as an abbreviation for the regularized representation; it is only in the simplest diagrams that the delta functions can be taken literally, see also \cite{FW}. Moreover, the photon propagator, Eq. (\ref{28}), becomes undetermined at point $k_0=0$ (the static limit), where $\delta(|\vec{k}|)$ arises on the semi-axis of integration.

Contrary to this, the form Eq. (\ref{24}) has no indeterminacies. What is more, it allows a simple introduction of gauges. Yet, both forms of thermal photon propagator contain a divergence, which originates from distribution function $n_\beta$ at point $\omega_k=0$. The paper \cite{DHR} demonstrated the renormalization procedure of such divergences for the case of free particles. The procedure will be shown to fail to work for bound states (there are no infrared divergences in the QED for bound states), and a regularization procedure to replace it will be described in section~\ref{cdc}.

\subsection{The thermal photon propagator: the Coulomb part}
\label{tppCp}
To clarify the contributions of the transverse and Coulomb parts of the thermal photon propagator, a set of general polarization vectors can be introduced for metric tensor $g^{\mu\nu}=e^\mu_{0,k}e^\nu_{0,k}-\sum\limits_{\lambda=1}^2e^\mu_{\lambda,k}e^\nu_{\lambda,k}-e^\mu_{3,k}e^\nu_{3,k}$ \cite{GrRe}. Namely, starting from an arbitrary time-like unit vector $n^\mu=(1,0,0,0)$, the scalar, transverse and longitudinal polarization vectors are
\begin{eqnarray}
\label{30}
e^\mu_{0,k}\equiv n^\mu = \left(\begin{array}{c}
1 \\
\vec{0}
\end{array}\right),
\\
\nonumber
e^\mu_{\lambda,k} = \left(\begin{array}{c}
0 \\
\vec{e}_{\lambda,k}
\end{array}\right),\,\,\, \lambda=1,2\,\, ,
\\
\nonumber
e^\mu_{3,k} = \frac{k^\mu-n^\mu(n\cdot k)}{\sqrt{(n\cdot k)^2-k^2}} =
\left(\begin{array}{c}
0 \\
\vec{k}/|\vec{k}|
\end{array}\right),
\end{eqnarray}
Two transverse polarization vectors $e^\mu_{1,k}$ and $e^\mu_{2,k}$ are purely spatial and orthogonal to $\vec{k}$. Longitudinal polarization vector $e^\mu_{3,k}$ is time-like positive, orthogonal to $\vec{k}$ as well as the transverse polarization vectors, and has unit negative norm.

Then, the metric tensor can be expressed as
\begin{eqnarray}
\label{31}
g^{\mu\nu} = -\sum\limits_{\lambda=1}^2e^\mu_{\lambda,k}e^\nu_{\lambda,k} - n^\mu n^\nu +
\\
\nonumber
\frac{(k^\mu -n^\mu (n\cdot k))(k^\nu-n^\nu(n\cdot k))}{(n\cdot k)^2-k^2}.
\end{eqnarray}
Regrouping terms in Eq. (\ref{31}), we obtain
\begin{eqnarray}
\label{32}
- g^{\mu\nu} = \sum\limits_{\lambda=1}^2e^\mu_{\lambda,k}e^\nu_{\lambda,k} + \frac{k^2n^\mu n^\nu}{(n\cdot k)^2-k^2}+
\\
\nonumber
\frac{k^\mu k^\nu-(n^\mu k^\nu+ n^\nu k^\mu)(n\cdot k)}{(n\cdot k)^2-k^2}
\end{eqnarray}
The last term here is immaterial for the following evaluation, since the photon propagator is to be contracted with the conserved current. According to the continuity equation $k^\mu j_\mu=0$, this term vanishes. Thus, the Coulomb part of metric tensor, $g^{00}$, is
\begin{eqnarray}
\label{33}
g^{\mu\nu} = -\frac{k^2n^\mu n^\nu}{\vec{k}^2}\delta_{\mu 0}\delta_{\nu 0},
\end{eqnarray}
and the transverse part, $g^{ij}$, is defined by
\begin{eqnarray}
\label{34}
g^{ij} = - \delta^{ij}+\frac{k^i k^j}{\vec{k}^2},\,\,\, i,j=1,2,3.
\end{eqnarray}

Finally, substitution of Eqs. (\ref{33}) and (\ref{34}) into Eq. (\ref{24}) results in
\begin{eqnarray}
\label{35}
D^{00}_{\beta}(x, x') =  4\pi i \int\limits_{C_1}\frac{d^4k}{(2\pi)^4}\frac{e^{i k (x-x')}}{\vec{k}^2}n_\beta(\omega),\qquad
\\
\label{36}
D^{ij}_{\beta}(x, x') = 4\pi i \int\limits_{C_1}\frac{d^4k}{(2\pi)^4}\frac{e^{i k (x-x')}}{k^2}n_\beta(\omega)\left(\delta^{ij}-\frac{k^i k^j}{\vec{k}^2}\right).
\end{eqnarray}
Here, we should note that the form Eq. (\ref{24}) represents the thermal photon propagator in the Feynman gauge, whereas the forms Eqs. (\ref{35}) and (\ref{36}) give the Coulomb and transverse parts of the thermal photon propagator in the Coulomb gauge. The result (\ref{35}) can be considered, in principle, as the BBR-induced Coulomb photons by the analogy with relation between the spontaneous and induced transition rates (transverse photons).

\subsection{The thermal Coulomb gauge}
\label{tcg}
The result, Eq. (\ref{24}), allows drawing an analogy with the ordinary photon propagator Eq. (\ref{7}). Thus, we can introduce the gauge for the thermal photon propagator in a conventional manner \cite{Berest}. The most general form of the ordinary photon propagator in the momentum space is
\begin{eqnarray}
\label{37}
D_{\mu \nu} (k) = -\frac{4\pi i g_{\mu \nu}}{k^2}+\chi_\mu k_\nu + \chi_\nu k_\mu + D(k^2)k_\mu k_\nu,\qquad
\end{eqnarray}
where $\chi_\mu$ and $D(k^2)$ are arbitrary functions. Gauge functions $\chi_\mu$ and $D(k^2)$ can be defined as
\begin{eqnarray}
\label{38}
\chi_0 = \frac{2\pi k_0 i}{\vec{k}^2(k_0^2-\vec{k}^2)},
\\
\nonumber
\chi_i = -\frac{2\pi k_i i}{\vec{k}^2(k_0^2-\vec{k}^2)}\,\,\, (i=1,2,3),
\\
\nonumber
D(k^2)=0.
\end{eqnarray}

In the thermal case, the functions can be modified by multiplying by Bose distribution $n_{\beta}(|\vec{k}|)$:
\begin{eqnarray}
\label{39}
\chi_0 = \frac{2\pi k_0 i}{\vec{k}^2(k_0^2-\vec{k}^2)}n_{\beta}(|\vec{k}|),
\\
\nonumber
\chi_i = - \frac{2\pi i k_i}{\vec{k}^2(k_0^2-\vec{k}^2)}n_{\beta}(|\vec{k}|)\,\,\, (i=1,2,3),
\\
\nonumber
D^\beta(k^2)=0.
\end{eqnarray}
Now, the zeroth (Coulomb) component of the thermal photon propagator is
\begin{eqnarray}
\label{40}
D^\beta_{0 0} (k) = -4\pi i n_{\beta}(|\vec{k}|)\left[\frac{1}{k^2}-\frac{k_0^2}{\vec{k}^2k^2}\right]
= \frac{4\pi i}{\vec{k}^2}n_{\beta}(|\vec{k}|)\qquad
\end{eqnarray}
and the transverse part is defined by
\begin{eqnarray}
\label{41}
D_{ij}^\beta (k) = \frac{4\pi i}{k^2}n_\beta(|\vec{k}|)\left(\delta_{ij}-\frac{k_ik_j}{\vec{k}^2}\right).
\end{eqnarray}

\section{The one-photon thermal exchange}
\label{exchange}
The derived expressions (Eq. (24) for the thermal photon propagator, which includes thermal scalar photons, and Eq. (42) for its Coulomb part) cannot be grounds to claim the existence of a physically sensible thermal correction. In the present section, we will formally derive an expression for the thermal correction to the Coulomb potential, which follows from the thermal Coulomb propagator Eq. (\ref{40}). Later, the expression will be compared with another one, see section~\ref{tCp}.

\subsection{Derivation of the Coulomb interaction via the photon propagator}
\label{ci}
\begin{figure}[hbtp]
	\centering
	\includegraphics[scale=0.2]{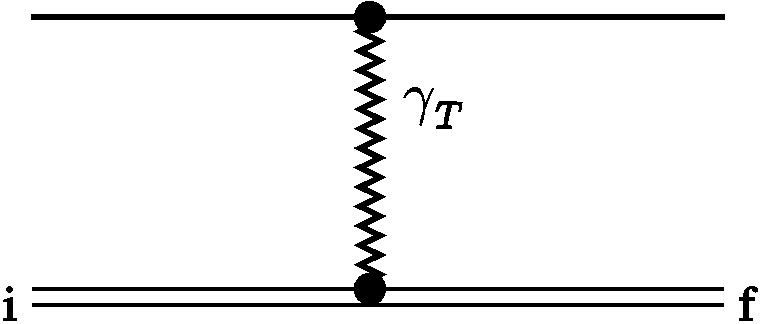}
	\caption{The Feynman graph corresponding to the one-photon exchange between a bound electron and the nucleus. The bold line denotes the propagation of the nucleus with charge $Ze$. The bold wavy line marked $\gamma_T$ denotes the thermal photon, and the double line corresponds to the bound electron in the Furry interaction picture. Indices $i$, $f$ denote the initial and final states of the bound electron, respectively.}
	\label{Fig-4}
\end{figure}

The Coulomb interaction of a bound electron and the nucleus is schematically shown in Fig.~\ref{Fig-4}.
The corresponding $S$-matrix element is given by
\begin{eqnarray}
\label{42}
S_{fi} = (-i e)(i e Z) \int d^4x_1 d^4x_2 \bar{\psi}_f(x_1)\gamma_\mu\psi_i(x_1)
\times
\\
\nonumber
i D_F^{\mu\nu}(x_1,x_2)j_\nu^{\rm ext}(x_2).
\end{eqnarray}
Here, $i$ and $f$ denote the full set of quantum numbers of the initial and final states, respectively. Wave functions $\psi(x)$ and their Dirac conjugated functions $\bar{\psi}(x)$ are the solutions of the Dirac equation in the nucleus field, $j_\nu^{\rm ext}$ represents the external nuclear current, and Feynman photon propagator is given by Eq. (\ref{7}).

Employing the Fourier transform for the nuclear current
\begin{eqnarray}
\label{43}
j_\nu^{\rm ext}(x_2) = \int\frac{d^4 q}{(2\pi)^4}e^{i q x_2}j_\nu^{\rm ext}(q),
\end{eqnarray}
we can integrate over $x_2$ variables in Eq. (\ref{42}). The result is the $\delta$-function, which cancels integration over $q$ variables. Then, with the use of definition Eq. (\ref{7}), we obtain
\begin{eqnarray}
\label{44}
S_{fi} = - 4\pi i Z e^2 \int d^4x \bar{\psi}_f(x)\gamma_\mu\psi_i(x)g^{\mu \nu}
\int \frac{d^4k}{(2\pi)^4} \frac{e^{ikx}}{k^2} j_\nu^{\rm ext}(k).\qquad
\end{eqnarray}

The expression can be substantially simplified in the static limit: for the zeroth component of the nuclear current, it is $j_0^{\rm ext}(k) = 2\pi\delta(k_0)\rho^{\rm ext}(\vec{k})$, where $\rho^{\rm ext}(\vec{k})$ is the charge distribution. Then, performing integration over $k_0$,
\begin{eqnarray}
\label{45}
S_{fi} = 4\pi i Z e^2 \int d^4x \bar{\psi}_f(x)\psi_i(x)\int \frac{d^3k}{(2\pi)^3} \frac{e^{i\vec{k}\vec{r}}}{\vec{k}^2} \rho^{\rm ext}(\vec{k}).\,\,\,
\end{eqnarray}
Integration over time variable yields $\delta$-function. Then, according to the definition, see, e. g., \cite{Andr},
\begin{eqnarray}
\label{46}
S_{fi} = -2\pi i\delta\left(E_f-E_i\right)U_{fi}
\\
\nonumber
\Delta E_a = \langle a| U |a\rangle,
\end{eqnarray}
we can write
\begin{eqnarray}
\label{47}
\Delta E_a = -4\pi Ze^2\int d^3r\psi^\star_a(\vec{r})\int \frac{d^3k}{(2\pi)^3} \frac{e^{i\vec{k}\vec{r}}}{\vec{k}^2} \rho^{\rm ext}(\vec{k})\psi_a(\vec{r}).\,\,\,
\end{eqnarray}

The further evaluation can be performed by assuming $\rho^{\rm ext}(\vec{k})\approx 1$, which corresponds to the point-like nucleus charge distribution. Now, the integration over angles in $d^3k$ yields
\begin{eqnarray}
\label{48}
\Delta E_a = -\frac{2Ze^2}{\pi}\int d^3r\psi^\star_a(\vec{r})\int\limits_0^\infty d\kappa\frac{\sin \kappa r}{\kappa r}\psi_a(\vec{r}),
\\
\nonumber
\Delta E_a =-Ze^2\langle a|\frac{1}{r}|a \rangle \equiv -\frac{Ze^2}{n_a^2}.\qquad
\end{eqnarray}
Here $\kappa\equiv |\vec{k}|$, $r\equiv |\vec{r}|$, and $n_a$ is the principal quantum number. The result Eq. (\ref{48}) corresponds to the hydrogen atom. It is two times the energy of the bound level in hydrogen, since the term represents the potential energy only. To find the complete level energy for the bound electron, its kinetic energy is to be taken into account, which can be found, e. g., via the virial relation. In case of the Coulomb potential, $\langle T\rangle=-\frac{1}{2}\langle U\rangle = \frac{Ze^2}{2 n_a^2}$, and, therefore, $\langle T\rangle+\langle U\rangle = -\frac{Ze^2}{2 n_a^2}$.

\subsection{Derivation of the thermal correction to the Coulomb potential in the Feynman and Coulomb gauges}
\label{tci}

The present part of work examines the Feynman diagram Fig.~\ref{Fig-4}, where the 'ordinary' photon propagator is replaced with the thermal part Eq. (\ref{24}) (denoted with $\gamma_T$). To derive the thermal interaction, we evaluate only the zeroth component of the thermal photon propagator. In accordance to the results of section~\ref{ci}, $S$-matrix element is
\begin{eqnarray}
\label{49}
S^\beta_{fi} = (-i e)(i e Z) \int d^4x_1 d^4x_2 \bar{\psi}_f(x_1)\gamma_\mu\psi_i(x_1)
\\
\nonumber
\times
 i D_\beta^{\mu\nu}(x_1,x_2)j_\nu^{\rm ext}(x_2).
\end{eqnarray}
Insertion of the zeroth component of the thermal photon propagator Eq. (\ref{24}) gives
\begin{eqnarray}
\label{50}
S_{fi}^\beta = -4\pi i Z e^2 \int d^4x\, \bar{\psi}_f(x)\psi_i(x)
\int\limits_{C_1} \frac{d^4k}{(2\pi)^4} \frac{e^{ikx}}{k^2}n_\beta(|\vec{k}|) j_0^{\rm ext}(k).\qquad
\end{eqnarray}

Employing the static limit for the point-like nucleus, we find the thermal interaction energy in the form:
\begin{eqnarray}
\label{51}
\Delta E^\beta_a = -8\pi Ze^2\int d^3r\psi^\star_a(\vec{r})\int \frac{d^3k}{(2\pi)^3} \frac{e^{i\vec{k}\vec{r}}}{\vec{k}^2} n_\beta(|\vec{k}|)\psi_a(\vec{r}).\,\,\,
\end{eqnarray}
Here, we have used again the approximation of the point-like nucleus. An additional factor $2$ has appeared as a result of contour integration Fig.~\ref{Fig3}. Then, integrating over $d^3k$, we arrive at
\begin{eqnarray}
\label{52}
\Delta E^\beta_a = -\frac{4Ze^2}{\pi}\int d^3r\psi^\star_a(\vec{r})\int\limits_0^\infty d\kappa\, n_\beta(\kappa) \frac{\sin\kappa r}{\kappa r}\psi_a(\vec{r}).\,\,\,
\end{eqnarray}
The same result can be easily found in the thermal Coulomb gauge, Eq. (\ref{40}), when we consider the scalar part of the thermal photon propagator in the form Eq. (\ref{35}).

The integral in Eq. (\ref{52}) diverges at $\kappa=0$ due to the singularity of Bose distribution function $n_\beta(\kappa)$. Applying the series expansion of $\sin\kappa r/\kappa r\sim 1 - \kappa^2r^2/6$, we find the divergence to correspond to the first term in the expansion: $\int_0^\infty d\kappa n_\beta(\kappa)$. However, it is independent of $r$ and is the same for any atomic state (unmeasurable). In principle, the divergence can be attributed to the heated vacuum infinite energy. In this way, the zeroth energy of the heated vacuum must be subtracted from the result (\ref{52}), i.e. the divergent term can be omitted from the consideration (see \S 23 \cite{Abr}).

\subsection{Cancellation of divergent contribution}
\label{cdc}
According to the QED theory, the divergences arising in the ordinary case (the zero vacuum) can be canceled out by the contribution of other diagrams (renormalization technique). In the heated vacuum case, the same procedure was described in \cite{Don,DHR} for free particles. That purpose was served by introducing the finite temperature counter-term to the Lagrangian.

To demonstrate the procedure, we evaluate briefly the thermal correction of the lowest order for free particles. The wave function of a free charge in simplest form can be written as
\begin{eqnarray}
\label{cdc.1}
\psi(\vec{r},t)=\int\frac{d^3q}{(2\pi)^{\frac{3}{2}}}\psi(\vec{q})e^{i\vec{q}\vec{r}-i \omega_q t},
\end{eqnarray}
where $\omega_q$ and $\vec{q}$ represent the energy and momentum of the particle, respectively.

Applying the $S$-matrix formalism to the diagram of Fig.~\ref{Fig-4} for the free particle case, one can find after the integration over time variables that
\begin{eqnarray}
\label{cdc.2}
S_{fi} = 2\pi i\delta(\omega_q-\omega_{q'})\frac{8\pi Ze^2}{(2\pi)^3}\int d^3q\int d^3q'\int d^3r
\times
\\
\nonumber
\bar{\psi}(\vec{q})\psi(\vec{q'})e^{i(\vec{q}-\vec{q'})\vec{r}}\int\frac{d^3k}{(2\pi)^3}\frac{e^{i\vec{k}\vec{r}}}{\vec{k}^2}n_\beta(|\vec{k}|).\,\,\,
\end{eqnarray}
Integration over $d^3r$ leads to $(2\pi)^3\delta(\vec{q}-\vec{q'}+\vec{k})$, which removes the integration over $d^3q'$. Then, the energy shift can be defined via Eqs. (\ref{46}):
\begin{eqnarray}
\label{cdc.3}
\Delta E = -8\pi Ze^2 \int d^3q\,\bar{\psi}(\vec{q})\int\frac{d^3k}{(2\pi)^3}\frac{n_\beta(|\vec{k}|)}{\vec{k}^2}\psi(\vec{q}+\vec{k}).\,\,\,
\end{eqnarray}
The series expansion of wave function $\psi(\vec{q}+\vec{k})$ in small values of $\vec{k}$ leads to
\begin{eqnarray}
\label{cdc.4}
\Delta E = -8\pi Ze^2 \int\frac{d^3k}{(2\pi)^3}\frac{n_\beta(|\vec{k}|)}{\vec{k}^2},
\end{eqnarray}
where the normalization property of a wave function was used.

Integration over angles in Eq. (\ref{cdc.4}) results in the same divergent contribution as the first term after the series expansion of $\sin$ in expression Eq. (\ref{52}):
\begin{eqnarray}
\label{cdc.5}
\Delta E = -\frac{4Ze^2}{\pi}\int\limits_0^\infty d\kappa\, n_\beta(\kappa).
\end{eqnarray}
The above expression diverges logarithmically at $\kappa\rightarrow 0$, which corresponds to the zero energy of thermal photons. However, there are no photons with the zero energy. Moreover, like in the result of the previous section, the divergence is independent of $\vec{r}$. Thus, the contribution should be ascribed to the 'dressed' particle, i.e. we are to subtract the counter-term from the Lagrangian, which is $\delta m\,\bar{\psi}\psi$ and corresponds to the mass renormalization of a particle.

While it is quite obvious that $\vec{r}$-independent contribution in Eq. (\ref{52}) vanishes if one subtracts Eq. (\ref{cdc.5}), a more rigorous proof consists in the introduction of regularization: the lower limit in the integrals of Eq. (\ref{cdc.5}) and Eq. (\ref{52}) should be replaced with $\epsilon$, which is to be set to zero after all calculations. Thus, in the lowest order, we see that the subtraction of expressions (\ref{cdc.4}) or (\ref{cdc.5}) leads to a finite result and corresponds to the renormalization procedure of the particle mass.

The same conclusion can be drawn in another way. Namely, to avoid the divergent or $\vec{r}$-independent constant contributions in all orders, we can consider the procedure described briefly in \S 23 in \cite{Abr}. Since we have to deal with the thermal photon part only, the  renormalization procedure can be attributed to regularization of the thermal photon Green's function.

In particular, the Lagrangian of the electron-photon interaction is written as
\begin{eqnarray}
\label{cdc.6}
L_{e\gamma}^{\rm int} \sim j_\mu(x)A_{\rm ext}^\mu(x),
\end{eqnarray}
where $j_\mu(x)$ is the electron current and $A_{\rm ext}^\mu(x)$ is the vector-potential of the field induced by external charge $J^{\rm ext}_{\nu}(x)$.
According to \cite{Akhiezer}, the external field can be defined via the photon Green's function (the thermal photon propagator in our case):
\begin{eqnarray}
\label{cdc.7}
A_{\rm ext}^\mu(x)\sim \int d^4y D^{\mu\nu}_\beta(x,y)J^{\rm ext}_{\nu}(x).
\end{eqnarray}
Thus, the renormalization procedure corresponds to the subtraction of the counter-term from Eq. (\ref{cdc.6}) or to the regularization of Eq. (\ref{cdc.7}).

It was shown in \cite{Abr} that the annihilation and creation operators can be modified as
\begin{eqnarray}
\label{r.a}
\hat{C}_{k\lambda} = \hat{\xi}_0 + \hat{C}'_{k\lambda}, \hat{C}^\dagger_{k\lambda} = \hat{\xi}_0^\dagger + \hat{C}'^\dagger_{k\lambda},
\end{eqnarray}
where the prime denotes the absence of the state with $\kappa=0$, and $\hat{\xi}_0$ and $\hat{\xi}_0^\dagger$ represent the annihilation and creation operators for particles in the state with $\kappa=0$. Such construction admits of the evaluation of the thermal photon propagator as in section~\ref{thev} but separately for the states with $\kappa=0$ and $\kappa\neq 0$.

Integration contour $C_1$ being closed, the part corresponding to $\hat{\xi}_0$, $\hat{\xi}_0^\dagger$ averaged over the heated vacuum states arises immediately from the expressions Eqs. (\ref{21}), (\ref{23}), where we can set $\vec{x}=\vec{x}'$ and $t=t'$. Then, we can write
\begin{eqnarray}
\label{59}
iD^{\mu\nu}_{\beta,\times} = -4\pi g^{\mu\nu} i\lim_{x_1\rightarrow x_2}\int\limits_{C_1}\frac{d^4k}{(2\pi)^4}\frac{e^{ik(x_1-x_2)}}{k^2}n_\beta(|\vec{k}|).\qquad
\end{eqnarray}
The expression can be compared with the result (\ref{cdc.4}). In the static limit, it is clear that i$iD^{\mu\nu}_{\beta,\times}$ leads precisely to the same contribution.

Since there are no photons with zero energies, the subtraction of the 'coincidence limit' Eq. (\ref{59}) provides the correct regularization of divergent or constant $\vec{r}$-independent contributions in all orders of the perturbation theory for the thermal photon part.

The coincidence limit for the thermal photon propagator in the Coulomb form (see Eqs. (\ref{35}), (\ref{36})) is
\begin{eqnarray}
\label{60}
D^{00}_{\beta,\times} = 4\pi i\lim_{x_1\rightarrow x_2}\int\limits_{C_1}\frac{d^4k}{(2\pi)^4}\frac{e^{i k (x_1-x_2)}}{\vec{k}^2}n_\beta(|\vec{k}|),\qquad\qquad
\\
\nonumber
D^{ij}_{\beta,\times} =  4\pi i\lim_{x_1\rightarrow x_2}\int\limits_{C_1}\frac{d^4k}{(2\pi)^4}\frac{e^{i k (x_1-x_2)}}{k^2}n_\beta(|\vec{k}|)\qquad
\\
\nonumber
\times\left(\delta^{ij}-\frac{k^i k^j}{\vec{k}^2}\right).\qquad
\end{eqnarray}
Below, we will demonstrate that the coincidence limit allows the exact cancellation of the $r$-independent contributions.

\subsection{The regularized thermal correction of the lowest order}
\label{rtCi}

To regularize expression (\ref{52}), we examine the coincident limit Eq. (\ref{59}). The $S$-matrix element for the zero-zero component of $D^{\mu\nu}_{\beta,\times}$ can be written as
\begin{eqnarray}
\label{61}
S^{\beta,\times}_{fi} = -i Ze^2 \int d^4x_1 d^4x_2 \bar{\psi}_f(x_1)\psi_i(x_1)
D_{\beta,\times}^{00}\,j_0^{\rm ext}(x_2).\,\,\,
\end{eqnarray}
Substitution of the Fourier transform of the nuclear current
\begin{eqnarray}
\label{62}
j_0^{\rm ext}(x_2) = \int \frac{d^4q}{(2\pi)^4}e^{iqx_2}j_0^{\rm ext}(k)
\end{eqnarray}
allows the integration over $x_2$ with the replacement of the coincidence limit with $x_1\rightarrow 0$. In other words, we have
\begin{eqnarray}
\label{63}
S^{\beta,\times}_{fi} =-4\pi\, i Ze^2 \int d^4x \bar{\psi}_f(x)\psi_i(x)
\qquad\qquad
\\
\nonumber
\times
\lim_{x\rightarrow 0}\int\limits_{C_1}\frac{d^4k}{(2\pi)^4}\frac{e^{i k x}}{k^2} n_\beta(|\vec{k}|)
\int\frac{d^4q}{(2\pi)^4}\delta^{(4)}(k-q)j_0^{\rm ext}(q).\,\,\,
\end{eqnarray}

Then, upon integration over $d^4q$ in Eq. (\ref{63}), the regularized $S$-matrix element Eq. (\ref{50}) can be written as
\begin{eqnarray}
\label{64}
S^{\beta,{\rm reg}}_{fi} = S_{fi}^\beta - S^{\beta,\times}_{fi} =
\qquad\qquad
\\
\nonumber
-4\pi i Ze^2 \int d^4x \bar{\psi}_f(x)\psi_i(x)\int\limits_{C_1}\frac{d^4k}{(2\pi)^4}\frac{e^{ikx}-1}{k^2}n_\beta\,j_0^{\rm ext}(k).\,\,\,
\end{eqnarray}
In the static limit for the point-like nucleus, $j_0^{\rm ext}(k) = 2\pi\delta(k_0)$, we find after the integration over $k_0$,
\begin{eqnarray}
\label{65}
S_{fi}^{\beta,{\rm reg}} =  8\pi i Z e^2 \int d^4x\, \bar{\psi}_f(x)\psi_i(x)
\int \frac{d^3k}{(2\pi)^3}\frac{1-e^{i\vec{k}\vec{r}}}{\vec{k}^2}n_\beta.\,\,\,
\end{eqnarray}
The integration over time variable in $d^4x$ gives a $\delta$-function, which is removed by the definition Eq. (\ref{46}).

The energy shift is defined by
\begin{eqnarray}
\label{66}
\Delta E_a^{\beta,{\rm reg}} =
\frac{Ze^2}{\pi^2}\int d^3r\left|\psi_a(\vec{r})\right|^2\int d^3k \frac{e^{i\vec{k}\vec{r}}-1}{\vec{k}^2} n_\beta(|\vec{k}|).\,\,
\end{eqnarray}

Integration over angles in $d^3k$ yields
\begin{eqnarray}
\label{67}
\Delta E_a^{\beta,{\rm reg}} = \frac{4Ze^2}{\pi}\int d^3r\left|\psi_a(\vec{r})\right|^2
\int\limits_0^\infty d\kappa\, n_\beta(\kappa)\left(\frac{\sin\kappa r}{\kappa r}-1\right).\qquad
\end{eqnarray}
The expression is regular and can be easily derived for the thermal photon propagator in the Coulomb form Eq. (\ref{35}) with the use of the coincidence limit Eq. (\ref{60}).

The thermal correction of the lowest order can be obtained with the estimate $\kappa r\ll 1$ at relevant temperatures. Employing the series expansion of $\sin\kappa r$, we arrive at
\begin{eqnarray}
\label{68}
\Delta E_a^{\beta,{\rm reg}} =
-\frac{2Ze^2}{3\pi}\int d^3r\left|\psi_a(\vec{r})\right|^2\int\limits_0^\infty d\kappa\, n_\beta(\kappa)\kappa^2r^2
\nonumber
\\
=-\frac{4Ze^2\zeta(3)}{3\pi\beta^3}\int d^3r\left|\psi_a(\vec{r})\right|^2r^2,\qquad
\end{eqnarray}
where $\zeta(s)$ is the Riemann zeta function.

\section{The thermal Coulomb potential}
\label{tCp}
\subsection{Derivation of the thermal Coulomb potential}
\label{dtCp}
We concentrate now on the Coulomb part, Eq. (\ref{35}), of the thermal photon propagator. Integrating over $k_0$, we obtain
\begin{eqnarray}
\label{69}
D^{00}_{\beta}(x, x') =  \frac{\delta(t-t')}{\pi^2}\int\,d^3k\frac{e^{i k (\vec{x}-\vec{x}')}}{\vec{k}^2}n_\beta(\omega_k),\qquad
\end{eqnarray}
and the integration over polar angles yields
\begin{eqnarray}
\label{70}
D^{00}_{\beta}(x, x') =  \frac{2\delta(t-t')}{\pi}\int\limits_0^\infty d\kappa \int\limits_{-1}^1 dy\, e^{i k r y}n_\beta(\kappa),\qquad
\end{eqnarray}
where $r\equiv |\vec{x}-\vec{x}'|$ and we made the substitution $\cos\theta = y$.

To regularize expression Eq. (\ref{70}), we subtract the coincident limit Eq. (\ref{60}) and also introduce the large distance regularization: $\lim\limits_{\alpha\rightarrow 0}exp(-\alpha \kappa)$. We find then that
\begin{eqnarray}
\label{71}
D^{00}_{\beta,{\rm reg}}(x, x')
 =  \frac{2\delta(\tau)}{\pi}\lim\limits_{\alpha\rightarrow 0}
\int\limits_{-1}^1 dy \int\limits_0^\infty d\kappa\, e^{-\alpha\kappa} \frac{e^{i \kappa r y}-1}{e^{\beta \kappa}-1}.\,\,
\end{eqnarray}

Integration over $\kappa$ and $y$ can be performed analytically with the use of $\psi(z)$ function (the logarithmic derivative of the Euler's gamma function, $d\ln[\Gamma(z)]/dz$, see \cite{abram}):
\begin{eqnarray}
\label{72}
\psi(z)+\gamma =\int\limits_0^\infty dt\frac{e^{-t}-e^{-zt}}{1-e^{-t}},
\end{eqnarray}
where $\gamma$ is the Euler-Mascheroni constant, $\gamma\simeq 0.577216$. Then
\begin{eqnarray}
\label{73}
\lim\limits_{\alpha\rightarrow 0}\int\limits_{-1}^1 dy \int\limits_0^\infty d\kappa\, e^{-\alpha\kappa} \frac{e^{i k r y}-1}{e^{\beta \kappa}-1} =
\nonumber
\\
\lim\limits_{\alpha\rightarrow 0}\int\limits_{-1}^1 dy \int\limits_0^\infty d\kappa\, \frac{e^{-(\alpha+\beta-i r y)\kappa}-e^{-(\alpha+\beta)\kappa}}{1-e^{-\beta \kappa}}.
\end{eqnarray}

Setting $\alpha=0$ in the second term of the numerator, we find
\begin{eqnarray}
\label{74}
\lim\limits_{\alpha\rightarrow 0}\int\limits_{-1}^1 dy \int\limits_0^\infty d\kappa\, \frac{e^{-(\alpha+\beta-i r y)\kappa}-e^{-\beta\kappa}}{1-e^{-\beta \kappa}}.
\end{eqnarray}
The substitution $\beta\kappa\equiv t$ in Eq. (\ref{74}) yields
\begin{eqnarray}
\label{75}
\lim\limits_{\alpha\rightarrow 0}\int\limits_0^\infty d\kappa\, e^{-\alpha\kappa} \frac{e^{i k r y}-1}{e^{\beta \kappa}-1} =
-\frac{1}{\beta}\lim\limits_{\alpha\rightarrow 0}\int\limits_0^\infty dt\frac{e^{-t}-e^{-z t}}{1-e^{-t}},\qquad
\end{eqnarray}
where $z\equiv 1+\frac{\alpha}{\beta}-i y\frac{r}{\beta}$. Now, in view of \cite{abram} and Eq. (\ref{72}), we obtain
\begin{eqnarray}
\label{76}
\lim\limits_{\alpha\rightarrow 0}\int\limits_0^\infty d\kappa\, e^{-\alpha\kappa} \frac{e^{i k r y}-1}{e^{\beta \kappa}-1} =
-\frac{1}{\beta}\lim\limits_{\alpha\rightarrow 0}\left[\psi\left(1+\frac{\alpha}{\beta}-iy\frac{r}{\beta}\right)+\gamma\right].\qquad
\end{eqnarray}
Integration over $y$ can be performed with definition $\psi(z) = \frac{d\ln \Gamma(z)}{dz}$:
\begin{eqnarray}
\label{77}
\lim\limits_{\alpha\rightarrow 0}\int\limits_{-1}^1 dy \int\limits_0^\infty d\kappa\, e^{-\alpha\kappa} \frac{e^{i k r y}-1}{e^{\beta \kappa}-1} =
\qquad
\\
\nonumber
-\lim\limits_{\alpha\rightarrow 0}\int\limits_{-1}^1 dy \frac{1}{\beta}\left[\gamma+\frac{d}{dy}\frac{dy}{dz}\ln\Gamma\left(1+\frac{\alpha}{\beta}-iy\frac{r}{\beta}\right)\right].
\end{eqnarray}
Insertion of $dy/dz=i\beta/r$ gives
\begin{eqnarray}
\label{78}
\lim\limits_{\alpha\rightarrow 0}\int\limits_{-1}^1 dy \int\limits_0^\infty d\kappa\, e^{-\alpha\kappa} \frac{e^{i k r y}-1}{e^{\beta \kappa}-1} =
\qquad
\\
\nonumber
-\lim\limits_{\alpha\rightarrow 0}\int\limits_{-1}^1 dy \frac{1}{\beta}\left[\gamma+i\frac{\beta}{r}\frac{d}{dy}\ln\Gamma\left(1+\frac{\alpha}{\beta}-iy\frac{r}{\beta}\right)\right].
\end{eqnarray}
Now integration over $y$ can be easily performed
\begin{eqnarray}
\label{79}
\lim\limits_{\alpha\rightarrow 0}\int\limits_{-1}^1 dy \int\limits_0^\infty d\kappa\, e^{-\alpha\kappa} \frac{e^{i k r y}-1}{e^{\beta \kappa}-1} =
\\
\nonumber
-\lim\limits_{\alpha\rightarrow 0}\left(\frac{2\gamma}{\beta}-\frac{i}{r}\ln \left[\frac{\Gamma \left(1+\frac{\alpha}{\beta}+\frac{i r}{\beta}\right)}{\Gamma \left(1+\frac{\alpha}{\beta}-\frac{i r}{\beta}\right)}\right]\right),
\end{eqnarray}
which leads to the expression for the thermal Coulomb potential:
\begin{eqnarray}
\label{80}
D^{00}_{\beta,{\rm reg}} = \frac{4\delta(t-t')}{\pi}
\left(-\frac{\gamma}{\beta}+\frac{i}{2 r}\ln \left[\frac{\Gamma \left(1+\frac{i r}{\beta}\right)}{\Gamma \left(1-\frac{i r}{\beta}\right)}\right]\right).\,\,\,
\end{eqnarray}

The result Eq. (\ref{80}) can be expanded into Taylor series at $r\rightarrow 0$. Then, the thermal corrections of the lowest order are
\begin{eqnarray}
\label{81}
D^{00}_{\beta,{\rm reg}} \approx \frac{4\delta(t-t')}{\pi} \left[-\frac{r^2 \zeta(3)}{3 \beta^3}+\frac{r^4 \zeta(5)}{5 \beta^5}+\dots\right].\,\,\,
\end{eqnarray}
The same result can be found for the low temperature regime $\beta\rightarrow\infty$. In both cases, one can find the asymptotic for Euler's gamma function at $z=r/\beta\ll 1$ (see \cite{abram}) with
\begin{eqnarray}
\label{82}
\Gamma(1+z) = \sqrt{\frac{\pi z}{\sin\pi z}\frac{1-z}{1+z}}e^\Lambda,
\end{eqnarray}
where $\Lambda = C_1 z - C_3 z^3 - C_5 z^5-\dots$ and
\begin{eqnarray}
\nonumber
\begin{array}{c c c}
C_1 = 0.422784335,& C_5 = 0.007385551,& C_9 = 0.000223155,\\
C_3 = 0.067352301,& C_7 = 0.001192754,& C_{11} = 0.000044926.
\end{array}
\end{eqnarray}

\subsection{The asymptotic of thermal Coulomb potential at large distances}
\label{asymp}

The asymptotic behavior of the potential (\ref{80}) at large distances and fixed temperatures can be found in a slightly different way. To this end, we should resume our derivation of $D^{00}_{\beta,{\rm reg}}(x, x')$. Namely, in this case, we integrate Eq. (\ref{70}) over $y$ at first:
\begin{eqnarray}
\label{83}
D^{00}_{\beta,{\rm reg}} =
\frac{4\delta(\tau)}{\pi}
\left[
\int\limits_0^\infty d\kappa \left(\frac{\sin\kappa r}{\kappa r}-1\right) \frac{e^{-\alpha\kappa}}{e^{\beta\kappa}-1}\right]_{\alpha\rightarrow 0}.\,\,\,
\end{eqnarray}
Writing $\sin$ as the imaginary part of exponential function, we have
\begin{eqnarray}
\label{84}
D^{00}_{\beta,{\rm reg}} \sim
\left[\frac{\Im}{r}\int\limits_0^\infty \frac{d\kappa }{\kappa} \frac{e^{-(\alpha-i r)\kappa}}{e^{\beta\kappa}-1} - \int\limits_0^\infty d\kappa\frac{e^{-\alpha\kappa}}{e^{\beta\kappa}-1}\right],\qquad
\end{eqnarray}
where $\Im$ denotes the imaginary part and the $\lim\limits_{\alpha\rightarrow 0}$ is assumed.

To evaluate the corresponding integrals, we employ the expression (3.427(4)) in \cite{GradRyzh}:
\begin{eqnarray}
\label{85}
\int\limits_0^\infty \frac{dx}{x}e^{-\mu x}\left(\frac{1}{2}-\frac{1}{x}+\frac{1}{e^x-1}\right)
 =
\\
\nonumber
=\ln\Gamma(\mu)-\left(\mu-\frac{1}{2}\right)\ln\mu+\mu-\frac{1}{2}\ln(2\pi)
\end{eqnarray}
for the positive real part of $\mu$, $\Re\mu>0$. Then, the integration over $\kappa$ in Eq. (\ref{84}) can be performed with substitution $x=\beta\kappa$:
\begin{eqnarray}
\label{86}
D^{00}_{\beta,{\rm reg}} \sim \frac{\Im}{r}\int\limits_0^\infty dx \frac{e^{-(\frac{\alpha}{\beta}-i \frac{r}{\beta})x}}{x}\left(\frac{1}{e^x-1}-\frac{1}{x}+\frac{1}{2}\right)
\nonumber
\\
+\Im\int\limits_0^\infty dx \frac{e^{-(\frac{\alpha}{\beta}-i \frac{r}{\beta})x}}{x}\left(\frac{1}{x}-\frac{1}{2}\right)-\frac{1}{\beta}\int\limits_0^\infty \frac{e^{-\frac{\alpha}{\beta} x}dx}{e^x-1}.\qquad
\end{eqnarray}
The latter can be transformed to
\begin{eqnarray}
\label{87}
D^{00}_{\beta,{\rm reg}} \sim \frac{\Im}{r}\int\limits_0^\infty dx \frac{e^{-(\frac{\alpha}{\beta}-i \frac{r}{\beta})x}}{x}\left(\frac{1}{e^x-1}-\frac{1}{x}+\frac{1}{2}\right)
\nonumber
\\
+\frac{1}{r}\int\limits_0^\infty dx \frac{\sin\frac{r\, x}{\beta}}{x}\left(\frac{1}{x}-\frac{1}{2}\right)-\frac{1}{\beta}\int\limits_0^\infty \frac{dx}{e^x-1},\,\,\,
\end{eqnarray}
where we set $\alpha$ zero in the last two terms.

The second integral in Eq. (\ref{87}) can be evaluated as follows. At first, we regroup it
\begin{eqnarray}
\label{88}
\frac{1}{r}\int\limits_0^\infty \frac{dx}{x}\sin\frac{r\, x}{\beta}\left(\frac{1}{x}-\frac{1}{2}\right) \equiv
\\
\nonumber
-\frac{1}{2r}\int\limits_0^\infty\frac{dx}{x}\sin\frac{r\, x}{\beta}+\frac{1}{r}\int\limits_0^\infty\frac{dx}{x^2}\sin\frac{r\, x}{\beta},
\end{eqnarray}
where the first integral in the right-hand side is $\pi/2$ and the second one is integrated by parts, ($\frac{1}{x^2}=-\frac{d}{dx}\frac{1}{x}$). Then,
\begin{eqnarray}
\label{89}
\int\limits_0^\infty \frac{dx}{r x}\sin\frac{r\, x}{\beta}\left(\frac{1}{x}-\frac{1}{2}\right) =
-\frac{\pi}{4r}+\frac{1}{\beta}\left(1+\int\limits_0^\infty dx \frac{\cos\frac{r\, x}{\beta}}{x}\right).\,\,\,
\end{eqnarray}
Substitution $t=\frac{rx}{\beta}$ in the last integral results in
\begin{eqnarray}
\label{90}
\int\limits_0^\infty \frac{dx}{r x}\sin\frac{r\, x}{\beta}\left(\frac{1}{x}-\frac{1}{2}\right) =
-\frac{\pi}{4r}+\frac{1}{\beta}\left(1+\int\limits_0^\infty dt \frac{\cos t}{t}\right).\,\,\,
\end{eqnarray}
With due regard for expressions Eqs. (\ref{85}) and (\ref{90}), we arrive at
\begin{eqnarray}
\label{91}
D^{00}_{\beta,{\rm reg}} \sim \Im\frac{1}{r}\left[\ln\Gamma\left(\frac{\alpha}{\beta}-i\frac{r}{\beta}\right)
\qquad\qquad
\right.
\\
\nonumber
\left.
-\left(\frac{\alpha}{\beta}-i\frac{r}{\beta}-\frac{1}{2}\right)\ln\left(\frac{\alpha}{\beta}-i\frac{r}{\beta}\right)+\frac{\alpha}{\beta}-i\frac{r}{\beta}-\frac{1}{2}\ln(2\pi)\right]
\\
\nonumber
-\frac{\pi}{4r}+\frac{1}{\beta}+\frac{1}{\beta}\int\limits_0^\infty dx \frac{\cos x}{x}-\frac{1}{\beta}\int\limits_0^\infty \frac{dx}{e^x-1}.
\end{eqnarray}
The remaining integrals yield Euler-Mascheroni constant:
\begin{eqnarray}
\label{92}
\frac{1}{\beta}\int\limits_0^\infty dt \left(\frac{\cos t}{t}-\frac{1}{e^t-1}\right) = -\frac{\gamma}{\beta}.
\end{eqnarray}

In the limit $\alpha\rightarrow 0$, we have
\begin{eqnarray}
\label{93}
D^{00}_{\beta,{\rm reg}}(x, x') = \frac{\delta(t-t')}{\pi^2}\left[
-\frac{\pi}{4r}+\frac{1-\gamma}{\beta} +
\qquad
\right.
\\
\nonumber
\left.
\frac{1}{r}\Im\left(\ln\Gamma\left(-i\frac{r}{\beta}\right)+\left(\frac{1}{2}+i\frac{r}{\beta}\right)\ln\left(-i\frac{r}{\beta}\right)-i\frac{r}{\beta}\right)
\right].
\end{eqnarray}
Now, using $\ln(-i)=-i\frac{\pi}{2}$ and
\begin{eqnarray}
\label{94}
\Im\left[\left(\frac{1}{2}+i\frac{r}{\beta}\right)\ln\left(-i\frac{r}{\beta}\right)-i\frac{r}{\beta}\right]=
\qquad\qquad
\\
\nonumber
\Im\left[\left(\frac{1}{2}+i\frac{r}{\beta}\right)\ln\left(-i\right)\right]+\frac{r}{\beta}\ln\frac{r}{\beta}-\frac{r}{\beta}
=-\frac{\pi}{4}+\frac{r}{\beta}\ln\frac{r}{\beta}-\frac{r}{\beta},
\end{eqnarray}
we find
\begin{eqnarray}
\label{95}
D^{00}_{\beta,{\rm reg}}(x, x') = \frac{4\delta(t-t')}{\pi}\left[
-\frac{\pi}{4r}+\frac{1-\gamma}{\beta} +
\qquad
\right.
\\
\nonumber
\left.
\frac{1}{r}\Im\left[\ln\Gamma\left(-i\frac{r}{\beta}\right)\right]-\frac{\pi}{4 r}+\frac{1}{\beta}\ln\frac{r}{\beta}-\frac{1}{\beta}
\right].
\end{eqnarray}
Below, we use the property of gamma function: $\Gamma(\overline{z})=\overline{\Gamma(z)}$ and employ the following asymptotic representation, \cite{Jahnke-eng}:
\begin{eqnarray}
\label{96}
\Gamma(i y) &=& h e^{i\omega},\qquad h\approx \sqrt{\frac{2\pi}{y}}e^{-\frac{\pi}{2}y},
\\
\nonumber
\omega &\approx& -\frac{\pi}{4} +y (\ln y -1) - \frac{1}{12y}-\frac{1}{360 y^3}-\dots,
\end{eqnarray}
where $y=r/\beta$.

The final result reads as
\begin{eqnarray}
\label{97}
D^{00}_{\beta,{\rm reg}}(x, x')= \delta(\tau)\left[-\frac{1}{r} + \frac{4}{\pi}\frac{1-\gamma}{\beta} + \frac{\beta}{3\pi r^2} + \dots
\right].\,\,\,
\end{eqnarray}
The presence of $r$-independent constant in Eq. (\ref{97}) causes the necessity of a renormalization procedure in the large-distance (free-particle) limit. The procedure can be found in \cite{DHR,Don}.

The results of the present section should be compared with the previous ones. Namely, in section~\ref{rtCi}, the thermal correction of the lowest order was found for Feynman diagram corresponding to the photon exchange between a bound electron and the nucleus, see Eq. (\ref{68}). Averaging the expression (\ref{79}) over state $a$ of the bound electron, one can arrive at the same result. Actually, evaluation of the matrix element (\ref{66}) without the series expansion of $\sin$ would repeat the derivations of section~\ref{dtCp}. Thus, we can conclude the potential (\ref{80}) to represent the renormalized external potential induced by the charge current. By analogy with the stimulated emission in the BBR field, the thermal Coulomb potential (\ref{80}) can be considered as the Bose-induced part of Coulomb photons.

\section{Thermal vertex corrections of next orders: the adiabatic $S$-matrix formalism}
\label{ThVC}

In the present section, the vertex correction is understood as the Feynman graph Fig.~\ref{Fig-4} describing the thermal Coulomb interaction of an electron with the nucleus in an atom. Then, the higher order vertices correspond to the Feynman graphs with two or three such vertices attached to the same electron line. The lowest order vertex represents, according to Eq. (\ref{67}), a correction of order $\alpha Z F_1(\beta m_e)$, where $\alpha$ is the fine structure constant and $F_1(\beta m_e)$ is a function of thermal parameter $1/\beta m_e$. The function is defined by the integral in Eq. (\ref{67}). The corrections of such low order $\alpha$, $\alpha Z$ are absent in the zero-temperature QED. The lowest order radiative corrections in the zero-temperature QED start with those of order $m_e\alpha(\alpha Z)^4$ in r.u. The smallness of the lowest order thermal Coulomb correction Eq. (\ref{67}) is determined by its dependence on thermal parameter $1/\beta m_e$. Therefore, it is necessary to check whether other vertex corrections will decrease in magnitude with growing number of vertices.

The evaluation of the energy corrections with the formula Eq. (\ref{48}) is possible only for the irreducible Feynman graphs \cite{LabKlim}. These are defined as the graphs, which cannot be separated into unconnected parts by cutting only one internal fermion line. Evaluation of the energy corrections represented by the reducible graphs is more involved and requires application of special methods. Historically, the first one was the adiabatic $S$ matrix approach, followed by the Green Function method \cite{Braun}, \cite{shabaev}, and the covariant evaluation operator method \cite{lindgren-review} and the Line Profile method \cite{Andr} were developed for the purpose. The Feynman graphs with several thermal interactions between an electron and the nucleus are reducible, and we choose the adiabatic $S$-matrix approach for their treatment.

Adiabatic $S$-matrix $\hat{S}_{\eta}$ differs from the standard one by presence of adiabatic (exponential) factor $e^{-\eta|t|}$ in each (interaction) vertex. This reduces to the notion of the interaction being adiabatically switched on and off, which is formally introduced by replacement $\hat{H}_{{\rm int}}(t) \longrightarrow \hat{H}^\eta_{{\rm int}}(t) = e^{-\eta|t|}\,\hat{H}_{{\rm int}}(t)$ \cite{Gell,Sucher}. A symmetrized version of the adiabatic formula containing $S_{\eta}(\infty,-\infty)$, which is more convenient for the QED calculations, was proposed by Sucher \cite{Sucher}. The first application to calculations within the framework of the bound-state QED theory was made in \cite{Lab}. It was shown there how to deal with the adiabatic exponential factor when evaluating the real part of corrections to the energy levels (see also \cite{LabKlim}). The same method was applied in \cite{LSP-sep} to evaluate the imaginary part of the corrections. The QED applications within the adiabatic $S$-matrix formalism for evaluation of energy shifts and level widths can be found also in \cite{bas78}.

\subsection{The thermal correction of the lowest order}
\label{tclo}
In terms of the adiabatic $S$-matrix formalism, the energy shift is defined by the Gell-Mann and Low adiabatic formula \cite{Gell}:
\begin{eqnarray}
\label{98}
\Delta E_a = \lim\limits_{\eta\rightarrow 0}\frac{i\eta}{2}\frac{e\frac{\partial}{\partial e}\langle a |\hat{S}_\eta|a\rangle}{\langle a |\hat{S}_\eta|a\rangle},
\end{eqnarray}
where the adiabatic $S$-matrix element can be expanded as a power series in $e$ and $\hat{S}_\eta = \sum\limits_{n=0}^\infty e^n\hat{S}_\eta^{(n)}$. Then, we expand the numerator and the denominator in Eq. (\ref{98}) in powers of $e$ and confine ourselves to the second order terms:
\begin{eqnarray}
\label{99}
\Delta E_a = \lim\limits_{\eta\rightarrow 0}\frac{ei\eta}{2} \left[ \langle a |\hat{S}_\eta^{(1)}|a \rangle + e\left(2\langle a |\hat{S}_\eta^{(2)}|a \rangle-\langle a |\hat{S}_\eta^{(1)}|a \rangle^2\right)\right].\,\,\,
\end{eqnarray}

The off-diagonal matrix element of the first order corresponds to the one-vertex diagram depicted in Fig.~\ref{Fig-4} and is
\begin{eqnarray}
\label{100}
e \langle f |\hat{S}_\eta^{(1)}|i \rangle = - Ze^2\int d^4x_1 d^4x_2 \bar{\psi}_f(x_1)\gamma_\mu e^{-\eta |t_1|}
\times
\nonumber
\\
 iD^{\mu\nu}_\beta(x_1,x_2)j_\nu(x_2)\psi_i(x_1).\qquad
\end{eqnarray}
Applying again the Fourier transform to the nuclei current and the static limit to its zeroth component, we arrive at
\begin{eqnarray}
\label{101}
e \langle f |\hat{S}_\eta^{(1)}|i \rangle = -8\pi i Ze^2 \int  d^3r\int\limits_{-\infty}^\infty dt e^{i(E_f-E_i)t-\eta|t|}
\nonumber
\\
\times
\bar{\psi}_f(\vec{r})\int\frac{d^3k}{(2\pi)^3}\frac{e^{i\vec{k}\vec{r}}}{\vec{k}^2}n_\beta(|\vec{k}|)\rho^{\rm ext}(\vec{k})\psi_i(\vec{r}).\qquad
\end{eqnarray}
Time integration yields
\begin{eqnarray}
\label{102}
\int\limits_{-\infty}^\infty dt e^{i(E_f-E_i)t-\eta|t|} = \int\limits_0^\infty dt e^{i(E_f-E_i)t-\eta t}
\nonumber
\\
+\int\limits_{-\infty}^0 dt e^{i(E_f-E_i)t+\eta t} = \frac{2 \eta}{(E_f-E_i)^2+\eta^2}.
\end{eqnarray}

Then, the diagonal matrix element $f=i=a$ gives the energy shift
\begin{eqnarray}
\label{103}
\Delta E_a^{(1)\beta} = -\lim\limits_{\eta\rightarrow 0}\frac{\eta}{2} 8\pi Ze^2\int d^3r\bar{\psi}_a(\vec{r})
\times
\nonumber
\\
\int\frac{d^3k}{(2\pi)^3}\frac{e^{i\vec{k}\vec{r}}}{\vec{k}^2}n_\beta(|\vec{k}|)\rho^{\rm ext}(\vec{k})\psi_a(\vec{r}) \frac{2}{\eta}.
\end{eqnarray}
The final result is obtained in the limit $\eta\rightarrow 0$. After the integration over angles, we find
\begin{eqnarray}
\label{104}
\Delta E_a^{(1)\beta} = \frac{4Ze^2}{\pi}\langle a|\int\limits_0^\infty d\kappa n_\beta(\kappa)\frac{\sin \kappa r}{\kappa r}\rho^{\rm ext}(\kappa)|a\rangle. \qquad
\end{eqnarray}
The energy shift Eq. (\ref{104}) coincides precisely with the result of the ordinary $S$-matrix formalism, Eq. (\ref{52}). Thus, applying the regularization procedure given in section~\ref{cdc}, we arrive at expression (\ref{68}).

\subsection{The double vertex diagram}
\label{dvd}
\begin{figure}[hbtp]
	\centering
	\includegraphics[scale=0.2]{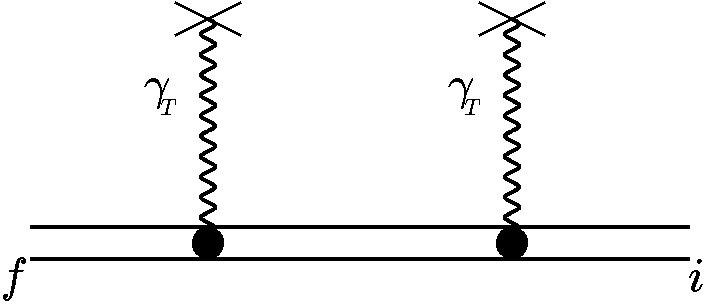}
	\caption{The Feynman graph corresponding to the double thermal photon exchange between a bound electron and the nucleus. All the notations are the same as in the previous Feynman diagrams. 
	}\label{dv}
\end{figure}

The double vertex diagram is depicted in Fig.~\ref{dv}.
The corresponding matrix element, $\langle a |\hat{S}_{\eta}^{(2)}|a \rangle$, can be found with the insertion of adiabatic factors $e^{-\eta|t_1|}$ and $e^{-\eta|t_2|}$ as
\begin{eqnarray}
\label{105}
\langle a |\hat{S}_\eta^{(2)}|a \rangle = Z^2e^2\int d^4x_1 d^4x_2 d^4x_3 d^4x_4 \bar{\psi}_a(x_1)e^{-\eta|t_1|}\gamma_\mu
\nonumber
\\
\times
S(x_1,x_2)e^{-\eta|t_2|}\gamma_\nu\psi_a(x_2)iD_\beta^{\mu\lambda}(x_1,x_3)
\times\qquad
\\
\nonumber
j_\lambda(x_3)iD_\beta^{\nu\sigma}(x_2,x_4)j_\sigma(x_4).
\qquad
\end{eqnarray}
Here, the Feynman propagator for the bound electron is
\begin{eqnarray}
\label{106}
S(x_1x_2) = \frac{i}{2\pi}\int\limits_{-\infty}^\infty d\omega e^{-i\omega(t_1-t_2)}\sum\limits_n\frac{\psi_n(\vec{r_1})\bar{\psi}_n(\vec{r_2})}{\omega-E_n(1-i0)}.\qquad
\end{eqnarray}

Applying again the Fourier transform for the zeroth component of the nuclear current and integrating over $d^4x_3$ and $d^4x_4$, we have in the static limit:
\begin{eqnarray}
\label{107}
\langle a |\hat{S}_\eta^{(2)}|a \rangle = i 32 \pi Z^2e^2\int d^4x_1 d^4x_2 \bar{\psi}_a(x_1)e^{-\eta|t_1|}
\qquad
\nonumber
\\
\times
\int\limits_{-\infty}^\infty d\omega\, e^{-i\omega(t_1-t_2)}\sum\limits_n \frac{\psi_n(\vec{r_1})\bar{\psi}_n(\vec{r_2})}{\omega-E_n(1-i0)}\psi_a(x_2)
\qquad
\\
\nonumber
\times
e^{-\eta|t_2|}\int\frac{d^3k}{(2\pi)^3}n_\beta(|\vec{k}|)\frac{e^{i\vec{k}\vec{r}_1}}{\vec{k}^2}   \int\frac{d^3q}{(2\pi)^3}n_\beta(|\vec{q}|)\frac{e^{i\vec{q}\vec{r}_2}}{\vec{q}^2}.\,\,\,
\end{eqnarray}
Then, with the use of Eq. (\ref{102}), this reduces to
\begin{eqnarray}
\label{108}
\langle a |\hat{S}_\eta^{(2)}|a \rangle= \frac{32 \pi i Z^2e^2}{(2\pi)^6}\sum\limits_n\int\limits_{-\infty}^\infty d\omega \frac{4\eta^2}{[(E_a-\omega)^2+\eta^2]^2}
\nonumber
\\
\times
\left|\langle a |\int\frac{d^3k}{\vec{k}^2}e^{i\vec{k}\vec{r}}n_\beta(|\vec{k}|)| n\rangle\right|^2\frac{1}{\omega-E_n(1-i0)}.
\qquad
\end{eqnarray}
Integration over $\omega$ (see in \cite{LSP-sep} for details) can be performed with
\begin{eqnarray}
\label{109}
\int\limits_{-\infty}^\infty\frac{\eta^2}{[(E_a-\omega)^2+\eta^2]^2}\frac{d\omega}{\omega-E_n(1-i0)}
 =
\\
\nonumber
=-\frac{2\pi}{\eta}
\frac{E_a-E_n+2i\eta}{[E_a-E_n+i\eta]^2}
\qquad
\end{eqnarray}
and, therefore,
\begin{eqnarray}
\label{110}
\langle a |\hat{S}_\eta^{(2)}|a \rangle = \frac{2^6\pi^2 i Z^2e^2}{\eta}
\times\qquad\qquad
\\
\nonumber
\sum\limits_n\left|\langle a|\int\frac{d^3k}{(2\pi)^3 \vec{k}^2}e^{i\vec{k}\vec{r}}n_\beta(|\vec{k}|)|n\rangle\right|^2
\frac{E_a-E_n+2i\eta}{\eta[E_a-E_n+i\eta]^2}.
\end{eqnarray}

Separating out the term $n=a$ (reference state) in expression (\ref{70}), we can write
\begin{eqnarray}
\label{111}
\langle a |\hat{S}_\eta^{(2)}|a \rangle = -\frac{2^7\pi^2 Z^2e^2}{\eta^2}\left|\langle a|\int\frac{d^3k}{(2\pi)^3\vec{k}^2}e^{i\vec{k}\vec{r}}n_\beta|a\rangle\right|^2+
\qquad
\\
\nonumber
\frac{2^6\pi^2 i Z^2e^2}{\eta}\sum\limits_{n\neq a}\left|\langle a|\int\frac{d^3k}{(2\pi)^3\vec{k}^2}e^{i\vec{k}\vec{r}}n_\beta|n\rangle\right|^2\frac{E_a-E_n+2i\eta}{[E_a-E_n+i\eta]^2}.
\end{eqnarray}
It follows from Eqs. (\ref{101}) and (\ref{102}) that
\begin{eqnarray}
\label{112}
\langle a |\hat{S}_\eta^{(1)}|a \rangle^2 = \frac{2^8 \pi^2 Z^2e^2}{\eta^2}\langle a|\int\frac{d^3k}{(2\pi)^3\vec{k}^2}e^{i\vec{k}\vec{r}}n_\beta(|\vec{k}|)| a \rangle^2.\,\,\,
\end{eqnarray}
Then, according to Eq. (\ref{99}), we obtain
\begin{eqnarray}
\label{113}
\Delta E_a^{(2)\beta} 
=-2^6 \pi^2 Z^2e^4\sum\limits_{n\neq a}\frac{\left|\langle a|\int\frac{d^3k}{(2\pi)^3\vec{k}^2}e^{i\vec{k}\vec{r}}n_\beta(|\vec{k}|)|n\rangle\right|^2}{E_a-E_n},\,\,\,
\end{eqnarray}
where the first term in Eq. (\ref{111}) (the summand $n=a$ in $2\langle a |\hat{S}_\eta^{(2)}|a \rangle$) cancels precisely $\langle a |\hat{S}_\eta^{(1)}|a \rangle^2$. The final expression for the second-order vertex thermal correction (see Fig.~\ref{dv}) can be obtained after the integration over angles in $d^3k$, so
\begin{eqnarray}
\label{114}
\Delta E_a^{(2)\beta} = -\frac{16 Z^2e^4}{\pi^2}\sum\limits_{n\neq a}\frac{\left|\langle a|\int\limits_0^\infty d\kappa\, n_\beta(\kappa)\frac{\sin\kappa r}{\kappa r}|n\rangle\right|^2}{E_a-E_n}.
\end{eqnarray}

In particular, it follows from the result (\ref{114}) that there is no divergence in the second-order correction. The term corresponding to the $\vec{r}$-independent contribution is zero due to the orthogonality of wave functions, and, therefore, the correction of the leading order is
\begin{eqnarray}
\label{115}
\Delta E_a^{(2)\beta} = \frac{64 Z^2e^4\zeta^2(3)}{9\pi^2\beta^6}\sum\limits_{n\neq a}\frac{\left|\langle a|r^2|n\rangle\right|^2}{E_a-E_n}.
\end{eqnarray}
The correction is negligibly small, as it is proportional to $(k_B T)^6$ and $k_B T\sim 9.50043\cdot 10^{-4}$ at the room temperature (in atomic units).

We should note here that the modification of the thermal photon propagator in accordance with Eq. (\ref{59}) does not change this result. The cancellation of the reference state in Eq. (\ref{113}) (the state $n=a$) means that all $r$-independent operators vanish in the off-diagonal matrix element. Thus, the counter-term in the scalar part of the thermal photon propagator for diagram Fig.~\ref{dv} is zero. The gauge invariance of the result can be easily accomplished with the use of Eq. (\ref{35}) for the thermal propagator and Eq. (\ref{60}) for the coincidence limit. The main conclusion of the present section is that corrections of the next orders do not make an additional renormalization procedure necessary. The regularization procedure described in section~\ref{cdc} in conjunction with two forms of the thermal photon propagator Eqs. (\ref{24}), (\ref{35}) yields the gauge-invariant and regular result.

\subsection{The third-order correction within the adiabatic S-matrix formalism}
\label{3d-order}
To substantiate the conclusion of the previous section, we consider briefly the third-order thermal vertex correction within the framework of the adiabatic $S$-matrix formalism. The corresponding energy shift can be written as
\begin{eqnarray}
\label{116}
\Delta E_a \sim 3\langle a |S^{(3)}_\eta|a\rangle
- 3\langle a |S^{(2)}_\eta|a\rangle \langle a |S^{(1)}_\eta|a\rangle+ \langle a |S^{(1)}_\eta|a\rangle^3.\qquad
\end{eqnarray}

Applying the Feynman rules, we find
\begin{widetext}
\begin{eqnarray}
	\label{117}
	e^3\langle a |S^{(3)}_\eta|a\rangle = iZ^3e^62^7\int d^3r_1 d^3r_2 d^3r_3\bar{\psi}_a(\vec{r}_1)\int\limits_{-\infty}^\infty d\omega_1\int\limits_{-\infty}^\infty d\omega_2\sum\limits_n \frac{\psi_n(\vec{r}_1)\bar{\psi}_n(\vec{r}_2)}{\omega_1-E_n(1-i0)}  \sum\limits_m \frac{\psi_m(\vec{r}_2)\bar{\psi}_m(\vec{r}_3)}{\omega_2-E_m(1-i0)}\psi_a(\vec{r}_3)\times\qquad
	\\
	\nonumber
	\frac{2\eta}{(E_a-\omega_1)^2+\eta^2}\frac{2\eta}{(\omega_1-\omega_2)^2+\eta^2}\frac{2\eta}{(E_a-\omega_2)^2+\eta^2}
	\int\frac{d^3k_1\,n_\beta(|\vec{k}_1|)}{(2\pi)^3}\frac{e^{i\vec{k}_1\vec{r}_1}}{\vec{k}_1^2}\int\frac{d^3k_2\,n_\beta(|\vec{k}_2|)}{(2\pi)^3}\frac{e^{i\vec{k}_2\vec{r}_2}}{\vec{k}_2^2}\int\frac{d^3k_3\,n_\beta(|\vec{k}_3|)}{(2\pi)^3}\frac{e^{i\vec{k}_3\vec{r}_3}}{\vec{k}_3^2}.
\end{eqnarray}
Then, considering the case $n=a$, $m\neq a$ along with $m=a$, $n\neq a$, we find that the terms vanish in view of the second summand in Eq. (\ref{116}). In turn, the case when $n=m=a$ cancels the third term in Eq. (\ref{116}). Thus, the expression for the third-order thermal correction can be written as
\begin{eqnarray}
	\label{118}
	e^3\langle a |S^{(3)}_\eta|a\rangle = iZ^3e^62^7\int d^3r_1 d^3r_2 d^3r_3\bar{\psi}_a(\vec{r}_1)\int\limits_{-\infty}^\infty d\omega_1\int\limits_{-\infty}^\infty d\omega_2\sum\limits_{n\neq a} \frac{\psi_n(\vec{r}_1)\bar{\psi}_n(\vec{r}_2)}{\omega_1-E_n(1-i0)}  \sum\limits_{m\neq a} \frac{\psi_m(\vec{r}_2)\bar{\psi}_m(\vec{r}_3)}{\omega_2-E_m(1-i0)}\psi_a(\vec{r}_3)\times\qquad
	\\
	\nonumber
	\frac{2\eta}{(E_a-\omega_1)^2+\eta^2}\frac{2\eta}{(\omega_1-\omega_2)^2+\eta^2}\frac{2\eta}{(E_a-\omega_2)^2+\eta^2}
	\int\frac{d^3k_1\,n_\beta(|\vec{k}_1|)}{(2\pi)^3\vec{k}_1^2}e^{i\vec{k}_1\vec{r}_1}\int\frac{d^3k_2\,n_\beta(|\vec{k}_2|)}{(2\pi)^3\vec{k}_2^2}e^{i\vec{k}_2\vec{r}_2}\int\frac{d^3k_3\,n_\beta(|\vec{k}_3|)}{(2\pi)^3\vec{k}_3^2}e^{i\vec{k}_3\vec{r}_3}.
\end{eqnarray}
\end{widetext}

Integration over $\omega_1$ and $\omega_2$ in Eq. (\ref{118}) results in
\begin{eqnarray}
\label{119}
\Delta E_a = -\frac{Z^3e^6 2^6}{\pi^4}\sum\limits_{n\neq a}\sum\limits_{m\neq a}\frac{\langle a|\int\limits_0^\infty d\kappa\, n_\beta(\kappa)\frac{\sin\kappa r}{\kappa r} |n\rangle }{(E_a-E_n)(E_a-E_m)}
\,\,\,
\\
\nonumber
\times
\langle n| \int\limits_0^\infty d\kappa\, n_\beta(\kappa)\frac{\sin\kappa r}{\kappa r} |m\rangle \langle m| \int\limits_0^\infty d\kappa\, n_\beta(\kappa)\frac{\sin\kappa r}{\kappa r} |a\rangle.
\end{eqnarray}
With the use of the series expansion of $\sin$, we find the final expression for the third-order correction in the form:
\begin{eqnarray}
\label{120}
\Delta E_a = \frac{64Z^3e^6\zeta^3(3)}{27\pi^4\beta^9}\sum\limits_{\mathop{n \neq a}\limits_{m\neq a}}\frac{\langle a|r^2 |n\rangle \langle n|r^2 |m\rangle \langle m|r^2 |a\rangle}{(E_a-E_n)(E_a-E_m)}.\:\:
\end{eqnarray}

Thus, we can conclude that the reference states $n=a$ and $m=a$ can be omitted when evaluating the thermal vertex correction of the third order. In turn, this means again that there are no divergences connected with Eq. (\ref{52}). The $r$-independent terms as well as the coincidence limit Eq. (\ref{59}) vanish by virtue of the orthogonality property of wave functions. The same conclusion can be drawn for the thermal Coulomb gauge. Therefore, the renormalization procedure given in section~\ref{cdc} and corresponding coincidence limit suffice in all orders of such diagrams.

\section{Evaluation of the thermal self-energy correction}
\label{SE-correction}

\begin{figure}[hbtp]
	\centering
	\includegraphics[scale=0.2]{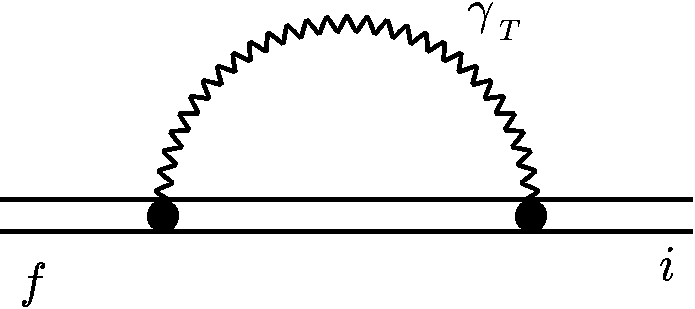}
	\caption{The Feynman graph corresponding to the thermal one-loop self-energy diagram. The double line denotes the bound electron. The wavy line represents the thermal photon exchange, where notation $\gamma_T$ corresponds to the thermal photon line and $i$, $f$ denote the initial and final states, respectively.}\label{SE}
\end{figure}

The present section deals with the one-loop self-energy (SE) correction, when the photon line is given by the thermal part of photon propagator (see Fig.~\ref{SE}).
The detailed description of the SE correction within the framework of the presented theory allows the accurate regularization of divergences in the Coulomb and Feynman gauges, thus avoiding the annoying mistakes made in \cite{SLP-QED}. In particular, we show the Hadamard representation of the thermal photon propagator to lead to the same final result as in \cite{SLP-QED}, where the thermal photon propagator in the form Eq. (\ref{29}) was used. To render this section more self-consistent, we repeat some expressions from the text above.

\subsection{The one-loop self-energy correction: the zeroth vacuum}
\label{SEa}
The self-energy correction with the one-photon thermal loop was first considered in \cite{SLP-QED}. In particular, the rigorous QED description was shown to allow the accurate accounting for the finite lifetimes of atomic levels. The significance of the finite lifetimes was noted in \cite{jentlach1,jentlach2}, this is what causes the enhancement of blackbody friction by several orders of magnitude. The same effect was found in \cite{SLP-QED}, where the blackbody-induced line broadening exceeded the corresponding quantum mechanical result by several orders of magnitude.

A generic non-diagonal second-order $S$-matrix element (Fig.~\ref{SE}) in the Furry picture for a bound atomic electron is
\begin{eqnarray}
\label{se.1}
\langle a'|\hat{S}^{{\rm SE}}|a\rangle =(-i e)^2\int d^4x_1d^4x_2\bar{\psi}_{a'}(x_1)\gamma_{\mu}
\\
\nonumber
\times
S(x_1,x_2)\gamma_{\nu}\psi_a(x_2)D^{\mu\nu}_F(x_2,x_1),
\end{eqnarray}
where integration is performed over space-time variables $x_1$, $x_2$, which designate spatial position vector $\vec{r}$ and time variable $t$. Here, we use relativistic units $m_e=\hbar=c=1$. The Dirac matrices are denoted as $\gamma_{\mu}$, where $\mu$ runs over values $\mu=(0,1,2,3)$; $\psi_a(x)=\psi_a(\vec{r})e^{-iE_a t}$ is the one-electron Dirac wave function, $\bar{\psi_a}$ is the conjugated Dirac wave function. The standard (zero-temperature) electron propagator defined as the vacuum-expectation value of the time-ordered product can be represented in terms of the eigenmode decomposition with respect to one-electron eigenstates:
\begin{eqnarray}
\label{se.2}
S(x,y) = \frac{i}{2\pi}\int\limits_{-\infty}^{\infty}d\omega e^{-i\omega(t_1-t_2)}\sum\limits_n\frac{\psi_n(\vec{x})\bar{\psi}_n(\vec{y})}{\omega-E_n(1-i0)},\qquad
\end{eqnarray}
where the summation is over the entire Dirac spectrum. Finally, the standard (zero-temperature) photon propagator $D^{\mu\nu}_F(x,y)$ in the Feynman gauge is given by Eqs. (\ref{7}), (\ref{8}).

Energy correction $\Delta E_a$ for an "irreducible" Feynman graph can be obtained via the relation \cite{LabKlim,Andr}:
\begin{eqnarray}
\label{se.3}
\langle a'|\hat{S}^{{\rm SE}}|a\rangle &=& - 2\pi i \langle a'| U | a \rangle\delta(E_{a'}-E_a)
\\
\nonumber
\Delta E_a &=& \langle a | U | a \rangle.
\end{eqnarray}

Integrating over time and $\omega$ variables in Eq. (\ref{se.3}), the energy shift can be written as
\begin{eqnarray}
\label{se.4}
\Delta E^{{\rm SE}}_a=\frac{e^2}{2\pi i}\sum\limits_n\left(\frac{1-\vec{\alpha_1}\vec{\alpha_2}}{r_{12}}I_{na}(r_{12})\right)_{anna},
\end{eqnarray}
where
\begin{eqnarray}
\label{se.5}
I_{na}(r_{12})=\int\limits_{-\infty}^{\infty}\frac{e^{i|\omega|r_{12}}d\omega}{E_n(1-i0)-E_a+\omega},
\end{eqnarray}
and the matrix element is to be understood as
\begin{eqnarray}
\label{se.6}
\left(\hat{A}(12) \right)_{abcd}\equiv \langle a(1) b(2)|\hat{A} | c(1) d(2)\rangle.
\end{eqnarray}

Correction $\Delta E_a$ can be split into the real and imaginary parts
\begin{eqnarray}
\label{se.7}
\Delta E^{{\rm SE}}_a=L_a^{{\rm SE}}-\frac{i}{2}\Gamma_a.
\end{eqnarray}
Here, $L_a^{{\rm SE}}$ denotes the lowest-order electron self-energy contribution to the Lamb shift, and $\Gamma_a$ denotes the lowest-order radiative width. The rigorous QED evaluation of (\ref{se.4}) can be found in \cite{LabKlim}, \cite{Andr} and lies beyond our interests. However, we should note that such description admits of the regularization of the divergent energy denominators in the photon scattering amplitudes, see \cite{Low,Andr}. The detailed analysis of the imaginary part of the higher order SE corrections can be found in \cite{ZSLP}.

\subsection{The self-energy correction with one thermal photon loop}
\label{SEth}

it was shown in \cite{SLP-QED} that introduction of the photon propagator in the form Eq. (\ref{29}) into Eq. (\ref{se.1}) leads to the Stark shift and depopulation rate induced by the blackbody radiation. Further on, we evaluate the thermal one-loop self-energy correction with the Hadamard representation, Eq. (\ref{24}), for the thermal photon propagator. The $S$-matrix element (\ref{se.1}) is
\begin{eqnarray}
\label{seth.1}
S_{fi}^{{\rm SE}_\beta} = -2e^2\int d^4x_1d^4x_2\bar{\psi}_{f}(x_1)\left(1-\vec{\alpha}_1\vec{\alpha}_2\right)
\times
\\
\nonumber
\int\limits_{-\infty}^\infty d\omega e^{-i\omega(t_1-t_2)}\sum\limits_n\frac{\psi_n(\vec{r}_1)\bar{\psi}_n(\vec{r}_2)}{\omega-E_n(1-i0)}
\times
\\
\nonumber
\int\limits_{C_1}\frac{d^4k}{(2\pi)^4}n_\beta(|\vec{k}|)\frac{e^{ik(x_1-x_2)}}{k^2}\psi_i(x_2).
\end{eqnarray}

Integrating in the thermal photon propagator first in $k_0$ plane, we obtain
\begin{eqnarray}
\label{seth.2}
S_{fi}^{{\rm SE}_\beta} = 2ie^2\int d^4x_1d^4x_2\bar{\psi}_{f}(x_1)\left(1-\vec{\alpha}_1\vec{\alpha}_2\right)
\times\qquad
\\
\nonumber
\int\limits_{-\infty}^\infty d\omega e^{-i\omega(t_1-t_2)}\sum\limits_n\frac{\psi_n(\vec{r}_1)\bar{\psi}_n(\vec{r}_2)}{\omega-E_n(1-i0)}
\times
\\
\nonumber
\int\frac{d^3k}{(2\pi)^3}\frac{\cos|\vec{k}|(t_1-t_2)}{|\vec{k}|}n_\beta(|\vec{k}|)e^{i\vec{k}(\vec{r}_1-\vec{r}_2)}\psi_i(x_2).
\end{eqnarray}
Then, integration over angles in $d^3k$ yields
\begin{eqnarray}
\label{seth.3}
S_{fi}^{{\rm SE}_\beta} = \frac{ie^2}{\pi^2}\int d^4x_1d^4x_2\bar{\psi}_{f}(x_1)\left(1-\vec{\alpha}_1\vec{\alpha}_2\right)
\times\qquad
\\
\nonumber
\int\limits_{-\infty}^\infty d\omega e^{-i\omega(t_1-t_2)}\sum\limits_n\frac{\psi_n(\vec{r}_1)\bar{\psi}_n(\vec{r}_2)}{\omega-E_n(1-i0)}
\times
\\
\nonumber
\int\limits_0^\infty d\kappa\,\cos[\kappa(t_1-t_2)]n_\beta(\kappa)\frac{\sin\kappa r_{12}}{r_{12}}\psi_i(x_2).
\end{eqnarray}

Integration over time variables and $\omega$ proceeds in usual manner and gives
\begin{eqnarray}
\label{seth.4}
S_{fi}^{{\rm SE}_\beta} = 2ie^2\int d^3r_1d^3r_2\bar{\psi}_{f}(\vec{r}_1)\left(1-\vec{\alpha}_1\vec{\alpha}_2\right)
\qquad
\\
\nonumber
\times
\int\limits_0^\infty d\kappa\,n_\beta(\kappa)\frac{\sin\kappa r_{12}}{r_{12}}\psi_i(\vec{r}_2)\delta(E_f-E_i)\times\qquad
\\
\nonumber
\left[\sum\limits_n\frac{\psi_n(\vec{r}_1)\bar{\psi}_n(\vec{r}_2)}{E_f+\kappa-E_n(1-i0)}+\sum\limits_n\frac{\psi_n(\vec{r}_1)\bar{\psi}_n(\vec{r}_2)}{E_f-\kappa-E_n(1-i0)}\right].
\end{eqnarray}
According to the definition (\ref{se.3}), we find the energy shift:
\begin{eqnarray}
\label{seth.5}
\Delta E^{{\rm SE}_\beta}_a = -\frac{e^2}{\pi}\int d^3r_1d^3r_2\bar{\psi}_a(\vec{r}_1)\left(1-\vec{\alpha}_1\vec{\alpha}_2\right)
\qquad
\\
\nonumber
\times
\int\limits_0^\infty d\kappa\,n_\beta(\kappa)\frac{\sin\kappa r_{12}}{r_{12}}\psi_a(\vec{r}_2)\times\qquad
\\
\nonumber
\left[\sum\limits_n\frac{\psi_n(\vec{r}_1)\bar{\psi}_n(\vec{r}_2)}{E_a+\kappa-E_n(1-i0)}+\sum\limits_n\frac{\psi_n(\vec{r}_1)\bar{\psi}_n(\vec{r}_2)}{E_a-\kappa-E_n(1-i0)}\right].
\end{eqnarray}

Expression (\ref{seth.3}) can be simplified with the use of series expansion for $\sin\kappa r_{12}/r_{12}\approx \kappa - \frac{1}{6}\kappa^3r_{12}^2+\dots$\,:
\begin{eqnarray}
\label{seth.6}
\Delta E^{{\rm SE}_\beta}_a = -\frac{e^2}{\pi}\int d^3r_1d^3r_2\bar{\psi}_a(\vec{r}_1)\int\limits_0^\infty d\kappa\,n_\beta(\kappa)
\qquad
\\
\nonumber
\times
\left(\kappa -\kappa(\vec{\alpha}_1\vec{\alpha}_2)- \frac{1}{6}\kappa^3r_{12}^2\right)\psi_a(\vec{r}_2)
\times
\qquad
\\
\nonumber
\left[\sum\limits_n\frac{\psi_n(\vec{r}_1)\bar{\psi}_n(\vec{r}_2)}{E_a+\kappa-E_n(1-i0)}+\sum\limits_n\frac{\psi_n(\vec{r}_1)\bar{\psi}_n(\vec{r}_2)}{E_a-\kappa-E_n(1-i0)}\right],
\end{eqnarray}
with retaining only the terms of lowest order. The following evaluation of Eq. (\ref{seth.6}) can be performed in the non-relativistic limit. To this end, we use relations $r_{12}^2=r_1^2+r_2^2-2(\vec{r}_1\vec{r}_2)$ and $(\vec{\alpha}_1\vec{\alpha}_2)_{anna} = (E_a-E_n)^2(\vec{r}_1\vec{r}_2)$, see \cite{LabKlim}. Upon omitting terms $r_1^2+r_2^2$, expression (\ref{seth.6}) transforms to
\begin{eqnarray}
\label{seth.7}
\Delta E^{{\rm SE}_\beta}_a = -\frac{e^2}{\pi}\int d^3r_1d^3r_2\bar{\psi}_a(\vec{r}_1)\int\limits_0^\infty d\kappa\,n_\beta(\kappa)
\qquad
\\
\nonumber
\times
\left( -\frac{2}{3}\kappa^3(\vec{r}_1\vec{r}_2)+ \kappa -\kappa(E_{an}^2-\kappa^2)(\vec{r}_1\vec{r}_2)\right)\psi_a(\vec{r}_2)\times
\qquad
\\
\nonumber
\left[\sum\limits_n\frac{\psi_n(\vec{r}_1)\bar{\psi}_n(\vec{r}_2)}{E_a+\kappa-E_n(1-i0)}+\sum\limits_n\frac{\psi_n(\vec{r}_1)\bar{\psi}_n(\vec{r}_2)}{E_a-\kappa-E_n(1-i0)}\right],
\end{eqnarray}
where we have added and substracted $\kappa^3(\vec{r}_1\vec{r}_2)$.

The first term in the parentheses represents the desired result:
\begin{eqnarray}
\label{seth.8}
\Delta E^{{\rm SE}_\beta}_a = \frac{2e^2}{3\pi}\int d^3r_1d^3r_2\bar{\psi}_a(\vec{r}_1)\int\limits_0^\infty d\kappa\,n_\beta(\kappa)\kappa^3(\vec{r}_1\vec{r}_2)\psi_a(\vec{r}_2)
\nonumber
\\
\left[\sum\limits_n\frac{\psi_n(\vec{r}_1)\bar{\psi}_n(\vec{r}_2)}{E_a+\kappa-E_n(1-i0)}+\sum\limits_n\frac{\psi_n(\vec{r}_1)\bar{\psi}_n(\vec{r}_2)}{E_a-\kappa-E_n(1-i0)}\right].
\qquad
\end{eqnarray}
The evaluation of the real and imaginary parts of the expression was reported in \cite{SLP-QED}, where they were shown to represent the BBR-induced Stark shift and BBR-induced level width, respectively.

Below, we consider the remaining terms in Eq. (\ref{seth.7}). The first of them is simplified substantially by virtue of the orthogonality property of wave functions:
\begin{eqnarray}
\label{seth.9}
\Delta E^{{\rm SE}_\beta^1}_a = -\frac{e^2}{\pi}\int\limits_0^\infty d\kappa\,n_\beta(\kappa)\kappa\left[\frac{1}{\kappa+i0}+\frac{1}{-\kappa+i0}\right].
\end{eqnarray}
In the second one, the imaginary parts in denominators can be dropped out. Then
\begin{eqnarray}
\label{seth.10}
\Delta E^{{\rm SE}_\beta^2}_a = \frac{2e^2}{\pi}\int\limits_0^\infty d\kappa\,n_\beta(\kappa)\kappa\sum\limits_n E_{an}\left|\langle a |\vec{r}|n\rangle\right|^2.
\end{eqnarray}
The summation over $n$ here can be performed with the use of Thomas-Reiche-Kuhn rule. As a result, these two expressions represent the divergent and constant contributions, which are independent of atomic states.

We should note also that the energy shift for the thermal self-energy correction can be obtained as in \cite{SLP-QED}, i.e. with the use of the thermal photon propagator in the form Eq. (\ref{29}). The result is
\begin{eqnarray}
\label{139}
\Delta E_a^{{\rm SE}_\beta}= \frac{e^2}{\pi}\sum\limits_n\left(\frac{1-\vec{\alpha}_1\vec{\alpha}_2}{r_{12}}I^\beta_{na}(r_{12})\right)_{anna},
\end{eqnarray}
where
\begin{eqnarray}
\label{140}
I^\beta_{na}(r_{12}) = \int\limits_0^{\infty} d\kappa\, n_\beta(\kappa) \sin{\kappa r_{12}} \times
\qquad
\\
\nonumber
\left[\frac{1}{E_n(1-i0) - E_a + \kappa} + \frac{1}{E_n(1-i0) - E_a - \kappa}\right]
\end{eqnarray}
or in the equivalent form
\begin{eqnarray}
\label{141}
I^\beta_{na}(r_{12})=\int\limits_{-\infty}^{+\infty}d\omega \frac{n_\beta(|\omega|)\sin{|\omega|r_{12}}}{E_n(1-i0)-E_a+\omega}.
\end{eqnarray}
The evaluation of the expression leads again to terms independent of $r _{1(2)}$. Such contributions are equivalent for any states and, therefore, can be considered as unphysical (unmeasurable). In \cite{SLP-QED}, such contributions were excluded from the consideration.

Nonetheless, the application of the coincidence limit allows cancelling such contributions precisely. To demonstrate this, one can start with Eq. (\ref{seth.5}), where $\sin\kappa r_{12}/r_{12}\rightarrow \kappa$. Repeating all the calculations as before, one can find the terms equal to Eqs. (\ref{seth.9}) and (\ref{seth.10}). Thus, the subtraction of the coincidence limit leads to the regular contribution Eq. (\ref{seth.8}).

Finally, the evaluation of (\ref{seth.8}) or (\ref{139}) can be reduced to the examination of the real and imaginary parts. With the use of Sokhotski-Plemelj theorem, see \cite{SLP-QED}, the results are
\begin{eqnarray}
\label{seth.11}
{\rm Re} \Delta E_a^{{\rm SE}_\beta}=\frac{4e^2}{3\pi}\int\limits_0^\infty d\kappa\,\kappa^3n_\beta(\kappa)\sum\limits_n\frac{\left|\langle a|\vec{r}|n\rangle\right|^2E_{an}}{E_{an}^2-\kappa^2},\,\,\,
\end{eqnarray}
\begin{eqnarray}
\label{seth.12}
{\rm Im} \Delta E_a^{{\rm SE}_\beta}=\frac{4e^2}{3}\sum\limits_n\left|\langle a|\vec{r}|n\rangle\right|^2 n_\beta(|E_{an}|)E_{an}^3.
\end{eqnarray}
Here, the real part of the thermal self-energy correction, ${\rm Re} \Delta E_a^{{\rm SE}_\beta}$, represents the ac-Stark shift induced by the blackbody radiation. The imaginary part, ${\rm Im} \Delta E_a^{{\rm SE}_\beta}$, is total BBR-induced depopulation rate $\Gamma_a^{BBR}$.

Expressions (\ref{seth.11}) and (\ref{seth.12}) are obtained disregarding the finite lifetimes of atomic levels and represent the well-known Quantum Mechanics results \cite{Farley}. In principle, the effect of finite lifetimes can be taken into account in Eq. (\ref{seth.8}) phenomenologically, i.e. by the inclusion of level widths into the energy denominators. However, a more rigorous procedure will be described in the next section.

\subsection{The QED regularization of resonant contributions}
\label{resonance}

\begin{figure}[hbtp]
	\centering
	\includegraphics[scale=0.2]{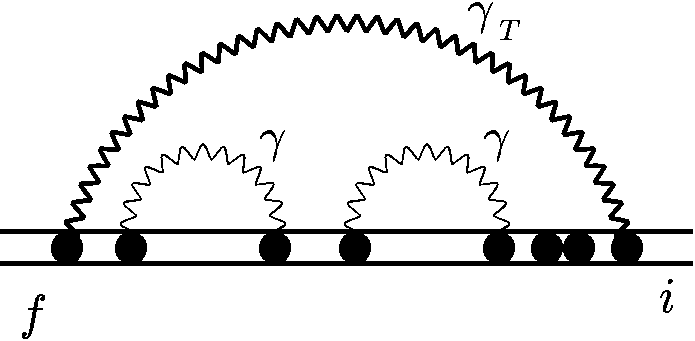}
	\caption{The Feynman graph corresponding to the regularization of resonant denominators in the thermal one-loop self-energy correction Eqs. (\ref{seth.8}), (\ref{139}). The bold wavy line represents the thermal photon, whereas the thin wavy lines correspond to the ordinary photons, remaining dots denoting the whole series of self-energy insertions. All other notations are the same as in Fig.~\ref{SE}.}\label{SE-reg}
\end{figure}

There is another divergence in Eq. (\ref{seth.8}) connected with the resonance in energy denominators. The corresponding regularization can be performed as in \cite{Low}, see also \cite{Andr, ZSLP-review}. The details of such regularization procedure in the thermal case can be found in \cite{SLP-QED}. To this end, the series of the ordinary SE-corrections should be included in the electron line, see Fig.~\ref{SE-reg}.

As a result of such procedure, we arrive at
\begin{eqnarray}
\label{152}
\widetilde{I}^\beta_{na}(r_{12})=\int\limits_{-\infty}^{+\infty}d\omega \frac{n_\beta(|\omega|)\sin{|\omega|r_{12}}}{E_n+\omega-E_a + L_n^{{\rm SE}} - \frac{i}{2}\Gamma_n},\qquad
\end{eqnarray}
where $L_n^{\rm SE}$ is the Lamb shift and $\Gamma_n$ is the natural level width of the $n$th level.

Then, the regularized expression of the thermal one-loop self-energy correction is
\begin{eqnarray}
\label{153}
\widetilde{\Delta E}_a^{{\rm SE}_\beta} = \frac{2e^2}{3\pi}\sum\limits_{n}|\langle a |\vec{r}| n \rangle|^2\int\limits_{0}^{\infty}d\kappa\, \kappa^3 n_\beta(\kappa)
\times\qquad
\\
\nonumber
\left[\frac{\tilde{E}_{an}+\kappa+\frac{i}{2}\Gamma_{an}}{(\tilde{E}_{an}+\kappa)^2 + \frac{1}{4}\Gamma_{an}^2} + \frac{\tilde{E}_{an}-\kappa + \frac{i}{2}\Gamma_{an}}{(\tilde{E}_{an}-\kappa)^2 + \frac{1}{4}\Gamma_{an}^2}\right],
\end{eqnarray}
where tilde stands for the inclusion of the Lamb shift.

The real part of the correction reduces to
\begin{eqnarray}
\label{154}
{\rm Re}\widetilde{\Delta E}_a^{{\rm SE}_\beta} =  \frac{2 e^2}{3\pi}\sum\limits_{n}|\langle a |\vec{r}| n \rangle|^2\int\limits_{0}^{\infty}d\kappa n_\beta(\kappa) \kappa^3
\times\qquad
\\
\nonumber
\left[\frac{\tilde{E}_{an}+\kappa}{(\tilde{E}_{an}+\kappa)^2 + \frac{1}{4}\Gamma_{an}^2} + \frac{\tilde{E}_{an}-\kappa}{(\tilde{E}_{an}-\kappa)^2 + \frac{1}{4}\Gamma_{an}^2}\right],\qquad
\end{eqnarray}
This differs from the quantum mechanical result \cite{Farley} by the inclusion of the Lamb shift and level width: in the limit of the zero Lamb shift and level width, the QM and QED results coincide precisely. The imaginary part of Eq. (\ref{153}) can be expressed as $2{\rm Im} \widetilde{\Delta E}_a^{{\rm SE}_\beta}=-\Gamma_a^{\rm BBR}$:
\begin{eqnarray}
\label{155}
\Gamma_a^{\rm BBR} = \frac{2e^2}{3\pi}\sum\limits_{n}
|\langle a |\vec{r}| n \rangle|^2
\int\limits_{0}^{\infty}d\kappa\, \kappa^3 n_\beta(\kappa)
\times\qquad
\\
\nonumber
\left[\frac{\Gamma_{an}}{(\tilde{E}_{an}+\kappa)^2 + \frac{1}{4}\Gamma_{an}^2} + \frac{\Gamma_{an}}{(\tilde{E}_{an}-\kappa)^2 + \frac{1}{4}\Gamma_{an}^2}\right].
\end{eqnarray}
The result was derived in \cite{SLP-QED} for the first time and transforms to the quantum mechanical expression in the limit $\Gamma_{an}\rightarrow 0$.

\subsection{The one-loop self-energy correction in the thermal Coulomb gauge}
\label{secg}

To verify the correctness of (\ref{seth.8}) and (\ref{153}), we evaluate here the self-energy correction in the thermal Coulomb gauge, see Eqs. (\ref{40}), (\ref{41}). First, we consider the Coulomb part (the zeroth component of the thermal photon propagator (\ref{40})) given by the $S$-matrix element:
\begin{eqnarray}
\label{cg.1}
S_{fi}^{{\rm SE}_\beta^C,C} = 2e^2\int d^4x_1 d^4x_2 \bar{\psi_f}(x_1)\int\limits_{C_1}\frac{d^4k}{(2\pi)^4}e^{ik(x_1-x_2)}
\qquad
\\
\nonumber
\times
\frac{n_\beta(|\vec{k}|)}{\vec{k}^2}\int\limits_{-\infty}^\infty d\omega\, e^{-i\omega(t_1-t_2)}\sum\limits_n\frac{\psi_n(\vec{r}_1)\bar{\psi}_n(\vec{r}_2)}{\omega-E_n(1-i0)}\psi_i(x_2).
\qquad
\end{eqnarray}

Here we can integrate over $k_0$, which leads to factor $4\pi\delta(t_1-t_2)$. Applying the definition (\ref{se.3}) after the integration over time variables, we find the thermal energy shift for the Coulomb part of the self-energy correction in the form:
\begin{eqnarray}
\label{cg.2}
\Delta E_a^{{\rm SE}_\beta^C,C} = 4e^2i\int d^3r_1 d^3r_2 \bar{\psi}_a(\vec{r}_1)\int\frac{d^3k}{(2\pi)^3}e^{i\vec{k}(\vec{r}_1-\vec{r}_2)}
\qquad
\nonumber
\\
\times
\frac{n_\beta(|\vec{k}|)}{\vec{k}^2}\int\limits_{-\infty}^\infty d\omega\sum\limits_n\frac{\psi_n(\vec{r}_1)\bar{\psi}_n(\vec{r}_2)}{\omega-E_n(1-i0)}\psi_a(\vec{r}_2).
\qquad\qquad
\end{eqnarray}
Then, integration over $\vec{k}$ angles yields
\begin{eqnarray}
\label{cg.3}
\Delta E_a^{{{\rm SE}_\beta^C}, C} = \frac{2e^2i}{\pi^2}\int d^3r_1 d^3r_2 \bar{\psi}_a(\vec{r}_1)
\times\qquad\qquad
\\
\nonumber
\int\limits_0^\infty d\kappa\frac{\sin\kappa r_{12}}{\kappa r_{12}}n_\beta(\kappa)\int\limits_{-\infty}^\infty d\omega\sum\limits_n\frac{\psi_n(\vec{r}_1)\bar{\psi}_n(\vec{r}_2)}{\omega-E_n(1-i0)}\psi_a(\vec{r}_2).
\qquad
\end{eqnarray}

Now, one can apply the Sokhotski-Plemelj theorem:
\begin{eqnarray}
\label{SEc.6}
\frac{1}{x\pm i0}= \wp\frac{1}{x}\mp i\pi\delta(x),
\end{eqnarray}
where $\wp$ denotes the Cauchy principal value. The evaluation of the part containing $\delta$-function presents no difficulty:
\begin{eqnarray}
\label{cg.4}
\Delta E_a^{{{\rm SE}_\beta^C}, C,\delta} = \frac{2e^2}{\pi}\int d^3r_1 d^3r_2 \bar{\psi}_a(\vec{r}_1)\int\limits_0^\infty d\kappa\frac{\sin\kappa r_{12}}{\kappa r_{12}}
\nonumber
\\
\times
n_\beta(\kappa)\sum\limits_n\, \psi_n(\vec{r}_1)\bar{\psi}_n(\vec{r}_2)\psi_a(\vec{r}_2)\qquad
\end{eqnarray}
and, with the completeness condition $\sum\limits _n\, \psi _n(\vec{r}_1)\bar{\psi}_n(\vec{r}_2)=\delta(\vec{r}_1-\vec{r}_2)$, it transforms to
\begin{eqnarray}
\label{cg.5}
\Delta E_a^{{{\rm SE}_\beta^C}, C,\delta} = \frac{2e^2}{\pi}\int d^3r  \left|\psi_a(\vec{r})\right|^2\int\limits_0^\infty d\kappa\,
n_\beta(\kappa)
\\
\nonumber
=\frac{2e^2}{\pi}\int\limits_0^\infty d\kappa\,
n_\beta(\kappa).
\end{eqnarray}
At the same time, the principal value integral is
\begin{eqnarray}
\label{cg.6}
\wp\int\limits_{-\infty}^\infty\frac{d\omega}{\omega-E_a} = \lim\limits_{A\rightarrow\infty}\left[ \int\limits_{-A}^{-\epsilon}\frac{dx}{x} + \int\limits_\epsilon^A\frac{dx}{x}\right]_{\epsilon\to 0}\equiv 0.
\qquad
\end{eqnarray}

Now let us consider the coincident limit for the Coulomb part. To this end, we apply expression (\ref{60}) to the $S$-matrix element Eq. (\ref{cg.1}). Repeating all the derivations above, one can find
\begin{eqnarray}
\label{cg.5a}
\Delta E_a^{{\rm SE}_\beta^C,C_\times} = \frac{2ie^2}{\pi^2}\int\limits_0^\infty d\kappa\,n_\beta(\kappa)\,\wp\int\limits_{-\infty}^\infty\frac{d\omega}{\omega-E_a}
\\
\nonumber
+\frac{2e^2}{\pi}\int\limits_0^\infty d\kappa\, n_\beta(\kappa).
\end{eqnarray}
Whereas the principal value integral is still zero, it is obvious that subtraction of $\Delta E_a^{{\rm SE}_\beta^C,C_\times}$ leads to the exact cancellation of (\ref{cg.5}) and (\ref{cg.6}) contributions. Thus, the Coulomb part of the thermal self-energy correction in the thermal Coulomb gauge is absent.

As the next step, we evaluate the transverse part of the self-energy correction. The corresponding $S$-matrix element is
\begin{eqnarray}
\label{cg.7}
S_{fi}^{{{\rm SE}_\beta^C},t} = 2e^2 \int d^4x_1 d^4x_2 \bar{\psi_f}(x_1)\int\limits_{C_1}\frac{d^4k}{(2\pi)^4}e^{ik(x_1-x_2)}
\nonumber
\\
\times
\frac{n_\beta(|\vec{k}|)}{k^2}\left(\vec{\alpha}_1\vec{\alpha}_2 - \frac{(\vec{\alpha}_1\vec{k})(\vec{\alpha}_2\vec{k})}{\vec{k}^2}\right)
\times\qquad
\\
\nonumber
\int\limits_{-\infty}^\infty d\omega e^{-i\omega(t_1-t_2)} \sum\limits_n\frac{\psi_n(\vec{r}_1)\bar{\psi}_n(\vec{r}_2)}{\omega-E_n(1-i0)}\psi_i(x_2).
\end{eqnarray}

Integration over $k_0$ can be carried out with Eq. (\ref{23}). Then, integrating over time variables and $\omega$ with the use of definition Eq. (\ref{se.3}), the energy shift can be written as
\begin{eqnarray}
\label{cg.8}
\Delta E_a^{{{\rm SE}_\beta^C},t} = 2\pi e^2\int d^3r_1 d^3r_2\bar{\psi}_a(\vec{r}_1)\int\frac{d^3k}{(2\pi)^3}e^{i\vec{k}(\vec{r}_1-\vec{r}_2)}
\nonumber
\\
\times
\frac{n_\beta(|\vec{k}|)}{|\vec{k}|}\left(\vec{\alpha}_1\vec{\alpha}_2 - \frac{(\vec{\alpha}_1\vec{k})(\vec{\alpha}_2\vec{k})}{\vec{k}^2}\right)\psi_a(\vec{r}_2)\times
\qquad
\\
\nonumber
\sum\limits_n\left[\frac{\psi_n(\vec{r}_1)\bar{\psi}_n(\vec{r}_2)}{E_a-|\vec{k}|-E_n(1-i0)}+\frac{\psi_n(\vec{r}_1)\bar{\psi}_n(\vec{r}_2)}{E_a+|\vec{k}|-E_n(1-i0)}\right].
\end{eqnarray}
The following simplification of the expression (see \cite{LabKlim}) can be performed using the substitution $(\vec{\alpha}_1\vec{k})(\vec{\alpha}_2\vec{k})\to (\vec{\alpha}_1\vec{\nabla}_1)(\vec{\alpha}_2\vec{\nabla}_2)$ and taking account of
\begin{eqnarray}
\label{cg.8a}
\left((\vec{\alpha}_1\vec{\nabla}_1)(\vec{\alpha}_2\vec{\nabla}_2)f(r_{12})\right)_{abba} = E_{a\,b}^2\left(f(r_{12})\right)_{abba}.
\end{eqnarray}
Then, the transverse part of the self-energy transforms to
\begin{eqnarray}
\label{cg.9}
\Delta E_a^{{{\rm SE}_\beta^C},t} = \frac{e^2}{\pi}\int d^3r_1 d^3r_2\bar{\psi}_a(\vec{r}_1)\int\limits_0^\infty d\kappa\, n_\beta(\kappa)\kappa
\times
\nonumber
\\
\left((\vec{\alpha}_1\vec{\alpha}_2) - \frac{(\vec{\alpha}_1\vec{\nabla}_1)(\vec{\alpha}_2\vec{\nabla}_2)}{\kappa^2}\right)\frac{\sin\kappa r_{12}}{\kappa r_{12}}\psi_a(\vec{r}_2)\times\qquad
\\
\nonumber
\sum\limits_n\left[\frac{\psi_n(\vec{r}_1)\bar{\psi}_n(\vec{r}_2)}{E_a-\kappa-E_n(1-i0)}+\frac{\psi_n(\vec{r}_1)\bar{\psi}_n(\vec{r}_2)}{E_a+\kappa-E_n(1-i0)}\right],
\end{eqnarray}
where we have integrated over $\vec{k}$-angles. Applying (\ref{cg.8a}) and the series expansion of $\sin$ with the relation $r_{12}^2 = r_1^2+r_2^2-2(\vec{r}_1\vec{r}_2)$, one can arrive at
\begin{eqnarray}
\label{cg.9a}
\Delta E_a^{{{\rm SE}_\beta^C},t} = \frac{e^2}{\pi}\int d^3r_1 d^3r_2\bar{\psi}_a(\vec{r}_1)\int\limits_0^\infty d\kappa\, n_\beta(\kappa)\kappa
\qquad
\nonumber
\\
\times
E_{an}^2\left(\frac{2}{3}(\vec{r}_1\vec{r}_2) - \frac{1}{\kappa^2}+\frac{1}{6}(r_1^2+r_2^2)\right)\psi_a(\vec{r}_2)\times\qquad
\\
\nonumber
\sum\limits_n\left[\frac{\psi_n(\vec{r}_1)\bar{\psi}_n(\vec{r}_2)}{E_a-\kappa-E_n(1-i0)}+\frac{\psi_n(\vec{r}_1)\bar{\psi}_n(\vec{r}_2)}{E_a+\kappa-E_n(1-i0)}\right].\qquad
\end{eqnarray}

First of all, one can note immediately that terms $- \frac{1}{\kappa^2}+\frac{1}{6}(r_1^2+r_2^2)$ yield a zero result due to the orthogonality property of wave functions and common factor $E_{an}^2$. To separate out the physical contribution, we make the substitution $\pm\kappa^2$ for the term $\frac{2}{3}(\vec{r}_1\vec{r}_2)$. So, we get
\begin{eqnarray}
\label{cg.10}
\Delta E_a^{{{\rm SE}_\beta^C},t} = \frac{2e^2}{3\pi}\int d^3r_1 d^3r_2\bar{\psi}_a(\vec{r}_1)\int\limits_0^\infty d\kappa\, n_\beta(\kappa)\kappa^3(\vec{r}_1\vec{r}_2)
\qquad
\\
\nonumber
\times
\sum\limits_n\left[\frac{\psi_n(\vec{r}_1)\bar{\psi}_n(\vec{r}_2)}{E_a-\kappa-E_n(1-i0)}+\frac{\psi_n(\vec{r}_1)\bar{\psi}_n(\vec{r}_2)}{E_a+\kappa-E_n(1-i0)}\right] \psi_a(\vec{r}_2)
\\
\nonumber
+\frac{2e^2}{3\pi}\int d^3r_1 d^3r_2\bar{\psi}_a(\vec{r}_1)\int\limits_0^\infty d\kappa\, n_\beta(\kappa)\kappa\,(\vec{r}_1\vec{r}_2)
\times\qquad
\\
\nonumber
\sum\limits_n\left[\frac{(E_{an}^2-\kappa^2)\psi_n(\vec{r}_1)\bar{\psi}_n(\vec{r}_2)}{E_a-\kappa-E_n(1-i0)}+\frac{(E_{an}^2-\kappa^2)\psi_n(\vec{r}_1)\bar{\psi}_n(\vec{r}_2)}{E_a+\kappa-E_n(1-i0)}\right]\psi_a(\vec{r}_2).
\end{eqnarray}
The first part of the above confirms Eq. (\ref{seth.8}). The results (\ref{seth.8}) and (\ref{cg.10}) correspond to the non-relativistic limit and coincide precisely with the expression derived within the quantum mechanical approach \cite{Farley}, see also \cite{SLP-QED}. Thus, the gauge-invariant result for the thermal self-energy correction of the lowest order is
\begin{eqnarray}
\label{cg.11}
\Delta E_a^{{{\rm SE}_\beta^C},t} = \frac{2e^2}{3\pi}\sum\limits_n \left|\langle a | \vec{r}|n\rangle\right|^2\int\limits_0^\infty d\kappa\,n_\beta(\kappa)\kappa^3
\times
\\
\nonumber
\left[\frac{1}{E_a-\kappa-E_n(1-i0)}+\frac{1}{E_a+\kappa-E_n(1-i0)}\right].
\end{eqnarray}

Now, we demonstrate the second part of Eq. (\ref{cg.10}) to be canceled by the corresponding transverse contribution of the coincidence limit. First, the contribution can be simplified by the dropping out the imaginary parts in energy denominators. Then, we can write
\begin{eqnarray}
\label{cg.12}
\Delta E_a^{{{\rm SE}_\beta^C},t(R)} = \frac{4e^2}{3\pi}\int\limits_0^\infty d\kappa\, n_\beta(\kappa)\kappa
\sum\limits_n E_{an}\left|\langle a | \vec{r}|n\rangle\right|^2.
\end{eqnarray}

The evaluation of the transverse part of the coincidence limit (see Eq. (\ref{64})) can start from Eq. (\ref{cg.7}). The energy shift reduces to
\begin{eqnarray}
\label{cg.13}
\Delta E_a^{{{\rm SE}_\beta^C},t_\times} = 4\pi e^2 \int d^3r_1 d^3r_2\bar{\psi}_a(\vec{r}_1)\int\frac{d^3k}{(2\pi)^3}\frac{n_\beta(|\vec{k}|)}{|\vec{k}|}
\nonumber
\\
\left(\vec{\alpha}_1\vec{\alpha}_2 - \frac{(\vec{\alpha}_1\vec{k})(\vec{\alpha}_2\vec{k})}{\vec{k}^2}\right)
\sum\limits_n\frac{\psi_n(\vec{r}_1)\bar{\psi}_n(\vec{r}_2)}{E_a-E_n(1-i0)}\psi_a(\vec{r}_2).
\qquad
\end{eqnarray}
To integrate over $\vec{k}$-angles, we can use the representation of the scalar product in spherical components $(\vec{A}\vec{B}) = \sum\limits_{q}(-1)^q A_qB_{-q}$, where the spherical component of vector is $A_q=\frac{4\pi}{3}|\vec{A}|Y_{1q}(\theta,\phi)$. Then, according to \cite{Varsh}, factor $\frac{8\pi}{3}$ can be easily found.
\begin{eqnarray}
\label{cg.14}
\Delta E_a^{{{\rm SE}_\beta^C},t_\times} =\frac{4 e^2}{3\pi} \int d^3r_1 d^3r_2\bar{\psi}_a(\vec{r}_1)\int\limits_0^\infty d\kappa\, n_\beta(\kappa)\kappa
\nonumber
\\
\times
(\vec{\alpha}_1\vec{\alpha}_2)\sum\limits_n\frac{\psi_n(\vec{r}_1)\bar{\psi}_n(\vec{r}_2)}{E_a-E_n(1-i0)}\psi_a(\vec{r}_2).
\qquad
\end{eqnarray}
Going back to the non-relativistic limit, $(\vec{\alpha}_1\vec{\alpha}_2)\rightarrow E^2_{an}(\vec{r}_1\vec{r}_2)$, we obtain exactly the result Eq. (\ref{cg.12}).

Thus, the divergent thermal contribution at $\kappa=0$ is regularized by the coincidence limits Eqs. (\ref{59}), (\ref{60}). The final result is gauge-invariant and equivalent for two forms of the thermal photon propagator, see Eqs. (\ref{24}) and (\ref{28}). Moreover, the results arising from the one-loop thermal self-energy correction have a clear physical sense, this was originally found within the quantum mechanical approach. The circumstance allows drawing extremely significant conclusions: a) the renormalization or regularization procedure described in section~\ref{cdc} is justified in this case also; b) besides the Stark shift and level broadening, there are no other temperature-dependent contributions.

Numerical values of the ac-Stark shift and level widths, Eqs. (\ref{154}), (\ref{155}) are given in Tables~\ref{tab:1}-\ref{tab:3} for the hydrogen atom. The analysis of the numerical evaluation and corresponding discussion of the results listed in Tables~\ref{tab:1} and \ref{tab:2} can be found in \cite{SLP-QED}. The values of the ac-Stark shifts obtained within the QED approach agree well with the quantum mechanical ones \cite{Farley}. The difference can be explained by the accurate account for the Lamb shift. The QED values for the ac-Stark shift coincide with the results of \cite{Jent-S}. The same conclusions can be drawn for the BBR-induced depopulation rates Eq. (\ref{cg.13}).

The most interesting result arises for the imaginary part of the one-loop self-energy correction within the QED approach, see \cite{SLP-QED}. As noted in \cite{jentlach1,jentlach2}, the accounting for the finite lifetimes of atomic states can lead to enhancement of the friction force in the BBR field by several orders of magnitude. The same situation was found in case of the BBR-induced level widths Eq. (\ref{155}) for the low-lying states. The analysis of the expression for $\Gamma_a^{\rm BBR}$ shows Eq. (\ref{155}) to represent a more general expression for the line broadening, since it contains the nonresonant contributions. The comparison of the values in Tables~\ref{tab:2} and \ref{tab:3} indicates that the nonresonant contribution can exceed the quantum mechanical result (pure resonant contribution) by several orders of magnitude. The detailed analysis of the numerical results for $\Gamma_a^{\rm BBR}$,  Eq. (\ref{155}), is given in \cite{Zal-19}, so we do not repeat it. Table~\ref{tab:3} has these values listed up to temperature $3000 $K, more accurate numerical methods should be applied for higher temperatures.

\begin{widetext}
\begin{center}
\begin{table}[h]
\caption{The BBR-induced ac-Stark shift in Hz for $ns$ states in the hydrogen atom for different temperatures $T$. The first column contains considered $n$ values. The second entries in the second column cells for each $n$ value present the results of \cite{Farley}. The asterix $*$ corresponds to the values evaluated without inclusion of the continuous spectrum.}\label{tab:1}
\begin{tabular}{c|c|c|c|c|c|c}
\hline
$a$ & $T=300$ K & $T=3$ K  $*$ & $T=5.5$ K  $*$ &  $T=3000$ K  & $T=4000$ K  & $T=5000$ K  \\
\hline
\multirow{2}{*}{1s} & -0.0387511 & $-3.1542\times 10^{-10}$ & $-3.56332\times 10^{-9}$ & $-391.455$ & $-1247.01$ & $-3079.38$ \\
 & -0.04128 & & & & & \\
\hline
\multirow{2}{*}{2s} & $-0.989702$ & $7.07128\times 10^{-7}$ & $2.27578\times 10^{-6}$ & $-22348.9$ & $-95953.1$ & $-2.52546\times 10^5$ \\
 & -1.077 & & & & & \\
 \hline
\multirow{2}{*}{3s} & $-8.93974$ & $1.28021\times 10^{-6}$ & $3.64341\times 10^{-6}$ & $-2.76704\times 10^5$ & $-5.72930\times 10^{5}$ & $-8.88017\times 10^{5}$ \\
& $-9.103$  & & & & & \\
 \hline
\multirow{2}{*}{4s} & $-50.1879$ & $1.33848\times 10^{-6}$ & $1.15004\times 10^{-6}$ & $-5.83784\times 10^4$ & $-5.79427\times 10^4$ & $8.4836\times 10^3$ \\
& $-51.19$  & & & & & \\
 \hline
\multirow{2}{*}{5s} & $-186.884$ & $6.24276\times 10^{-7}$ & $-9.94188\times 10^{-6}$ & $1.93186\times 10^5$ & $3.80032\times 10^5$ & $6.19669\times 10^5$ \\
& $-209.5$  & & & & & \\
 \hline
  \hline
\end{tabular}
\end{table}

\begin{table}[h]
\caption{The BBR-induced depopulation rates in $s^{-1}$ for the $ns$ states in the hydrogen atom as evaluated within the QM approach (the imaginary part of Eq. (\ref{cg.11})) for different temperatures $T$. The second entries in the second column cells for each $n$ value present the results of \cite{Farley}. The asterix $*$ corresponds to the values evaluated without inclusion of the continuum spectrum. The last column lists the natural level widths in $s^{-1}$.}
\label{tab:2}
\begin{tabular}{c|c|c|c|c|c|c|c}
\hline
$a$ & $T=300$ K & $T=3$ K  $*$ & $T=5.5$ K  $*$ &  $T=3000$ K  & $T=4000$ K  & $T=5000$ K  &  $\Gamma_a$ \\
\hline
\multirow{2}{*}{2s} & $1.4228\times 10^{-5}$ & $1.4348\times 10^{-7}$ & $2.6203\times 10^{-7}$ &
$4.7038\times 10^4$ & $3.0727\times 10^5$ & $9.7739\times 10^5$ & $8.22935$ \\
 & $1.42\times 10^{-5}$ & & & & & & \\
\hline
\multirow{2}{*}{3s} & $8.0354\times 10^{-5}$ & $9.0978\times 10^{-8}$ & $1.6658\times 10^{-7}$ & $9.8933\times 10^5$ & $2.2813\times 10^6$ & $4.0318\times 10^6$ & $6.3169\times 10^6$\\
 & $7.97\times 10^{-5}$ & & & & & & \\
 \hline
\multirow{2}{*}{4s} & $15.9454$ & $4.5077\times 10^{-8}$ & $8.2601\times 10^{-8}$ & $1.7563\times 10^6$ & $3.1844\times 10^6$ & $4.861\times 10^6$ & $4.4159\times 10^6$\\
& $16.02$  & & & & & &  \\
 \hline
\multirow{2}{*}{5s} & $1196.44$ & $2.6374\times 10^{-8}$ & $4.8341\times 10^{-8}$ & $1.9336\times 10^6$ & $3.1602\times 10^6$ & $4.5265\times 10^6$ & $2.8399\times 10^6$\\
& $1199$  & & & & & & \\
 \hline
  \hline
\end{tabular}
\end{table}

\begin{table}[h]
\caption{The dynamic Stark-mixing level widths of Eq. (\ref{155}) in $s^{-1}$ for the $ns$ states in the hydrogen atom for different temperatures $T$. The second entries in each row show the values in $s^{-1}$ without the account for finite lifetimes.}\label{tab:3}
\begin{tabular}{c|c|c|c|c|c|c}
\hline
$a$ & $T=300$ K & $T=270$ K & $T=77$ K & $T=3$ K  $*$ & $T=5.5$ K  $*$ & $T=3000$ K  \\
\hline
$2s$ & $4.15854\times 10^{-3}$ & $3.36964\times 10^{-3}$ & $2.7664\times 10^{-4}$ &$5.54615\times 10^{-7}$ & $1.65129\times 10^{-6}$ & $4.6982\times 10^4$  \\
 & $3.42136\times 10^{-7}$ & $2.2403\times 10^{-7}$ & $1.47047\times 10^{-9}$ & $3.38597\times 10^{-15}$ & $3.82517\times 10^{-14}$ & $4.6981\times 10^4$   \\
 \hline
$3s$ & $7.86448\times 10^{-3}$ & $6.31671\times 10^{-3}$ & $5.14843\times 10^{-4}$ & $8.68118\times 10^{-7}$ & $2.7804\times 10^{-6}$ & $9.71198\times 10^5$ \\
 & $7.58576\times 10^{-5}$ & $7.10358\times 10^{-6}$ & $1.82895\times 10^{-8}$ & $4.18565\times 10^{-14}$ & $4.72864\times 10^{-13}$ & $9.71197\times 10^5$  \\
 \hline
$4s$ & $15.9569$ & $4.28494$ & $7.58125\times 10^{-4}$ & $1.19441\times 10^{-6}$ & $3.94504\times 10^{-6}$ & $1.64562\times 10^6$ \\
& $15.9455$ & $4.27576$ & $1.02324\times 10^{-5}$ & $1.60527\times 10^{-8}$ & $5.28938\times 10^{-8}$ & $1.64562\times 10^6$ \\
 \hline
$5s$ & $1196.44$ & $580.985$ & $9.88488\times 10^{-4}$ & $1.51061\times 10^{-6}$ & $5.037\times 10^{-6}$ & $1.72221\times 10^6$ \\
& $1196.42$ & $580.973$ & $1.0057\times 10^{-5}$ & $1.12694\times 10^{-12}$ & $1.27348\times 10^{-11}$ & $1.72221\times 10^6$ \\
 \hline
  \hline
\end{tabular}
\end{table}
\end{center}
\end{widetext}

\subsection{The one-loop self-energy: Donoghue-Holstein-Robinett correction}
\label{DHRc}

Another thermal correction (see Eq. (185)) to the Lamb shift was found in \cite{DHR}. According to it, the dominant low temperature modifications to the $2s_{1/2}$, $2p_{1/2}$ energy splitting (to the Lamb shift) involve Bethe's non-relativistic calculation of the low-energy virtual photon 'bubble'. Below, we consider the remaining terms in the thermal self-energy correction Eqs. (\ref{139}), (\ref{140}). Namely, correction Eq. (185) in \cite{DHR} can arise from the terms in Taylor series of $\sin$ and subsequent representation of $r_{12}^2$ as $r_1^2+r_2^2-2(\vec{r}_1\vec{r}_2)$.

In the previous sections of the chapter, the contribution $r_1^2+r_2^2$ was omitted from the consideration in the Feynman gauge and was noted to be zero in the Coulomb gauge. Now, we evaluate it briefly within the Feynman gauge. To this end, we should consider the $\vec{\alpha}$-independent term in Eqs. (\ref{139}), (\ref{140}). Then, the arising thermal correction looks as
\begin{eqnarray}
\label{dhrc.1}
\Delta E_a^{\rm DHR} = -\frac{e^2}{6\pi}\sum\limits_n\int\limits_0^\infty d\kappa\, n_\beta(\kappa)\kappa^3
\times\qquad
\\
\nonumber
\left[\frac{\langle an|r_1^2+r_2^2 |n a\rangle}{E_a-\kappa-E_n(1-i0)}+\frac{\langle an|r_1^2+r_2^2 |n a\rangle}{E_a+\kappa-E_n(1-i0)}\right].
\end{eqnarray}
In view of the orthogonality property of wave functions, the only state left in the sum over $n$ is $n=a$. Since there is no pole in this case, the imaginary part of energy denominators can be dropped out. As a consequence, two terms in the brackets cancel each other. We should note that the same result (equal to zero) can be obtained for the ordinary case (the zero vacuum) of the self-energy correction. Contrary to the conclusion. the authors of \cite{DHR} found the four times larger contribution than the first term in Eq. (\ref{dhrc.1}). This results from the not quite correct approximation.

\section{The recoil, vacuum polarization and combined vertex-SE corrections}
\label{RZ-VP}
\subsection{The recoil effect}
\label{RE}
Before discussing the recoil effect, we should recall that of the finite mass of the nucleus. It arises immediately from Eq. (\ref{64}), when we replace the electron mass ($m_e=1$ in our units) with the reduced mass $\mu = m_e M/(m_e+M)$, where $M$ is the nuclear mass. The series expansion of $\mu$ over $m_e/M$ yields the same thermal correction plus the term proportional to the ratio $-m_e^2/M$. The expansion brings no changes to the regularization procedure, see subsection~\ref{cdc} and Eq. (\ref{59}), and produces the finite nuclear mass correction:
\begin{eqnarray}
\label{fM}
\Delta E_{a,{\rm fm}}^{\beta,{\rm reg}} = -\frac{4Ze^2\zeta(3)}{3\pi\beta^3 M}\int d^3r\left|\psi_a(\vec{r})\right|^2r^2,
\end{eqnarray}
which is three orders of magnitude less than the result Eq. (\ref{68}) for the hydrogen atom ($m_e^2/M\sim 1836^{-1}$ in our units \cite{BS,Sal}).

Expression (\ref{fM}) holds for a one-electron atom or an ion in the non-relativistic approximation, i.e. for the low $Z$ values. In the ordinary QED, there are also recoil corrections of the order $m_e/M$ containing an additional smallness in relativistic parameter $\alpha Z$ ($\alpha$ is the fine structure constant). The Coulomb corrections of that kind were evaluated in \cite{Lab1972}, see also \cite{Lab1999}. The transverse part of these corrections was evaluated in \cite{Braun1973, shabaev1985,shabaev1988,artemyev1995}, see also \cite{shabaev}.

Such thermal QED corrections can arise due to the exchange with transverse thermal photons, which takes place between the electrons and the nucleus in an atom. However, in the standard QED, there are no recoil corrections apart from the reduced mass for the one-electron atom in the non-relativistic limit. Below, we show the same to hold for the thermal QED. To this end, we consider the diagram Fig.~\ref{Fig-4} with the exchange with the transverse thermal photons.

The corresponding $S$-matrix element is
\begin{eqnarray}
\label{RE.1}
S^{\beta, {\rm tr}}_{fi}  = -4\pi i Ze^2\int d^4x \bar{\psi}_f(x)\gamma_l\psi_i(x)
\times\\
\nonumber
\int\limits_{C_1}\frac{d^4k}{(2\pi)^4}\frac{e^{ikx}}{k^2}n_\beta(|\vec{k}|)j^l(k),
\end{eqnarray}
where indices $l,m$ run over $(1,2,3)$ and we have used the Fourier transform for the nuclear current (see the corresponding procedure in section~\ref{exchange}).

In the non-relativistic limit, Dirac matrix $\gamma_l$ reduces to the matrix element for momentum operator $p$, see \cite{LabKlim}. In turn, momentum operator $p$ acting upon wave functions gives the zero result for the diagonal matrix element in the non-relativistic limit. Thus, we find
\begin{eqnarray}
\label{RE.2}
S^{\beta, {\rm tr}}_{fi}  = -4\pi i Ze^2\int d^4x \bar{\psi}_f(x)
\times
\\
\nonumber
\int\limits_{C_1}\frac{d^4k}{(2\pi)^4}\frac{e^{ikx}}{k^2}n_\beta(|\vec{k}|)\left(\vec{k}\,\vec{j}(k)\right)\psi_i(x),
\end{eqnarray}
where the scalar product $\left(\vec{k}\,\vec{j}(k)\right)$ results from the action of the operator $\vec{p}=-i\vec{\nabla}$ upon the exponential. The scalar product can be transformed into $k_0\, j^0(k)$ in the momentum space via the continuity equation. Applying the static limit to the zeroth component of current $j^0(k) = 2\pi\delta(k_0)\rho^{\rm ext}(\vec{k})$, we can conclude that the transverse contribution to the vertex diagram Fig.~\ref{Fig-4} vanishes. To verify the result, we consider also the thermal Coulomb gauge.

The $S$-matrix element for the transverse part of the thermal photon exchange diagram can be written using Eq. (\ref{35}):
\begin{eqnarray}
\label{RE.3}
S^{\beta, {\rm tr}(C)}_{fi}  = -4\pi i Ze^2\int d^4x \bar{\psi}_f(x)\vec{\alpha}^l\psi_i(x)\int\limits_{C_1}\frac{d^4k}{(2\pi)^4}\frac{e^{ikx}}{k^2}
\nonumber
\\
\times
n_\beta(|\vec{k}|)\left(\delta_{lm} - \frac{k_lk_m}{\vec{k}^2}\right)j^m(k),\qquad
\end{eqnarray}
which gives
\begin{eqnarray}
\label{RE.4}
S^{\beta, {\rm tr}(C)}_{fi}  = -4\pi i Ze^2\int d^4x \bar{\psi}_f(x)\int\limits_{C_1}\frac{d^4k}{(2\pi)^4}\frac{e^{ikx}}{k^2}
\nonumber
\\
\times
n_\beta(|\vec{k}|)\left((\vec{\alpha}\, \vec{j}(k)) - \frac{(\vec{\alpha}\, \vec{k})(\vec{k}\, \vec{j}(k)}{\vec{k}^2}\right)\psi_i(x).
\end{eqnarray}
Passing to the non-relativistic limit for the $\vec{\alpha}$-matrix, we find
\begin{eqnarray}
\label{RE.5}
S^{\beta, {\rm tr}(C)}_{fi}  = -4\pi i Ze^2\int d^4x \bar{\psi}_f(x)\int\limits_{C_1}\frac{d^4k}{(2\pi)^4}\frac{e^{ikx}}{k^2}
\nonumber
\\
\times
n_\beta(|\vec{k}|)\left((\vec{k}\, \vec{j}(k)) - \frac{(\vec{k}\, \vec{k})(\vec{k}\, \vec{j}(k)}{\vec{k}^2}\right)\psi_i(x).
\end{eqnarray}
Thus, two terms in brackets of Eq. (\ref{RE.5}) cancel each other, and we obtain the gauge-invariant zero result. The same can be shown within the adiabatic $S$-matrix formalism.

For completeness' sake, we should consider the transverse part of the coincident limit, Eqs. (\ref{59}), (\ref{60}), which obviously should produce the zero result. The corresponding evaluation starts with the $S$-matrix element
\begin{eqnarray}
\label{RE.6}
S^{\beta, {\rm tr},\times}_{fi}  = -4\pi i Ze^2\int d^4x \bar{\psi}_f(x)\psi_i(x)
\times
\\
\nonumber
\lim\limits_{x\rightarrow 0}\int\limits_{C_1}\frac{d^4k}{(2\pi)^4}\frac{e^{ikx}}{k^2}n_\beta(|\vec{k}|)\left(\vec{k}\,\vec{j}(k)\right),
\end{eqnarray}
where the limit $x\rightarrow 0$ means that we set the origin of coordinates at the nucleus. Then, applying the continuity equation $(\vec{k}\,\vec{j}(k))=k_0\, j^0(k)$ again, we arrive at the zero result in the static limit for the nuclear current. The proof for the thermal Coulomb gauge repeats equation (\ref{RE.5}).

\subsection{The thermal vacuum polarization correction}
\label{TVP}

\begin{figure}[h]
	\centering
	\includegraphics[scale=0.15]{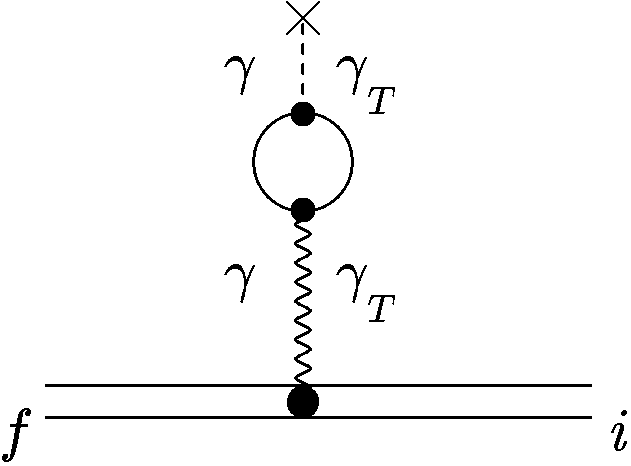}
	\caption{The Feynman graph corresponding to the vacuum polarization. The double line denotes the bound electron. The wavy line represents the photon exchange, where notation $\gamma$ corresponds to the ordinary photon line and $\gamma_T$ corresponds to the thermal one. The dashed line with a cross represents the Coulomb part of the photon propagator and shows the interaction with the nucleus via the ordinary or thermal photon exchange.}\label{vp}
\end{figure}

The section deals with the vacuum polarization (VP) correction, which is due to the external blackbody radiation field. In compliance to the standard QED methods, the fermion loop is to be presented as the expansion in the Coulomb interaction of a bound electron with the nucleus \cite{Berest}. The lowest order diagram of such expansion is shown in Fig.~\ref{vp}. Then, the Uehling potential can be derived from given Feynman graph. Employing the same procedure, we derive here the thermal Uehling potential, which corresponds to the ordinary photon lines (depicted in Fig.~\ref{vp} as wavy lines with $\gamma$) being replaced with the thermal photon lines marked with $\gamma_T$.

The $S$-matrix element for Fig.~\ref{vp} is
\begin{eqnarray}
\label{vp.1}
S_{f i} = Z e^2\int d^4x d^4x' \bar{\psi}_f(x)\gamma^\mu \psi_i(x) i D^U_{\mu \nu}(x,x') j_{\rm ext}^\nu(x'),
\qquad
\end{eqnarray}
where $D^U_{\mu \nu}(x,x')$ represents the Uehling photon propagator:
\begin{eqnarray}
\label{vp.2}
iD^{U}_{\mu \nu}(x,x') = \int d^4x_1 d^4x_2 iD_{\mu \lambda}^{F,\beta}(x,x_1)
\times
\\
\nonumber
 i\Pi^{\lambda \sigma}(x_1,x_2)iD_{\sigma \nu}^{F,\beta}(x_2,x').
 \qquad
\end{eqnarray}
Here, $\Pi^{\lambda \sigma}(x_1,x_2)$ is the tensor corresponding to the fermion loop in Fig.~\ref{vp} (see, e. g., \cite{Greiner}). We do not modify the fermion propagator, assuming the fermion part to be suppressed by factor $exp(-\beta m_e)$ ($m_e$ is the fermion mass) and, therefore, negligible in the thermal QED description of bound states \cite{DHR}.

The photon propagator $D^{\mu \nu}_{F,\beta}(x,x')$ is the sum of ordinary and thermal propagators, see Eq. (\ref{17}). Allowing for the subtraction of the coincidence limit Eq. (\ref{59}), we have
\begin{eqnarray}
\label{vp.3}
iD^{F,\beta}_{\mu \nu}(x,x') = -4\pi i g_{\mu \nu}\left[\int \frac{d^4k}{(2\pi)^4}\frac{e^{ik(x-x')}}{k^2}+
\right.
\\
\nonumber
\left.
\int\limits_{C_1} \frac{d^4k}{(2\pi)^4}n_{\beta}(|\vec{k}|)\frac{e^{i k(x-x')}-1}{k^2}\right].
\end{eqnarray}

The $S$-matrix element Eq. (\ref{vp.1}) can be evaluated by parts. First, we consider the integral
\begin{eqnarray}
\label{vp.4}
iD^U_{\mu}(x)= \int d^4x'\int d^4x_1 d^4x_2 iD^{F,\beta}_{\mu \lambda}(x,x_1)
\\
\nonumber
\times
i\Pi^{\lambda \sigma}(x_1,x_2) iD^{F,\beta}_{\sigma \nu}(x_2,x') j_{\rm ext}^\nu(x').
\end{eqnarray}
Applying the Fourier transform to the nuclear current with the subsequent integration over $d^4x'$, we find
\begin{eqnarray}
\label{vp.5}
iD^U_{\mu}(x)= \int d^4x_1 d^4x_2 iD^{F,\beta}_{\mu \lambda}(x,x_1)i\Pi^{\lambda \sigma}(x_1,x_2)
\times
\qquad
\\
\nonumber
(-4\pi i g_{\sigma\nu})\left[\int \frac{d^4k}{(2\pi)^4}\frac{e^{ikx_2}}{k^2}+\int\limits_{C_1} \frac{d^4k}{(2\pi)^4}n_{\beta}(|\vec{k}|)\frac{e^{i kx_2}}{k^2}
\right.
\\
\nonumber
\left.
-\lim_{x_2\rightarrow 0}\int\limits_{C_1}\frac{d^4k}{(2\pi)^4}\frac{e^{ikx_2}}{k^2}n_\beta(|\vec{k}|)\right]j^\nu(k),
\end{eqnarray}
where the $\lim_{x_2\rightarrow 0}$ means that the nucleus is placed at the origin of coordinates. Then, we use the Fourier transform for the vacuum polarization tensor:
\begin{eqnarray}
\label{vp.6}
i\Pi^{\lambda \sigma}(x_1,x_2) = i\int \frac{d^4q}{(2\pi)^4}e^{i q(x_1-x_2)}\Pi^{\lambda \sigma}(q)
\equiv
\\
\nonumber
\equiv i\int \frac{d^4q}{(2\pi)^4}e^{i q(x_1-x_2)}\left(q^2g^{\lambda\sigma}-q^\lambda q^\sigma\right)\Pi^R(q).
\end{eqnarray}
Here $\Pi^R(q)$ is the 'ordinary' regularized vacuum polarization tensor, see \cite{Greiner,Berest}.

We can now integrate over $d^4x_2$ to get the $\delta$-function that removes the integration over $d^4k$. In view of the continuity equation for the current $q^\nu j_\nu(k)=0$, only the first term in parentheses of Eq. (\ref{vp.6}) is left. The further consideration can be restricted to the scalar part of nuclear current $j_{\nu=0}(k)$. In this case, the thermal part of the photon propagator in Eq. (\ref{vp.5}) reduces precisely to the coincident limit. Thus, we conclude that there is no contribution of the thermal photon interaction in the external line of the VP correction (Fig.~\ref{vp}). The Uehling potential is expressed as
\begin{eqnarray}
\label{vp.7}
iD^U_{0}(x)= 4\pi \int d^4x_1 iD^{F,\beta}_{00}(x,x_1)
\int\frac{d^4q}{(2\pi)^4}e^{iqx_1}\Pi^R(q)j^{0}(q).\qquad
\end{eqnarray}
Then, substituting Eq. (\ref{vp.3}) into Eq. (\ref{vp.7}) with subsequent integration over $d^4x_1$ and $d^4q$, we get
\begin{eqnarray}
\label{vp.8}
iD^U_{0}(x)= -i(4\pi)^2\left[\int \frac{d^4k}{(2\pi)^4}\frac{e^{ikx}}{k^2}+\int\limits_{C_1} \frac{d^4k}{(2\pi)^4}\frac{e^{i kx}}{k^2}n_{\beta}
\right.
\nonumber
\\
\left.
-\lim_{x\rightarrow 0}\int\limits_{C_1}\frac{d^4k}{(2\pi)^4}\frac{e^{ikx}}{k^2}n_\beta\right]
\Pi^R(k)j_0(k).
\end{eqnarray}

The first term in brackets of Eq. (\ref{vp.8}) corresponds to the ordinary QED result for the zero vacuum, and we ignore it in the following evaluation. Applying the static limit for the point-like nucleus $j_0(k) = 2\pi\delta(k_0)\rho(\vec{k})\approx 2\pi\delta(k_0)$, the thermal Uehling potential can be reduced to
\begin{eqnarray}
\label{vp.9}
iD^{U,\beta}_{0}(x)= 2i(4\pi)^2\left[\int \frac{d^3k}{(2\pi)^3}\frac{e^{i\vec{k}\vec{r}}}{\vec{k}^2}n_{\beta}(|\vec{k}|)
\right.
\\
\nonumber
\left.
-\lim_{\vec{r}\rightarrow 0}\int\frac{d^3k}{(2\pi)^3}\frac{e^{i\vec{k}\vec{r}}}{\vec{k}^2}n_\beta(|\vec{k}|)\right]
\Pi^R(\vec{k}).
\end{eqnarray}
The integrand in Eq. (\ref{vp.9}) can be simplified substantially with the use of the approximation of small $|\vec{k}|$ values for vacuum polarization function $\Pi^R(\vec{k})$. The condition holds due to the presence of Bose distribution function $n_\beta(|\vec{k}|)$.

The vacuum polarization function in the lowest order, see \cite{Greiner}, is given by
\begin{eqnarray}
\label{vp.10}
\Pi^R(\vec{k}) \approx \frac{e^2\vec{k}^2}{15\pi m_e^2},
\end{eqnarray}
where we wrote down electron mass $m_e$ for clarity. Integrating over angles in Eq. (\ref{vp.9}), we find
\begin{eqnarray}
\label{vp.11}
iD^{U,\beta}_{0}(r)= \frac{16i}{15\pi m_e^2}\int\limits_0^\infty d\kappa\, \kappa^2 n_\beta(\kappa)\left(\frac{\sin\kappa r}{\kappa r}-1\right).
\end{eqnarray}

The $S$-matrix element Eq. (\ref{vp.1}) transforms to
\begin{eqnarray}
\label{vp.12}
S_{f i}^\beta = Z e^2\int d^4x \bar{\psi}_f(x)\gamma^{\mu=0} \psi_i(x) iD^{U,\beta}_{\mu=0}(r).\qquad
\end{eqnarray}
Then, after the time integration with the use of definition Eq. (\ref{46}), we obtain the energy shift
\begin{eqnarray}
\label{vp.13}
\Delta E_a^{{\rm VP},\beta} = -\frac{16 Ze^4}{15\pi m_e^2}\int d^3r\left|\psi_a(\vec{r})\right|^2
\times
\nonumber
\\
\int\limits_0^\infty d\kappa\, \kappa^2 n_\beta(\kappa)\left(\frac{\sin\kappa r}{\kappa r}-1\right).
\end{eqnarray}

Within the non-relativistic limit, we can write approximately
\begin{eqnarray}
\label{vp.14}
\Delta E_a^{{\rm VP},\beta}\approx \frac{8 Ze^4}{45\pi m_e^2}\int d^3r\,r^2\left|\psi_a(\vec{r})\right|^2\int\limits_0^\infty d\kappa\, \kappa^4 n_\beta(\kappa),
\end{eqnarray}
so the final result is
\begin{eqnarray}
\label{vp.15}
\Delta E_a^{{\rm VP},\beta}\approx \frac{64\zeta(5) Ze^4}{15\pi m_e^2\beta^5}\int d^3r\,r^2\left|\psi_a(\vec{r})\right|^2.
\end{eqnarray}
The gauge invariance of the result can be easily shown with the procedures described above. The thermal vacuum polarization correction for the hydrogen atom yields
\begin{eqnarray}
\label{vp.16}
\Delta E_a^{{\rm VP (H)},\beta}= \frac{32\zeta(5) Ze^4}{15\pi m_e^2\beta^5}n_a^2(5n_a^2+1-3l_a(l_a+1)),
\end{eqnarray}
where $n_a$, $l_a$ are the principal quantum number and the orbital momentum of the bound state $a$, respectively.

We should note that the evaluation of the VP correction without the inclusion of the coincidence limit would lead to the thermal correction, which is constant and independent of the atomic state. Thus, the regularization suggested in section~\ref{cdc} for the thermal photon propagator works in this case as well. We have used here the approximation of the point-like nucleus, which suffices for our purposes in view of the small temperature factor. At the room temperature, the thermal vacuum-polarization correction is negligible in respect to the leading thermal photon exchange correction Eq. (\ref{68}) or the self-energy correction, which are proportional to the third and fourth degrees of the temperature, respectively. The suppression factor in the thermal VP correction is $1/\beta^5\sim 10^{-16}$ in atomic units. In principle, the VP nuclear finite-size correction can be found from the considerations above, see section~\ref{FSVP}; however, it is negligibly small. The thermal nuclear finite-size correction of the leading order will be derived in the next chapter.

\subsection{The combined vertex-self-energy corrections}
\label{seR}
\begin{figure}[h]
	\centering
	\includegraphics[scale=0.2]{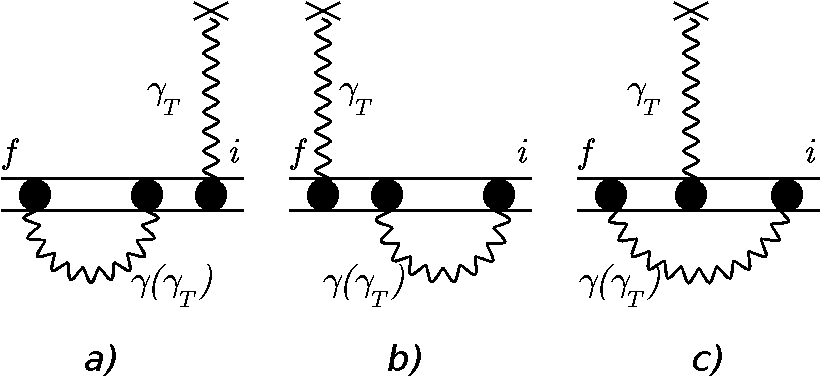}
	\caption{The Feynman graphs corresponding to the introduction of the self-energy loops into the thermal vertex diagram Fig.~\ref{Fig-4}. All the notations are the same as in the previous figures.}\label{SEinV}
\end{figure}

The part of work deals briefly with the Vertex diagram Fig.~\ref{Fig-4} 'dressed' with the self-energy loops. Such diagrams in the lowest order are shown in Fig.~\ref{SEinV}.
After the time integration within the $S$-matrix formalism, it can be shown that the term corresponding to the vertex Fig.~\ref{Fig-4} can be separated out. As before, the contribution is divergent. However, two graphs a) and b) will in this case compensate each other precisely. The same will be concluded within the adiabatic $S$-matrix formalism for the reducible graphs, where the 'reference' state is canceled by the terms arising from the previous orders of the perturbation theory, see section~\ref{ThVC}.

The expression for the sum of graphs a) and b) (its irreducible part) can be found as
\begin{widetext}
\begin{eqnarray}
\label{ab.1}
\Delta E_a^{\rm irr\,\,\, a)+b)} = 4iZe^4\sum\limits_{\mathop{n,m}\limits_{m\neq a}}\int\limits_{-\infty}^\infty d\omega\frac{\langle an| \hat{\Sigma}^{(\beta)}(|\omega|,r_{12}) |nm \rangle}{E_a-\omega-E_n(1-i0)}
\frac{\langle m | \int\frac{d^3k}{(2\pi)^3}\frac{e^{i\vec{k}\vec{r}}}{\vec{k}^2}n_\beta(|\vec{k}|) | a \rangle}{E_a-E_m} +
\\
\nonumber
4iZe^4\sum\limits_{\mathop{n,m}\limits_{n\neq a}}\int\limits_{-\infty}^\infty d\omega\frac{\langle a | \int\frac{d^3k}{(2\pi)^3}\frac{e^{i\vec{k}\vec{r}}}{\vec{k}^2}n_\beta(|\vec{k}|) | n \rangle}{E_a-E_n}
\frac{\langle nm | \hat{\Sigma}^{(\beta)}(|\omega|,r_{12}) |ma \rangle}{E_a-\omega-E_m(1-i0)}+
\\
\nonumber
4iZe^4\langle a| \int\frac{d^3k}{(2\pi)^3}\frac{e^{i\vec{k}\vec{r}}}{\vec{k}^2}n_\beta(|\vec{k}|) | a \rangle
 \sum\limits_n \frac{\partial}{\partial E_a} \int\limits_{-\infty}^\infty d\omega\frac{\langle a n|\hat{\Sigma}^{(\beta)}(|\omega|,r_{12})| n a\rangle}{E_a-\omega-E_n(1-i0)}.
\end{eqnarray}
This is a well-known result, which can also be found within the two-time Green's function method, see \cite{shabaev}. The first two contributions can be considered as the corrections to the wave function of the state $a$, and the last one gives the correction to the interaction, which arises within the framework of the perturbation theory in the presence of an external potential.

The result for graph c) is given by
\begin{eqnarray}
\label{ab.2}
\Delta E_a^{\rm c)} = 4iZe^4\sum\limits_{n,m}\langle n|\int\frac{d^3k}{(2\pi)^3}\frac{e^{i\vec{k}\vec{r}}}{\vec{k}^2}n_\beta(|\vec{k}|)|m\rangle
\int\limits_{-\infty}^\infty \frac{\langle a m|\hat{\Sigma}^{(\beta)}(|\omega|,r_{12})|n a\rangle  d\omega}{[E_a-\omega-E_n(1-i0)][E_a-\omega-E_m(1-i0)]}.
\end{eqnarray}
\end{widetext}
Here $\hat{\Sigma}^{(\beta)}(|\omega|,r_{12})$ is the self-energy operator for the zero and none-zero temperatures:
\begin{eqnarray}
\label{ab.3}
\hat{\Sigma}^{(\beta)}(|\omega|,r_{12}) =
\begin{cases}
\frac{1-\vec{\alpha}_1\vec{\alpha}_2}{r_{12}}e^{i|\omega|r_{12}},\; T=0\\
\frac{1-\vec{\alpha}_1\vec{\alpha}_2}{r_{12}}\sin|\omega|r_{12}\,n_\beta(|\omega|),\; T\neq 0.\\
\end{cases}
\end{eqnarray}

The following evaluation of Eqs. (\ref{ab.1}) and (\ref{ab.2}) with the thermal self-energy loop can be excluded from the consideration, since it produces an additional factor of temperature $1/\beta^4$ in the final result, see the previous sections and \cite{SLP-QED}. The first two terms in Eq. (\ref{ab.1}) are easily seen to contain no divergences corresponding to the thermal interaction: since $n,m\neq a$, the $\vec{r}$-independent contribution is canceled by the orthogonality property of wave functions. Moreover, the coincidence limit can be easily shown to vanish in the case. Rough estimates of these summands are about $-3.05 \cdot 10^{-4}$ $s^{-1}$ in magnitude for the ground state and not in excess of $-0.1$ $s^{-1}$ for the $6s$ state of the hydrogen atom at the room temperature.

Nonetheless, the regularization of the expression (\ref{ab.2}) is called for. The divergence of the thermal part is partially canceled by the remaining third summand in Eq. (\ref{ab.1}): the contribution $n=m=a$ in Eq. (\ref{ab.2}) occurs with the opposite sign to the third term in (\ref{ab.1}) with $n=a$. There are also divergences related with the 'ordinary' self-energy loop. Here, we assume the regularization in this case to arise via the known renormalization procedure, and, therefore, the operator $\hat{\Sigma}(|\omega|,r_{12})$ is to be replaced with the renormalized one.

An accurate evaluation of Eqs. (\ref{ab.1}), (\ref{ab.2}) is a separate task, and the result should be rather small in respect to the lowest-order correction Eq. (\ref{68}) or to the Stark shift Eq. (\ref{seth.11}). To demonstrate this, we restrict ourselves with the estimation of (\ref{ab.2}) only. In particular, we consider the diagonal matrix element $\langle a n|\hat{\Sigma}^{(\beta)}(|\omega|,r_{12})|n a\rangle$ under the assumption of the off-diagonal elements being smaller. Then, the thermal correction (\ref{ab.2}) with $n=m\neq a$ and the third term in (\ref{ab.1}) with $n\neq a$ can be expressed with the formula:
\begin{eqnarray}
\label{ab.4}
\Delta E_a = \frac{e^2}{2\pi i}\sum\limits_{n\neq a} \frac{\langle a n|\hat{\Sigma}^{(R)}(|\omega|,r_{12})|n a\rangle}{\left(E_a-\omega-E_n(1-i0)\right)^2}(-8\pi Ze^2)
\qquad
\\
\nonumber
\times
\left[\langle n|\int\frac{d^3k}{(2\pi)^3}\frac{e^{i\vec{k}\vec{r}}n_\beta}{\vec{k}^2}|n\rangle -\langle a|\int\frac{d^3k}{(2\pi)^3}\frac{e^{i\vec{k}\vec{r}}n_\beta}{\vec{k}^2}|a\rangle\right].
\end{eqnarray}
Applying again the coincidence limit to the thermal part of Eq. (\ref{ab.4}), one can find the contribution of the order of Eq. (\ref{68}). The order of magnitude of Eq. (\ref{ab.4}) can be found by the way of the parametric estimations: $\omega\sim\kappa\sim m_e(\alpha Z)^2$, $r^2\sim 1/(m_e\alpha Z)^2$, and $\langle\hat{\Sigma}^{(R)}\rangle/\Delta E\sim m_e\alpha(\alpha Z)^4$ in relativistic units, see \cite{LabKlim}:
\begin{eqnarray}
\label{ab.5}
\Delta E_a\sim \frac{m_e\alpha(\alpha Z)^4}{m_e(\alpha Z)^2}\frac{Z\alpha m_e^3(\alpha Z)^6}{(m_e \alpha Z)^2\beta^3} = \frac{1}{\beta^3} m_e\alpha(\alpha Z)^7,
\end{eqnarray}
which is $\alpha^2$ times as small as the thermal correction Eq. (\ref{68}). The estimate Eq. (\ref{ab.5}) can also be compared with the contribution arising for the thermal self-energy correction, which is proportional to $m_e\alpha(\alpha Z)^4/\beta^4$. We leave the accurate evaluation of Eqs. (\ref{ab.1}), (\ref{ab.2}) for the future.

\section{The nuclear finite-size correction}
\label{FSC}
\subsection{The nuclear finite-size correction: the quantum mechanical approach}
\label{QM}

At first, we consider the nuclear finite-size correction to the bound electron energy without thermal effects. The correction can be found in different ways. Here, we demonstrate these options in order to resort to them for the derivation and checking the correctness of the corresponding thermal contribution. One of methods rests on the use of an approximate expression, see, e. g., \cite{Eides}, for the charge distribution:
\begin{eqnarray}
\label{r1}
\rho^{\rm ext}(\vec{k})\approx 1 -\frac{\vec{k}^2}{6} r^2_p,
\end{eqnarray}
where $r^2_p$ denotes the average value of the squared radius of the nuclear charge. Then, the substitution of (\ref{r1}) into the $S$-matrix element corresponding to the graph Fig.~\ref{Fig-4} produces the part dependent on the nuclear radius
\begin{eqnarray}
\label{r2}
\Delta E_a^{({\rm fs})} = \frac{2\pi Ze^2}{3}\int d^3r\left|\psi_a(\vec{r})\right|^2\int\frac{d^3k}{(2\pi)^3} e^{i\vec{k}\vec{r}}r^2_p.\qquad
\end{eqnarray}
Integration over $d^3k$ yields a $\delta$-function,
\begin{eqnarray}
\label{r3}
\Delta E_a^{(\rm fs)} = \frac{2\pi Ze^2}{3}\int d^3r\left|\psi_a(\vec{r})\right|^2\delta(\vec{r})r^2_p
\end{eqnarray}
and we arrive at the expression
\begin{eqnarray}
\label{r4}
\Delta E_a^{(\rm fs)} = \frac{2\pi Ze^2}{3}\left|\psi_a(0)\right|^2r^2_p,
\end{eqnarray}
which represents the well-known result, i.e. the correction for the finite size of the nucleus.

The same can be achieved within the framework of the quantum mechanics (QM), see \cite{Landau}. True electrostatic potential $\varphi(r)$ obeys the Poisson equation $\vartriangle\varphi(r) = -4\pi \rho(r)$, where $\rho(r)$ represents the nuclear charge distribution. Then, the energy shift of the bound electron in respect to the Coulomb interaction $1/r$ is
\begin{eqnarray}
\label{r7}
\Delta E_a^{({\rm fs})} = -Ze^2 \int d^3r \left|\psi_a(\vec{r})\right|^2\left(\varphi(r)-\frac{1}{r}\right),\qquad
\end{eqnarray}
where the difference of potentials is other than zero in the nucleus volume. Therefore, the squared modulus of the wave function can be taken out at point $\vec{r}=0$.

The following evaluation is performed with the use of relation $\vartriangle r^2 = 6$ and integration by parts:
\begin{eqnarray}
\label{r8}
\Delta E_a^{({\rm fs})} = -\frac{Ze^2}{6}\left|\psi_a(0)\right|^2 \int d^3r\vartriangle r^2\left(\varphi(r)-\frac{1}{r}\right)
\nonumber
\\
= -\frac{Ze^2}{6}\left|\psi_a(0)\right|^2 \int d^3r\, r^2\vartriangle \left(\varphi(r)-\frac{1}{r}\right),\qquad
\end{eqnarray}
Employing the Poisson equation, we find
\begin{eqnarray}
\label{r9}
\Delta E_a^{({\rm fs})}= \frac{2\pi Ze^2}{3}\left|\psi_a(0)\right|^2
\int d^3r\, r^2 \left(\rho(r) - \delta(\vec{r})\right).\qquad
\end{eqnarray}
The second term in Eq. (\ref{r9}) yields zero, and we obtain again
\begin{eqnarray}
\label{r10}
\Delta E_a^{({\rm fs})} = \frac{2\pi  Ze^2}{3}\left|\psi_a(0)\right|^2 r_p^2,
\end{eqnarray}
where the notation $r_p^2=\int d^3r\, r^2 \rho(r)$ for the average value of the squared nuclear charge radius is introduced.

\subsection{The thermal nuclear finite-size correction}
\label{tfs}
In the previous section, we have used the QM approach to evaluate the nuclear finite-size correction to the energy of the bound electron disregarding the thermal effects. Now, we employ the procedure to evaluate the thermal nuclear finite-size correction. The energy shift Eq. (\ref{64}) admits the introduction of the nuclear finite-size correction as
\begin{eqnarray}
\label{200}
\Delta E^\beta_a = \frac{4Ze^2}{\pi}\int d^3r\left|\psi_a(\vec{r})\right|^2
\times
\\
\nonumber
\int\limits_0^\infty d\kappa\, n_\beta(\kappa) \left(\frac{\sin\kappa r}{\kappa r}-1\right)\rho^{\rm ext}(\vec{k}),\qquad
\end{eqnarray}
Making use of the charge distribution Eq. (\ref{r1}) yields
\begin{eqnarray}
\label{201}
\Delta E^\beta_a = \frac{4Ze^2}{\pi}\int d^3r\left|\psi_a(\vec{r})\right|^2\int\limits_0^\infty d\kappa\, n_\beta(\kappa) \left(\frac{\sin\kappa r}{\kappa r}-1\right)
\qquad
\nonumber
\\
-\frac{2Ze^2r_p^2}{3\pi}\int\limits_{V_n} d^3r\left|\psi_a(\vec{r})\right|^2\int\limits_0^\infty d\kappa\, n_\beta(\kappa) \left(\frac{\kappa\sin\kappa r}{ r}-\kappa^2\right).\qquad
\end{eqnarray}
The first term here was examined in section~\ref{exchange}. The second one results in the desired correction, where the space integration is over nuclear volume $V_n$, see \cite{Salpeter-fs}.

Integrating over $\kappa$ in Eq. (\ref{201}), we obtain
\begin{eqnarray}
\label{202}
\Delta E^{\beta, (fs)}_a = -\frac{2Ze^2r_p^2}{3\pi}\left|\psi_a(0)\right|^2
\times\qquad\qquad
\\
\nonumber
\int\limits_{V_n} d^3r\left(-\frac{2\zeta(3)}{\beta^3}+\frac{i}{2\beta^2 r}\left[\psi'\left(1+i\frac{r}{\beta}\right)-\psi'\left(1-i\frac{r}{\beta}\right)\right]\right).
\end{eqnarray}
Here, as before, we took $\left|\psi_a(\vec{r})\right|^2$ at $\vec{r}\rightarrow 0$ out of the integral. The expression can be transformed to
\begin{eqnarray}
\label{203}
\Delta E^{\beta, (fs)}_a = -\frac{2Ze^2r_p^2}{3\pi}\left|\psi_a(0)\right|^2\int\limits_{V_n} d^3r\left(-\frac{2\zeta(3)}{\beta^3}+
\qquad
\right.
\\
\nonumber
\left.
+\frac{1}{2\beta r}\frac{d}{dr}\left[\psi\left(1+i\frac{r}{\beta}\right)+\psi\left(1-i\frac{r}{\beta}\right)\right]\right).
\end{eqnarray}

According to \cite{abram}, the real part $\Re\psi\left(1+iy\right)=\Re\psi\left(1-iy\right)$ and the imaginary part $\Im\psi\left(1+iy\right) = -\Im\psi\left(1-iy\right)$
. Thus,
\begin{eqnarray}
\label{204}
\Delta E^{\beta, (fs)}_a = -\frac{2Ze^2r_p^2}{3\pi}\left|\psi_a(0)\right|^2
\times\qquad\qquad
\\
\nonumber
\int\limits_{V_n} d^3r\left[-\frac{2\zeta(3)}{\beta^3}+\frac{1}{\beta r}\frac{d}{dr}\Re\psi\left(1+i\frac{r}{\beta}\right)\right].
\end{eqnarray}
The real part of the psi (digamma) function, see Eq. (\ref{72}), can be presented in the form \cite{abram}:
\begin{eqnarray}
\label{205}
\Re\psi\left(1+iy\right) = -\gamma+\sum\limits_{n=1}^\infty\frac{y^2}{n(n^2+y^2)},\qquad
\end{eqnarray}
for $(-\infty<y<\infty)$. Then, differentiating with respect to $r$, we have
\begin{eqnarray}
\label{206}
\Delta E^{\beta, (fs)}_a = -\frac{2Ze^2r_p^2}{3\pi}\left|\psi_a(0)\right|^2
\times\qquad\qquad
\\
\nonumber
\int\limits_{V_n} d^3r\left[-\frac{2\zeta(3)}{\beta^3}+2\beta\sum\limits_{n=1}^\infty\frac{n}{(\beta^2 n^2+r^2)^2}\right].
\end{eqnarray}

The last expression can be integrated over angles, which results in factor $4\pi$. Then, expanding in Taylor series to $r\leqslant r_p$, one can obtain
\begin{eqnarray}
\label{207}
\Delta E^{\beta, (fs)}_a = -\frac{8Ze^2r_p^2}{3}\left|\psi_a(0)\right|^2
\times\qquad\qquad
\\
\nonumber
\int\limits_0^{r_p}dr\left[-\frac{2\zeta(3)r^2}{\beta^3}+2\beta\sum\limits_{n=1}^\infty\left(\frac{r^2}{\beta^4n^3}-\frac{2r^4}{\beta^6n^5}+\frac{3r^6}{\beta^8n^7}+\dots\right)\right].
\end{eqnarray}
The summation over $n$ gives the corresponding Riemann zeta functions. Then, the first term vanishes, and the leading thermal nuclear finite-size correction is
\begin{eqnarray}
\label{208}
\Delta E^{\beta, (fs)}_a = \frac{32 Ze^2}{15}\frac{\zeta(5)}{\beta^5}r_p^7\left|\psi_a(0)\right|^2.
\end{eqnarray}
The correction is proportional to $r_p^7$ and is rather negligible.

To verify the correctness of the result (\ref{208}), one can resort to the description given in \cite{Landau}, see Eqs. (\ref{r7})-(\ref{r10}). To this end, however, it is necessary to find the thermal averaged charge distribution, which corresponds to the Poisson equation $\vartriangle\varphi^\beta(r) = -4\pi \rho^\beta(r)$. According to Eq. (\ref{80}):
\begin{eqnarray}
\label{210}
\vartriangle D^{00}_{\beta, {\rm reg}} = \frac{8i}{\pi^2r}\delta(t-t')
\left[\psi'\left(1-i\frac{r}{\beta}\right)-\psi'\left(1+i\frac{r}{\beta}\right)\right],\qquad
\end{eqnarray}
Series expansion of the last expression yields
\begin{eqnarray}
\label{211}
\vartriangle D^{00}_{\beta, {\rm reg}} \approx -\frac{8 \zeta(3)}{\pi\beta^3}+\frac{16 r_p^2 \zeta(5)}{\pi\beta^5}-\frac{24 r_p^4 \zeta(7)}{\pi\beta^7}+\dots
\end{eqnarray}
Then, we should write
\begin{eqnarray}
\label{214}
\Delta E^{\beta, (fs)}_a = \frac{2\pi Ze^2}{3}\left|\psi_a(0)\right|^2\int d^3r\, r^2\left(\rho^\beta(r)-\rho^\beta(0)\right),\,\qquad
\end{eqnarray}
where $\rho^\beta(0)$ represents the result Eq. (\ref{211}) at $r_p=0$. Then, integrating over angles and restricting integration over $r$ with $r_p$, the leading-order correction can be found as
\begin{eqnarray}
\label{215}
\Delta E^{\beta, (fs)}_a = \frac{2 Ze^2}{3}\left|\psi_a(0)\right|^2\int\limits_0^{r_p}dr \frac{16 r^4 r_p^2 \zeta(5)}{\pi\beta^5}
\\
\nonumber
 = \frac{32 Ze^2}{15}\frac{\zeta(5)}{\beta^5}r_p^7\left|\psi_a(0)\right|^2.
\end{eqnarray}
This result coincides precisely with Eq. (\ref{208}).

\subsection{The nuclear finite-size correction to the vacuum polarization effect}
\label{FSVP}
To evaluate the thermal correction to the vacuum polarization effect for the finite size of the nucleus, we should enter the charge distribution (\ref{r1}) into Eq. (\ref{vp.13}):
\begin{eqnarray}
\label{fsvp.1}
\Delta E_a^{{\rm VP},\beta} = -\frac{16 Ze^4}{15\pi m_e^2}\int d^3r\left|\psi_a(\vec{r})\right|^2
\times
\nonumber
\\
\int\limits_0^\infty d\kappa\, \kappa^2 n_\beta(\kappa)\left(\frac{\sin\kappa r}{\kappa r}-1\right)\rho^{\rm ext}(\kappa).
\end{eqnarray}
Then, dropping out the unity, which corresponds to the case considered in section~\ref{TVP}, we obtain
\begin{eqnarray}
\label{fsvp.2}
\Delta E_a^{{\rm FSVP},\beta} = -\frac{16 Ze^4}{15\pi m_e^2}\int\limits_{V_n} d^3r\left|\psi_a(\vec{r})\right|^2
\times
\nonumber
\\
\int\limits_0^\infty d\kappa\, \kappa^2 n_\beta(\kappa)\left(\frac{\sin\kappa r}{\kappa r}-1\right)\left(-\frac{\kappa^2 r_p^2}{6}\right).
\end{eqnarray}

Using series expansion $\sin\kappa r/\kappa r-1\approx -\kappa^2r^2/6$ and taking out $\left|\psi_a(\vec{r})\right|^2$ at $r=0$, one can find
\begin{eqnarray}
\label{fsvp.3}
\Delta E_a^{{\rm FSVP},\beta} = -\frac{4 Ze^4r_p^2}{135\pi m_e^2}\left|\psi_a(0)\right|^2\int\limits_{V_n} d^3r\,r^2
\int\limits_0^\infty d\kappa\, \kappa^6 n_\beta(\kappa).\qquad
\end{eqnarray}
Integration over $d^3r$ yields the factor $\frac{4\pi}{5}r_p^5$, and integration over $\kappa$ results in $720\zeta(7)/\beta^7$. Thus, the thermal nuclear finite-size correction of the leading order to the vacuum polarization is
\begin{eqnarray}
\label{fsvp.4}
\Delta E_a^{{\rm FSVP},\beta} = -\frac{256 Ze^4}{15 m_e^2}\left|\psi_a(0)\right|^2\frac{\zeta(7)}{\beta^7}r_p^7.\qquad
\end{eqnarray}
The correction is $e^2/\beta^2$ times as small as the result Eq. (\ref{215}) and, thus, is completely negligible.

\section{Discussion and applications}
\label{conclusions}

\subsection{Discussion}

In the present paper, we have evaluated the thermal corrections of the lowest order in fine structure constant, $\alpha$, for the non-relativistic atomic system. To this end, we have examined the Feynman diagrams corresponding to the one thermal photon exchange between a bound electron and the nucleus, the one-loop self-energy, the vacuum polarization diagrams, and the thermal corrections for the finite size of the nucleus, which arise within the one thermal photon exchange and vacuum polarization. In all these cases, the thermal correction emerges by the way of the thermal part of the photon propagator, see section~\ref{pp}. In view of very accurate experiments \cite{Mat}, the hydrogen atom is the most attractive system to study thermal effects of this type. Therefore, all final results are given in the non-relativistic limit, which allows evaluating the thermal corrections analytically.

To find thermal corrections, the Hadamard form of thermal photon propagator was introduced in Eq. (\ref{24}). The form is more suitable for the corresponding derivations, since it allows the direct analogy with the 'ordinary' (zeroth vacuum) Feynman photon propagator. We have also shown the form to be equivalent to the thermal photon propagator derived in \cite{Dol,Don}. Unlike the result of \cite{Dol,Don}, the Hadamard representation has no problems with the order of integration \cite{FW} or analytical properties beyond the mass shell, but it leads to a divergence at $|\vec{k}|=0$. The divergence corresponds to the case, when the number of Bose-particles in the condensate state can be arbitrarily large, see \cite{Abr,FG}. For the photons, however, we should take into account that there are no particles with zero momentum; therefore, the divergence can be taken out from the consideration just by subtracting the divergent term. It was noted in \cite{Bel} that the infinite result should be attributed to the zero-point energy of the vacuum, which can be subtracted. Here, we have suggested a regularization procedure, which represents the subtraction of the coincident limit. This results in all derived thermal corrections being convergent.

Moreover, the thermal photon propagator in the form  (\ref{24}) admits simple introduction of gauges. The gauge invariance allows verifying the correctness of found thermal effects. For example, the thermal self-energy correction was evaluated in section~\ref{SE-correction}, where, as distinct from \cite{SLP-QED}, the Hadamard representation of the thermal photon propagator was used. In particular, the application of the coincident limit was shown to lead to the convergent final result. Its real part represents the ac-Stark shift and imaginary part yields the level width induced by the blackbody radiation. As it should be, the results perfectly coincide with \cite{SLP-QED} and are gauge invariant. In addition, the QED description of these thermal effects allows accurate accounting for the finite lifetimes of atomic levels. In the limit of zero level widths, the QED results reduce to the well-known expressions derived within the QM approach. Numerical values for the Stark shift, BBR-induced depopulation rates, and level widths found with the account for finite lifetimes along with the corresponding discussion can be found in section~\ref{SE-correction}.

As the next step, the vacuum polarization effect was described within the QED theory framework, see section~\ref{TVP}. The final gauge-invariant result is given by Eq. (\ref{vp.15}), which leads to Eq. (\ref{vp.16}) for the hydrogen atom. This correction is proportional to the fifth power of the temperature and is of the next order in the fine structure constant. Taking into account that $\beta^{-1}\sim 10^{-4}$ at the room temperature (in atomic units), we can conclude the correction to be negligibly small and lie beyond the accuracy of modern experiments \cite{Mat}. The same conclusion can be made for other corrections: the recoil one, those for finite mass ans size of the nucleus, as well as the combined vertex and self-energy corrections, the thermal photon exchange corrections with vertices more than one. We should note also that the Feynman diagrams corresponding to the thermal photons exchange were evaluated within the framework of the adiabatic $S$-matrix formalism, which confirms the correctness of our results, see section~\ref{ThVC}.

The most attractive result of the present paper emerges for the thermal one-photon exchange diagram. In the zero vacuum case, the diagram (see Fig.~\ref{Fig-4}) corresponds to the Coulomb interaction, which is represented by the zeroth component of the 'ordinary' Feynman propagator. In case of the heated vacuum states, we also found the potential, derived from the Coulomb part of the thermal photon propagator, Eq. (\ref{80}). The potential is gauge invariant and has an asymptotic tending to a constant at large distances (the case of free particles), Eq. (\ref{97}).

\begin{figure}[hbtp]
	\centering
	\includegraphics[scale=0.3]{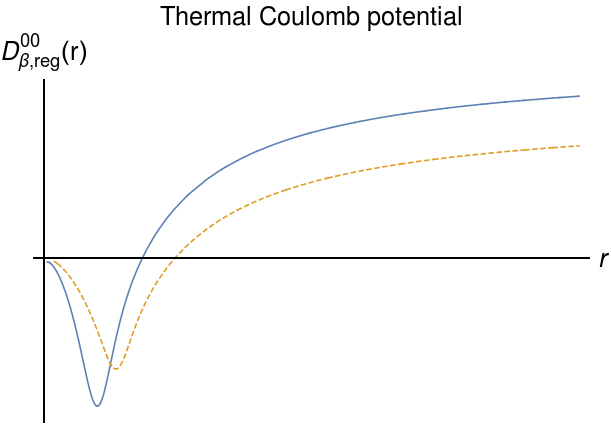}
	\caption{Schematic behavior of the thermal Coulomb potential Eq. (\ref{80}). The bold line corresponds to the potential at a higher temperature as compared to the dashed one.}\label{TP}
\end{figure}

The behavior of the thermal Coulomb potential, Eq. (\ref{80}), is schematically illustrated in Fig.~\ref{TP}. Two graphs correspond to different temperatures, which shows the potential well to become deeper and closer to the nucleus with the growing temperature. The asymptotic behaves as defined by Eq. (\ref{97}) and tends to different constants at different temperatures.

However, the direct evaluation of the potential Eq. (\ref{80}) is possible rather numerically or analytically within the non-relativistic approximation, i.e. when we employ the approximation of small distances (on the order of the Bohr radius). In this way, one can find the thermal correction of the lowest order, see Eq. (\ref{68}). Then, with the use of Schr\"odinger wave functions, one can obtain for the hydrogen atom:
\begin{eqnarray}
\label{dc.1}
\Delta E_a^{\beta} = -\frac{4Ze^2\zeta(3)}{3\pi\beta^3}\frac{n_a^2}{2}\left[5n_a^2+1-3l_a(l_a+1)\right]a_0^2,\qquad
\end{eqnarray}
where $n_a$ is the principal quantum number of the hydrogenic state $a$, $l_a$ is the corresponding orbital momentum and $a_0$ is the Bohr radius. The energy difference for the states with the same principal quantum number is
\begin{eqnarray}
\label{dc.2}
\Delta E_{(n_al_a)-(n_al_a-1)}^{\beta} = \frac{4Ze^2\zeta(3)}{\pi\beta^3}n_a^2l_a.
\end{eqnarray}
As should be expected, the external field removes degeneracy of atomic states, which corresponds in the non-relativistic limit to the dependence of the energy shift on orbital momentum $l_a$.

The parametric estimation of the correction follows from the series expansion of $\sin$ in Eq. (\ref{68}), where we are to  set $\kappa\sim m_e(\alpha Z)^2$ and $r\sim 1/m_e \alpha Z$ in relativistic units. Thus, $\Delta E_a^{\beta}\sim Z\alpha m_e(\alpha Z)^4/\beta^3$ r.u. or $Z^5\alpha^3/\beta^3$ in atomic units. The estimate is of the same order in the fine structure constant as the BBR-induced Stark shift, see Eq. (\ref{cg.11}), but is larger due to the temperature factor. Thus, it can be expected that the thermal correction Eq. (\ref{dc.1}) should cause a more substantial energy shift of atomic levels than the well-known BBR-induced Stark shift.

The spectroscopic hydrogen experiments of the type \cite{Mat}, \cite{H-exp} are carried out at cryogenic temperatures (understood as the boiling point of oxygen (about $90$ K) and lower). The numerical values of energy shift Eq. (\ref{dc.1}) at the nitrogen temperature $77$ K for different hydrogenic $n_al_a$ states are listed in Table~\ref{tab:4}. Table~\ref{tab:5} shows the values of the thermal energy shift for the difference of $2p-2s$ states in the hydrogen atom at different temperatures. The thermal energy shifts Eq. (\ref{dc.1}) for the $1s$ and $2s$ states in hydrogen at different temperatures are given in Table~\ref{tab:5} in $s^{-1}$.

\begin{table}
\caption{Numerical values of the energy shift Eq. (\ref{dc.1}). The first column lists the principal quantum number $n_a$ and orbital momentum $l_a$. The second and third columns give the energy shift values (\ref{dc.1}) at the temperature $77$ K in atomic units and $s^{-1}$, respectively.}
\label{tab:4}
\begin{tabular}{|| l | c | c ||}
\hline \hline
$n_a, l_a$ &$T=77$ K,  $\Delta E_a $ in a.u. &$\Delta E_a $ in $s^{-1}$ \\ \hline
  1,0 ($1s$ state) &  $-8.6232\cdot 10^{-18}$ &  $-0.356$  \\
  2,0 ($2s$ state) &  $-1.2072\cdot 10^{-16}$ &  $-4.991$ \\
  2,1 ($2p$ state) &  $-8.6232\cdot 10^{-17}$ &  $-3.565$ \\
  3,0 ($3s$ state) &  $-5.95002\cdot 10^{-16}$ &  $-24.598$ \\
  3,1 ($3p$ state) &  $-5.1739\cdot 10^{-16}$ &  $-21.389$ \\
  3,2 ($3d$ state) &  $-3.6217\cdot 10^{-16}$ &  $-14.973$ \\
  4,0 ($4s$ state) &  $-1.8626\cdot 10^{-15}$ &  $-77.002$ \\
  4,1 ($4p$ state) &  $-1.7246\cdot 10^{-15}$ &  $-71.298$ \\
  4,2 ($4d$ state) &  $-1.4487\cdot 10^{-15}$ &  $-59.891$ \\
  4,3 ($4f$ state) &  $-1.0348\cdot 10^{-15}$  &  $-42.779$ \\
\hline \hline
\end{tabular}
\end{table}
\begin{table}
\caption{Numerical values of Eqs. (\ref{dc.1}), (\ref{dc.2}) in the hydrogen atom at different temperatures. The first column lists the temperatures in Kelvins. The second and third columns give the energy shift values (\ref{dc.1}) in atomic units and $s^{-1}$, respectively. Numerical values of Eq. (\ref{dc.1}) at different temperatures are given in the last two columns.}
\label{tab:5}
\begin{tabular}{|| l | c | c | c | c ||}
\hline \hline
$T$, K & $\Delta E_{2p-2s}^{\beta(1)}$ in a.u. & $\Delta E_{2p-2s}^{\beta}$, $s^{-1}$ & $\Delta E_{1s}^{\beta}$, $s^{-1}$ & $\Delta E_{2s}^{\beta}$, $s^{-1}$ \\ \hline
  300 & $2.0399\cdot 10^{-15}$ & $84.334$ & $-21.08$ & $-295.17$\\
 270 &  $1.4871\cdot 10^{-15}$ & $61.479$ & $-15.37$ & $-215.18$\\
 77 &  $3.4493\cdot 10^{-17}$ & $1.426$ & $-0.356$ & $-4.991$\\
 4 &  $4.8354\cdot 10^{-21}$ & $1.999\cdot 10^{-4}$  & $-4.998\cdot 10^{-5}$ & $-6.997\cdot 10^{-4}$ \\
\hline \hline
\end{tabular}
\end{table}

At the nitrogen temperature, the thermal correction Eq. (\ref{dc.1}) is twice as small, see Table~\ref{tab:4}, as the declared experimental error ($10$ $Hz$) \cite{Mat} for the $2s-1s$ transition frequency. Nonetheless, the correction at the room temperature (see Table~\ref{tab:5}) is about $300$ $s^{-1}$ for the $2s$ state in hydrogen. Thus, the thermal correction Eq. (\ref{dc.1}) can be measured in modern experiments at varying temperature and with the use, for example, of the thermal radiation shield \cite{Beloy}. The relative energy shift at different temperatures is $\delta\left(\Delta E_a^{\beta}\right)/\Delta E_a^{\beta} = 3\delta T/T_0$ ($\delta T = T-T_0$ and $T_0$ is some fixed temperature), whereas the Stark shift will give the ratio $4\delta T/T_0$ at low temperatures, i.e. these shifts can be observed as straight lines with different slopes.

We should note here that the thermal correction Eq. (\ref{dc.1}) has no significant influence on the muonic hydrogen experiments \cite{Pohl}, \cite{Anton}. The corresponding estimate can be given as $m_\mu \Delta E_a^{\beta}$, where muon mass $m_\mu\approx 207 m_e$. At the room temperature, it is about $1.22\times 10^{-12}$ $eV$, which is negligibly small. Along with this, the nuclear finite-size correction Eq. (\ref{215}) is proportional to $m_\mu^3$, but is suppressed by the temperature factor $\beta^{-5}\sim 10^{-20}$ at the room temperature. The same conclusion ensues for all other thermal corrections described in the paper.

\subsection{The application to the determination of Rydberg constant and proton's charge radius}
\label{Rc-pr}

As a most intriguing corollary, one can estimate the effect of the thermal correction on determination of the Rydberg constant and proton radius. The root-mean-square charge radius, $r_p$, has been found by electronic hydrogen spectroscopic experiments and the bound-state QED calculations \cite{Eides} with 1 per cent uncertainty. The present value quoted by CODATA is $r_p=0.8775(51)$ fm \cite{Mohr-2012}. At the same time, the muonic hydrogen experiment produces the value $r_p=0.84184(67)$ fm \cite{Pohl}. The difference of these values is still under discussion today. However, quite recently, a new value $0.8335(91)$ fm of the proton radius derived from the electronic hydrogen experiments was reported in \cite{H-exp}. The deviation of $r_p=0.8758(77)$ and $r_p=0.8335(91)$ fm was explained by the interfering transitions between $2s$ and $4p$ states in the hydrogen atom. The corresponding theory can be found in \cite{LKG} and subsequent works \cite{LSPS,PRL-LSSP}, where the nonresonant corrections caused by such interference were introduced.

To determine the Rydberg constant and the proton radius, we employ the expression for the energy levels of a hydrogen atom
\begin{eqnarray}
\label{dc.3}
E_{nlj} = R_\infty\left(-\frac{1}{n^2}+f_{nlj}\left(\alpha,\frac{m_e}{m_p},r_p\dots\right)\right),
\end{eqnarray}
where $n$, $l$ and $j$ are the principal, orbital and total angular momentum quantum numbers, respectively; $R_\infty=m_e \alpha^2 c/2h$ is the Rydberg constant ($c$ is the speed of light and $h$ is Planck's constant), $m_e$ and $m_p$ designate the electron and proton masses. Function $f_{nlj}$ denotes all the possible corrections arising within the relativistic QED theory, see \cite{Mohr-2016}.

According to (\ref{dc.3}), we can write $E_{nlj}-E_{n'l'j'}=\Delta E^{\rm exp}_{nlj-n'l'j'}$, where we compare the theoretical result (the left-hand side) with the experimental one, which can be found in \cite{Mohr-2016}. Solving two equations for two independent transitions produces the values of $R_\infty$ and $r_p$. Our calculations used the values of fine structure constant and speed of light $\alpha = 1/137.035999136$ and $c = 299792458$ $m/s$. The conversion factor from atomic to SI units for energies (\ref{dc.3}) is $2R_\infty c/\alpha^2$. To evaluate the root-mean-squared value, the definition $x^{\rm rms}=\left(\sum\limits_{i=1}^N x_i^2/N\right)^{1/2}$ was employed. At zero temperature, the rms deviation was defined as the value of the transition frequency $\pm$ experimental uncertainty. The standard deviation of the overall result was found with the use of $\Delta x^{\rm rms}=\left(\sum\limits_{i=1}^N (x_i-\bar{x})^2/(N(N-1))\right)^{1/2}$, where $\bar{x}$ is defined by means of the arithmetic average. The results of numerical calculations for different transitions and temperatures are listed in Table~\ref{tab:6}.

\begin{widetext}
\begin{center}
\begin{longtable}{|| l | l | l | l | l | l | l ||}
\caption{The values of Rydberg constant $R_\infty$ and proton radius $r_p$, defined using Eq. (\ref{dc.3}), with allowance for the thermal correction Eq. (\ref{dc.1}) in the hydrogen atom. The second and the third columns list the temperatures in Kelvins and the used transitions, respectively. The values in parentheses show the root-mean-squared deviations in last digits, defined via the experimental inaccuracy.}\label{tab:6}
\\
\hline
 &$T$, K & Transitions & $R_\infty$ in $m^{-1}$ & $r_p$ in fm \\ \hline
\multirow{4}{*}{1$^a$}&0 & $2s_{1/2}-2p_{1/2}$, $1s_{1/2}-2s_{1/2}$& $10973731.5685501(2804)$ & $0.879224(26185)$ \\

&300 & $2s_{1/2}-2p_{1/2}$, $1s_{1/2}-2s_{1/2}$ & $10973731.568549$ & $0.879185$ \\

&77 & $2s_{1/2}-2p_{1/2}$, $1s_{1/2}-2s_{1/2}$ & $10973731.568550$ & $0.879223$ \\

&35 & $2s_{1/2}-2p_{1/2}$, $1s_{1/2}-2s_{1/2}$ & $10973731.568550$ & $0.879224$  \\

&5.8$^b$ & $2s_{1/2}-2p_{1/2}$, $1s_{1/2}-2s_{1/2}$ & $10973731.568550$ & $0.879224$  \\
	\hline
	\hline

\multirow{4}{*}{2$^a$}&0 & $2s_{1/2}-2p_{1/2}$, $1s_{1/2}-3s_{1/2}$& $10973731.5685701(3091)$ & $0.879224(26185)$ \\

&300 & $2s_{1/2}-2p_{1/2}$, $1s_{1/2}-2s_{1/2}$ & $10973731.568569$ & $0.879185$ \\

&77 & $2s_{1/2}-2p_{1/2}$, $1s_{1/2}-2s_{1/2}$ & $10973731.568570$ & $0.879223$ \\

&35 & $2s_{1/2}-2p_{1/2}$, $1s_{1/2}-2s_{1/2}$ & $10973731.568570$ & $0.879224$ \\

&5.8 & $2s_{1/2}-2p_{1/2}$, $1s_{1/2}-2s_{1/2}$ & $10973731.568570$ & $0.879224$ \\
	\hline
	\hline

\multirow{4}{*}{3}&0 &$2s_{1/2}-2p_{3/2}$, $1s_{1/2}-2s_{1/2}$& $10973731.5684420(3737)$ & $0.869069(35322)$ \\

&300 & $2s_{1/2}-2p_{3/2}$, $1s_{1/2}-2s_{1/2}$ & $10973731.568441$ & $0.869030$ \\

&77 & $2s_{1/2}-2p_{3/2}$, $1s_{1/2}-2s_{1/2}$ & $10973731.568442$ & $0.869069$ \\

&35 & $2s_{1/2}-2p_{3/2}$, $1s_{1/2}-2s_{1/2}$ & $10973731.568442$ & $0.869069$ \\

&5.8 & $2s_{1/2}-2p_{3/2}$, $1s_{1/2}-2s_{1/2}$ & $10973731.568442$ & $0.869069$ \\

	\hline
	\hline

\multirow{4}{*}{4}&0 &$2s_{1/2}-2p_{3/2}$, $1s_{1/2}-3s_{1/2}$& $10973731.5684698(2983)$ & $0.869069(35322)$ \\

&300 & $2s_{1/2}-2p_{3/2}$, $1s_{1/2}-3s_{1/2}$ & $10973731.568468$ & $0.869030$ \\

&77 & $2s_{1/2}-2p_{3/2}$, $1s_{1/2}-3s_{1/2}$ & $10973731.568469$ & $0.869069$ \\

&35 & $2s_{1/2}-2p_{3/2}$, $1s_{1/2}-3s_{1/2}$ & $10973731.568469$ & $0.869069$ \\

&5.8 & $2s_{1/2}-2p_{3/2}$, $1s_{1/2}-3s_{1/2}$ & $10973731.568469$ & $0.869069$ \\

	\hline
	\hline

\multirow{4}{*}{5}&0 &$2s_{1/2}-8s_{1/2}$, $1s_{1/2}-2s_{1/2}$& $10973731.5684174(2226)$ & $0.866737(21089)$ \\

&300 & $2s_{1/2}-8s_{1/2}$, $1s_{1/2}-2s_{1/2}$ & $10973731.568121$ & $0.838254$ \\

&77 & $2s_{1/2}-8s_{1/2}$, $1s_{1/2}-2s_{1/2}$ & $10973731.568412$ & $0.866284$ \\

&35 & $2s_{1/2}-8s_{1/2}$, $1s_{1/2}-2s_{1/2}$ & $10973731.568417$ & $0.866714$ \\

&5.8 & $2s_{1/2}-8s_{1/2}$, $1s_{1/2}-2s_{1/2}$ & $10973731.568417$ & $0.866758$ \\

	\hline
	\hline

\multirow{4}{*}{6}&0 &$2s_{1/2}-8s_{1/2}$, $1s_{1/2}-3s_{1/2}$& $10973731.5683873(1917)$ & $0.860376(14689)$ \\

&300 & $2s_{1/2}-8s_{1/2}$, $1s_{1/2}-3s_{1/2}$ & $10973731.568072$ & $0.817419$ \\

&77 & $2s_{1/2}-8s_{1/2}$, $1s_{1/2}-3s_{1/2}$ & $10973731.568382$ & $0.859830$ \\

&35 & $2s_{1/2}-8s_{1/2}$, $1s_{1/2}-3s_{1/2}$ & $10973731.568387$ & $0.860325$ \\

&5.8 & $2s_{1/2}-8s_{1/2}$, $1s_{1/2}-3s_{1/2}$ & $10973731.568387$ & $0.860376$ \\

	\hline
	\hline

\multirow{4}{*}{7}&0 &$2s_{1/2}-4p$, $1s_{1/2}-2s_{1/2}$& $10973731.5680753(955)$ & $0.833701(9399)$ \\

&300 & $2s_{1/2}-4p$, $1s_{1/2}-2s_{1/2}$ & $10973731.568050$ & $0.831188$ \\

&77 & $2s_{1/2}-4p$, $1s_{1/2}-2s_{1/2}$ & $10973731.568075$ & $0.833658$ \\

&35 & $2s_{1/2}-4p$, $1s_{1/2}-2s_{1/2}$ & $10973731.568075$ & $0.833697$ \\

&5.8 & $2s_{1/2}-4p$, $1s_{1/2}-2s_{1/2}$ & $10973731.568075$ & $0.833701$ \\

	\hline
	\hline

\multirow{4}{*}{8}&0 &$2s_{1/2}-4p$, $1s_{1/2}-3s_{1/2}$& $10973731.5679848(283)$ & $0.817951(2213)$ \\

&300 & $2s_{1/2}-4p$, $1s_{1/2}-3s_{1/2}$ & $10973731.567957$ & $0.815018$ \\

&77 & $2s_{1/2}-4p$, $1s_{1/2}-3s_{1/2}$ & $10973731.567984$ & $0.817902$ \\

&35 & $2s_{1/2}-4p$, $1s_{1/2}-3s_{1/2}$ & $10973731.567985$ & $0.817946$ \\

&5.8 & $2s_{1/2}-4p$, $1s_{1/2}-3s_{1/2}$ & $10973731.567985$ & $0.817951$ \\

	\hline
	\hline

\multirow{4}{*}{9}&0 &$2s_{1/2}-8d_{3/2}$, $1s_{1/2}-2s_{1/2}$& $10973731.5685481(2177)$ & $0.879045(20348)$ \\

&300 & $2s_{1/2}-8d_{3/2}$, $1s_{1/2}-2s_{1/2}$ & $10973731.568264$ & $0.852088$ \\

&77 & $2s_{1/2}-8d_{3/2}$, $1s_{1/2}-2s_{1/2}$ & $10973731.568543$ & $0.878596$ \\

&35 & $2s_{1/2}-8d_{3/2}$, $1s_{1/2}-2s_{1/2}$ & $10973731.568548$ & $0.879003$ \\

&5.8 & $2s_{1/2}-8d_{3/2}$, $1s_{1/2}-2s_{1/2}$ & $10973731.568548$ & $0.879045$ \\

	\hline
	\hline

\multirow{4}{*}{10}&0 &$2s_{1/2}-8d_{3/2}$, $1s_{1/2}-3s_{1/2}$& $10973731.5685285(3157)$ & $0.875030(40541)$ \\

&300 & $2s_{1/2}-8d_{3/2}$, $1s_{1/2}-3s_{1/2}$ & $10973731.568225$ & $0.843903$ \\

&77 & $2s_{1/2}-8d_{3/2}$, $1s_{1/2}-3s_{1/2}$ & $10973731.568523$ & $0.874513$ \\

&35 & $2s_{1/2}-8d_{3/2}$, $1s_{1/2}-3s_{1/2}$ & $10973731.568528$ & $0.874981$ \\

&5.8 & $2s_{1/2}-8d_{3/2}$, $1s_{1/2}-3s_{1/2}$ & $10973731.568528$ & $0.875030$ \\

	\hline
	\hline

\multirow{4}{*}{11}&0 &$2s_{1/2}-8d_{5/2}$, $1s_{1/2}-2s_{1/2}$& $10973731.5686806(1678)$ & $0.891334(1546)$ \\

&300 & $2s_{1/2}-8d_{5/2}$, $1s_{1/2}-2s_{1/2}$ & $10973731.568396$ & $0.864759$ \\

&77 & $2s_{1/2}-8d_{5/2}$, $1s_{1/2}-2s_{1/2}$ & $10973731.568676$ & $0.890891$ \\

&35 & $2s_{1/2}-8d_{5/2}$, $1s_{1/2}-2s_{1/2}$ & $10973731.568680$ & $0.891292$ \\

&5.8 & $2s_{1/2}-8d_{5/2}$, $1s_{1/2}-2s_{1/2}$ & $10973731.568681$ & $0.891334$ \\

	\hline
	\hline

\multirow{4}{*}{12}&0 &$2s_{1/2}-8d_{5/2}$, $1s_{1/2}-3s_{1/2}$& $10973731.5686702(1322)$ & $0.889229(8291)$ \\

&300 & $2s_{1/2}-8d_{5/2}$, $1s_{1/2}-3s_{1/2}$ & $10973731.568367$ & $0.858617$ \\

&77 & $2s_{1/2}-8d_{5/2}$, $1s_{1/2}-3s_{1/2}$ & $10973731.568665$ & $0.888720$ \\

&35 & $2s_{1/2}-8d_{5/2}$, $1s_{1/2}-3s_{1/2}$ & $10973731.568669$ & $0.889182$ \\

&5.8 & $2s_{1/2}-8d_{5/2}$, $1s_{1/2}-3s_{1/2}$ & $10973731.568670$ & $0.889229$ \\

	\hline
	\hline

\multirow{4}{*}{13}&0 &$2s_{1/2}-12d_{3/2}$, $1s_{1/2}-2s_{1/2}$& $10973731.5682973(2038)$ & $0.855289(19577)$ \\

&300 & $2s_{1/2}-12d_{3/2}$, $1s_{1/2}-2s_{1/2}$ & $10973731.566908$ & $0.709526$ \\

&77 & $2s_{1/2}-12d_{3/2}$, $1s_{1/2}-2s_{1/2}$ & $10973731.568274$ & $0.853031$ \\

&35 & $2s_{1/2}-12d_{3/2}$, $1s_{1/2}-2s_{1/2}$ & $10973731.568295$ & $0.855078$ \\

&5.8 & $2s_{1/2}-12d_{3/2}$, $1s_{1/2}-2s_{1/2}$ & $10973731.568295$ & $0.855288$ \\

	\hline
	\hline

\multirow{4}{*}{14}&0 &$2s_{1/2}-12d_{3/2}$, $1s_{1/2}-3s_{1/2}$& $10973731.5682628(2611)$ & $0.847749(32329)$ \\

&300 & $2s_{1/2}-12d_{3/2}$, $1s_{1/2}-3s_{1/2}$ & $10973731.566785$ & $0.676319$ \\

&77 & $2s_{1/2}-12d_{3/2}$, $1s_{1/2}-3s_{1/2}$ & $10973731.568238$ & $0.845139$ \\

&35 & $2s_{1/2}-12d_{3/2}$, $1s_{1/2}-3s_{1/2}$ & $10973731.568260$ & $0.847505$ \\

&5.8 & $2s_{1/2}-12d_{3/2}$, $1s_{1/2}-3s_{1/2}$ & $10973731.568263$ & $0.847748$ \\

	\hline
	\hline

\multirow{4}{*}{15}&0 &$2s_{1/2}-12d_{5/2}$, $1s_{1/2}-2s_{1/2}$& $10973731.5683920(1719)$ & $0.864333(16339)$ \\

&300 & $2s_{1/2}-12d_{5/2}$, $1s_{1/2}-2s_{1/2}$ & $10973731.567003$ & $0.720402$ \\

&77 & $2s_{1/2}-12d_{5/2}$, $1s_{1/2}-2s_{1/2}$ & $10973731.568368$ & $0.862099$ \\

&35 & $2s_{1/2}-12d_{5/2}$, $1s_{1/2}-2s_{1/2}$ & $10973731.568389$ & $0.864124$ \\

&5.8 & $2s_{1/2}-12d_{5/2}$, $1s_{1/2}-2s_{1/2}$ & $10973731.568392$ & $0.864332$ \\

	\hline
	\hline

\multirow{4}{*}{16}&0 &$2s_{1/2}-12d_{5/2}$, $1s_{1/2}-3s_{1/2}$& $10973731.5683636(2271)$ & $0.858201(28433)$ \\

&300 & $2s_{1/2}-12d_{5/2}$, $1s_{1/2}-3s_{1/2}$ & $10973731.566886$ & $0.689374$ \\

&77 & $2s_{1/2}-12d_{5/2}$, $1s_{1/2}-3s_{1/2}$ & $10973731.568339$ & $0.855623$ \\

&35 & $2s_{1/2}-12d_{5/2}$, $1s_{1/2}-3s_{1/2}$ & $10973731.568361$ & $0.857959$ \\

&5.8 & $2s_{1/2}-12d_{5/2}$, $1s_{1/2}-3s_{1/2}$ & $10973731.568364$ & $0.858199$ \\

	\hline
	\hline

\multirow{4}{*}{rms}&0 &  & $10973731.568415(235)(61)$ & $0.8649(240)(62)$ \\

&300 &  & $10973731.567941(162)$ & $0.8163(179)$ \\

&77 &  & $10973731.568407(50)$ & $0.8640(49)$ \\

&35 &  & $10973731.568414(48)$ & $0.8650(49)$ \\

 &5.8 & & $10973731.568415(48)$ & $0.8651(49)$ \\

	\hline
	\hline












\end{longtable}
\end{center}
\end{widetext}

$^a$The Lamb shift value measured in Harvard laboratory and experimental inaccuracy $9$ kHz for determination of the rms deviation in the first subrow were used \cite{Mohr-2016}. $^b$The nozzle temperature denoted in \cite{H-exp}.

We should note that the thermal correction Eq. (\ref{dc.1}) increases with the growing principal quantum number and becomes very noticeable for highly excited states. In case of $8s$ state in hydrogen, the correction is $-11489.4$ Hz at the room temperature $300$ K and is substantially larger than the experimental uncertainty $8.6$ kHz. For the $4p$ state, it is $-671.107$ Hz at the same room temperature, whereas the experimental uncertainty of $2s-4p$ frequency measurement is about $2.3$ kHz. Thus, we can expect a strong impact of temperature environment on the determination of Rydberg constant and proton's charge radius. This fact can be easily seen from the results in Table~\ref{tab:6}.

The last row of Table~\ref{tab:6} gives the rms values of the Rydberg constant and proton's charge radius along with the corresponding standard deviations. The value in the first parentheses for the Rydberg constant at zero temperature represents the result of deviation defined via the experimental error. All results were found by the 'straightforward' averaging of data in the table. It is obvious, however, that some results fall out. For example, the proton radius value for the transition pair $2s_{1/2}-12d_{5/2}$ and $1s_{1/2}-3s_{1/2}$ (see row 16) is $0.689374$ fm. It means the only thing: formulas for such transitions should not take into account the thermal correction at the room temperature. Nonetheless, since the blackbody radiation is poorly shielded, one can assume the thermal radiation to have an effect on the corresponding transition frequency measurements depending on the external conditions of the experiment.
Then, combining the results in Table~\ref{tab:6}, which are close to the data of $\mu H$ experiment, 1-6 at $300$ K, 7,8 at $5.8$ K, 9-12 at $300$ K and 13-16 at $77$ K, one can find
\begin{eqnarray}
\label{dc.5}
R_\infty&=&10973731.568296(43)\,m^{-1},
\\
\nonumber
r_p&=&0.8526(47)\, fm.
\end{eqnarray}

Moreover, it was demonstrated in \cite{H-exp} that the experimental accuracy in $eH$ measurements have reached the level, when the nonresonant effects caused by the interfering transitions become important. The most noticeable effect is observed for closely lying states with allowance for the fine or hyperfine level structure, see  \cite{PRL-LSSP}. In this case, the process-dependent corrections lead to the line profile asymmetry, which does not allow an accurate determination of the line profile maximum, see \cite{LSPS}. Discarding transitions $2s-2p_{1/2}$ and $2s-2p_{3/2}$ from the analysis, we get
\begin{eqnarray}
\label{dc.6}
R_\infty&=&10973731.568227(44)\,m^{-1},
\\
\nonumber
r_p&=&0.8453(51)\, fm,
\end{eqnarray}
which is in perfect agreement ($0.5\%$) with the $\mu H$ result for the proton's charge radius.

Completing this part of work, one can conclude that experiments with the hydrogen atom should take into account not only the nonresonant effects, but also the influence of the thermal radiation. It has been found that the Stark effect induced by the blackbody radiation does not represent the dominant contributor to the energy shift of atomic levels. In turn, correction Eq. (\ref{dc.1}) leads to the dominant effect of the thermal field on the atomic level energies. This becomes most significant for highly excited states in the hydrogen atom. An accurate analysis of thermal correction Eq. (\ref{dc.1}) is required for each measured transition. In turn, the muonic hydrogen experiment is not subjected to the effect.

\section{Conclusions}

This work presents the thermal QED theory for bound states. To this end, the photon propagator was derived when the vacuum state was replaced with the occupied (heated) one. The treatment of the heated vacuum was restricted to the photon part only, since the fermion part is here suppressed exponentially. In particular, the Feynman photon propagator was shown to contain the thermal part, which is given by the Hadamard propagation function. In turn, this allows finding an alternative form of the thermal photon propagator, which can of course be reduced to the well-known result. However, the new form presented by means of the contour integration in $k_0$ plane admits a simple introduction of gauges. The gauge invariance serves as a tool to validate results.

As opposed to the thermal photon propagator derived in \cite{Dol,Don}, the new form allows also a simple consideration of the static limit for the external charge current and, hence, the description of the thermal interaction for two charged particles. However, as in case of \cite{DHR}, there are divergences in various radiative corrections including the lowest order (one-photon exchange correction). All divergences arise from the Planck's distribution function included in the thermal photon propagator and are infrared. The renormalization procedure of such divergences was described in \cite{DHR} within the QED theory for free particles.

The paper examined the thermal correction of the lowest order within the QED theory for bound states. The correction corresponds to the one thermal photon exchange diagram and is divergent. To renormalize the correction, a new procedure was proposed. In particular, it was shown first that the infrared divergence for bound states is canceled precisely by the contribution of free particles, which is state-independent. The subtraction of the thermal contribution for free particles corresponds to the renormalization by the 'unphysical' (unmeasurable) contribution, which represents the counter-term. Then, the same result was demonstrated to arise, where the regularization procedure of the thermal photon propagator by the way of the coincidence limit was used. The final result for the lowest order thermal correction is convergent and gauge invariant.

Further on, the regularization procedure of thermal photon propagator was extended onto the thermal radiative corrections of the next orders. The gauge invariance of the results was verified. Using the thermal self-energy correction as an example, it was established that the regularization procedure proposed in the work produces a physical result. The latter bears a simple analogy to the QM results and the 'ordinary' zero vacuum QED case, when the real part of the one-loop self-energy correction represents the energy shift and the imaginary parts corresponds to the lifetime of atomic level, see \cite{LabKlim}.

The most important result of the work is the derivation of the thermal Coulomb potential. The potential follows immediately from the Coulomb part of the thermal photon propagator with the use of the regularization procedure proposed in the paper, although the subtraction of the counter-term for free particles could be used equally well. The potential is gauge invariant and can be regarded as the Bose potential induced by the thermal Coulomb photons. Two asymptotics are found for the potential: at small and large distances (temperatures). The former corresponds to the bound states and laboratory experiments with atomic systems. Then, the thermal correction of the lowest order (the thermal one-photon exchange) follows immediately from the corresponding Taylor series.

As an application of the found thermal correction, the problem of finding the Rydberg constant and proton radius was discussed. It was demonstrated that the thermal effects can lead to a substantial contribution in measurements of transition frequencies. In the work, such influence was taken into account for the 'directly' measured transitions only, see Table~\ref{tab:6}. Finally, the perfect agreement of the proton's charge radius values defined from the $eH$ and $\mu H$ spectroscopic experiments can be established.

In addition, the thermal effects described in the work can be important in studies of highly-charged ions and atomic clocks, the research of time variation of fundamental constants in laboratory and astrophysical experiments. Perhaps, the thermal Coulomb potential will find application in other fields of physics.

\section*{Acknowledgements}
The author is grateful to Prof. L.N. Labzowsky, Dr. G. Plunien, Dr. A. Volotka, Dr. D. Glazov, and Dr. O. Andreev for helpful discussions in private communications.

\bibliography{mybibfile}
\end{document}